\documentclass[12pt]{book}
\usepackage{amsfonts,amsmath,amssymb,amsthm}
\usepackage{graphicx}
\usepackage[top=3em,bottom=3em,rmargin=3cm,lmargin=3cm]{geometry}
\usepackage{fancyhdr}
\usepackage{ebgaramond}
\usepackage[T1]{fontenc}
\usepackage{lmodern}
\usepackage{bigints}
\usepackage{float}
\usepackage{graphicx}
\usepackage[percent]{overpic}
\usepackage{color}

\fancypagestyle{plain}{%
}
\usepackage[colorlinks=true,bookmarks=false]{hyperref}
\definecolor{myurlcolor}{rgb}{0,0,0.8}
\definecolor{mycitecolor}{rgb}{0,0,0.8}
\definecolor{myrefcolor}{rgb}{0,0,0.8}
\hypersetup{linkcolor=myrefcolor,citecolor=mycitecolor,urlcolor=myurlcolor}
\usepackage{blindtext}

\usepackage[dvipsnames]{xcolor}
\definecolor{lightblue}{rgb}{0.5,0.5,0.8}
\definecolor{darkblue}{rgb}{0,0,0.8}
\definecolor{darkgreen}{rgb}{0,0.7,0}
\definecolor{darkpurple}{rgb}{0.4,0,0.5}
\definecolor{brickred}{rgb}{0.7,0,0}
\definecolor{darkred}{rgb}{0.5,0,0}
\definecolor{verydarkred}{rgb}{0.4,0,0}

\newcommand{\define}[1]{{\bf \color{darkred} \boldmath{#1}}}

\usepackage{mfirstuc}
\newcommand\entry[1]{\begin{center}
\define{\MakeUppercase{#1}}\phantomsection\addcontentsline{toc}{section}{\capitalisewords{#1}}
\end{center} \vskip 1em}

\usepackage{tikz}
\usetikzlibrary{matrix,cd,arrows,decorations.pathmorphing,positioning}
\usetikzlibrary{intersections,decorations.markings}
\usetikzlibrary{arrows,positioning,fit,matrix,shapes.geometric,external}
\usepackage{array,mathtools}
\usepackage{pgfplots}

\newcolumntype{B}[1]{r*{#1}{@{\,}r}}
\tikzstyle{empty}=[circle,fill=none, draw=none]
\pgfdeclarelayer{edgelayer}
\pgfdeclarelayer{nodelayer}
\pgfsetlayers{edgelayer,nodelayer,main}
 
\newcommand\R{{\mathbb R}}
\newcommand{\maps}{\colon}

\theoremstyle{definition}
\newtheorem{puzzle}{Puzzle}

\begin{document}

\pagestyle{empty}

  \begin{center}
\includegraphics[width = 25 em]{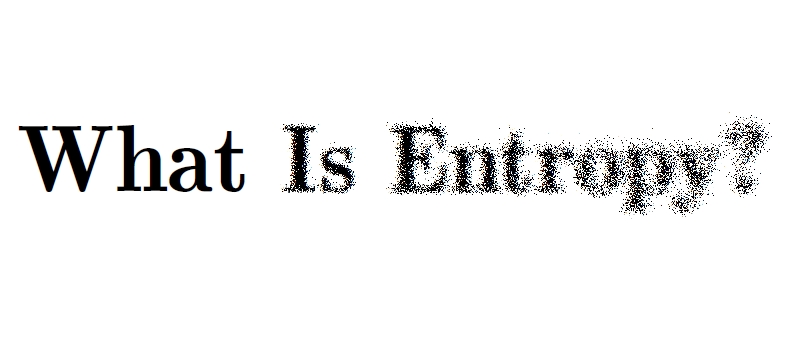}
 \\[20em]
  {\textbf{John C.\ Baez \\
  2024}}
  \end{center}
  
\newpage

\vskip 20em
\begin{center}
\href{https://creativecommons.org/licenses/by-sa/4.0/}{
\includegraphics{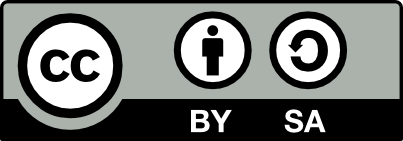} \break
Attribution-ShareAlike 4.0 International}
\end{center}

You are free to:

\begin{itemize}
\item   
Share --- copy and redistribute the material in any medium or format for any purpose, even commercially.
 \item
 Adapt --- remix, transform, and build upon the material for any purpose, even commercially.  The licensor cannot revoke these freedoms as long as you follow the license terms.
\end{itemize}

under the following terms:

\begin{itemize}
\item
Attribution --- You must give appropriate credit, provide a link to the license, and indicate if changes were made.  You may do so in any reasonable manner, but not in any way that suggests the licensor endorses you or your use.
\item
ShareAlike --- If you remix, transform, or build upon the material, you must distribute your contributions under the same license as the original.
\item
No additional restrictions --- You may not apply legal terms or technological measures that legally restrict others from doing anything the license permits.
\end{itemize}

\newpage

\vbox{\vskip 20 em}
  \begin{center} 
  \textbf{To my wife, Lisa Raphals }
  \end{center}
\newpage

  \begin{center}
  \textbf{Foreword}
  \end{center}

Once there was a thing called Twitter, where people exchanged short messages called `tweets'.    While it had its flaws, I came to like it and eventually decided to teach a short course on entropy  in the form of tweets.    This little book is a slightly expanded version of that course.  

It's easy to wax poetic about entropy, but what is it?   I claim it's the amount of information we don't know about a situation, which in principle we could learn.     But how can we make this idea precise and quantitative?  To focus the discussion I  decided to tackle a specific puzzle: why does hydrogen gas at room temperature and pressure have an entropy corresponding to about 23 unknown bits of information per molecule?   This gave me an excuse to explain these subjects:

\begin{itemize}
\item information
\item Shannon entropy and Gibbs entropy
\item the principle of maximum entropy
\item the Boltzmann distribution
\item temperature and coolness
\item the relation between entropy, expected energy and temperature
\item the equipartition theorem
\item the partition function
\item the relation between entropy, free energy and expected energy
\item the entropy of a classical harmonic oscillator
\item the entropy of a classical particle in a box
\item the entropy of a classical ideal gas.
\end{itemize}

I have largely avoided the second law of thermodynamics, which says that entropy always increases.  While fascinating, this is so problematic that a good explanation would require another book!   I have also avoided the role of entropy in biology, black hole physics, etc.   Thus, the aspects of entropy most beloved by physics popularizers will not be found here.  I also never say that entropy is `disorder'.

I have tried to say as little as possible about quantum mechanics, to keep the 
physics prerequisites low.   However, Planck's constant shows up in the formulas
for the entropy of the three classical systems mentioned above.  The reason for this is fascinating: 
Planck's constant provides a  unit of volume in position-momentum space, which
is necessary to define the entropy of these systems.  Thus, we need a tiny
bit of quantum mechanics to get a good approximate formula for the entropy of hydrogen, 
even if we are trying our best to treat this gas classically.

Since I am a mathematical physicist, this book is full of math.  I spend more time trying to make concepts precise and looking into strange counterexamples than an actual `working' physicist would.   If at any point you feel I am sinking into too many technicalities, don't be shy about jumping to the next tweet.
The really important stuff is in the boxes.  It may help to reach the end before going back and
learning all the details.  It's up to you.

\newpage

\renewcommand{\baselinestretch}{1.5}\normalsize
\tableofcontents
\renewcommand{\baselinestretch}{1.0}\normalsize

\newpage

\setcounter{page}{1}
\pagestyle{plain}

\begin{center}\noindent\fbox{\begin{minipage}{30em}

\begin{center}
\entry{THE ENTROPY OF THE OBSERVABLE UNIVERSE}
\end{center} \vskip 1em

\vskip 1em
\noindent \textbf{\boldmath
In 2010, Chas A.\ Egan and Charles H.\ Lineweaver estimated the biggest contributors to the entropy of the observable universe.   Measuring entropy in bits, these are:
\begin{itemize}
\item
 stars: $10^{81}$ bits.
\item
interstellar and intergalactic gas and dust: $10^{82}$ bits.
\item
gravitons: $10^{88}$ bits.
\item
neutrinos: $10^{90}$ bits.
\item
photons: $10^{90}$ bits.
\item
stellar black holes: $10^{98}$ bits.
\item
supermassive black holes: $10^{105}$ bits. 
\end{itemize}
So, almost all the entropy is in supermassive black holes!
}

\end{minipage}} \end{center} \vskip 1em 
 
 In 2010, \href{https://arxiv.org/abs/0909.3983}{Chas A.\ Egan and Charles H.\ Lineweaver} estimated the entropy of the observable universe.  Entropy corresponds to unknown information, so there's a heck of a lot we don't know!    For stars, most of this unknown information concerns the details of every single electron and nucleus zipping around in the hot plasma.   There's more entropy in interstellar and intergalactic gas and dust.   Most of the gas here is hydrogen---some in molecular form $\mathrm{H}_2$, some individual atoms, and some ionized.    For all this stuff, the unknown information again mostly concerns the details, like the position and momentum, of each of these molecules, atoms and ions.
 
There's a lot more we don't know about the precise details of other particles whizzing through the
universe, like gravitons, neutrinos and photons.  But there's even more entropy in black holes!   One reason Stephen Hawking is famous is that he figured out how to compute the entropy of black holes.
To do that you need a combination of statistical mechanics, general relativity and quantum physics.
Statistical mechanics is the study of physical systems where there's unknown information, which you study using probability theory.    I'll explain some of that in these tweets.    General relativity is Einstein's theory of gravity, and while I've explained that \href{https://arxiv.org/abs/gr-qc/0103044}{elsewhere}, I don't want to get into it here---so I will say nothing about the entropy of black holes. 

Quantum physics was also necessary for Hawking's calculation, as witnessed by the fact that his answer involves Planck's constant, which sets the scale of quantum uncertainty in our universe.   I will try to steer 
clear of quantum mechanics in these tweets, but in the end we'll need a tiny bit of it.   There's a funny sense in which statistical mechanics is somewhat incomplete without quantum mechanics.   You'll
eventually see what I mean.

 \vfill \eject \begin{center}\noindent\fbox{\begin{minipage}{30em}

\begin{center}
\entry{THE ENTROPY OF HYDROGEN}
\end{center} \vskip 1em

\noindent \textbf{\boldmath At standard temperature and pressure, hydrogen gas has an entropy of
\begin{center}
     $130.68$ joule/kelvin per mole 
\end{center} 
But a joule/kelvin of entropy is about
\begin{center}
        $1.0449 \cdot 10^{23}$ bits of unknown information
\end{center} 
and a mole of any chemical is about
\begin{center}
        $6.0221 \cdot 10^{23}$ molecules
\end{center}
So the unknown information about the precise microscopic state of hydrogen is
\begin{center}
$ \displaystyle{ 130.68 \cdot \frac{1.0449 \cdot 10^{23}}{6.0221 \cdot 10^{23}} \quad \approx \quad {\color{darkred} 23} }$ {\color{darkred} bits per molecule!}
\end{center} \vskip 1em
}

 \end{minipage}} \end{center} \vskip 1em

Egan and Lineweaver estimated the entropy of all the 
interstellar and intergalactic gas and dust in the observable universe.  Entropy corresponds to information we don't know.  Their estimate implies that there are $10^{82}$ bits of
information we don't know about all this gas and dust.   

Most of this stuff is hydrogen.  Hydrogen is very simple stuff.   So it would be good to understand the entropy of hydrogen.   You can measure \emph{changes} in entropy by doing experiments.  If you assume hydrogen has no entropy at absolute zero, you can do experiments to figure out the entropy of hydrogen under other conditions.    From this you can calculate that each molecule in a container of hydrogen gas at standard temperature and pressure has about 23 bits of information that we don't know.

You can see a sketch of the calculation above.  But \emph{everything about it} is far from obvious!   What does `missing information' really mean here?   Joules are a unit of energy; kelvin is a unit of temperature.  So why is entropy measured in joules per kelvin?  Why does one joule per kelvin correspond to  $1.0449 \cdot 10^{23}$ bits of missing information?   How can we do experiments to measure changes in entropy?   And why is missing information the same as---or more precisely proportional to---entropy?

The good news: all these questions have answers!  You can learn them here.   However, you will have to persist.     Since I'm starting from scratch it won't be quick.  It takes some math---but luckily, nothing much more than calculus of several variables.   When you can calculate the entropy of hydrogen from first principles, and understand what it means, that will count as true success.

See how it goes!   Partial success is okay too.  
 
 \vfill \eject \begin{center}\noindent\fbox{\begin{minipage}{30em}
 
\begin{center}
\entry{WHERE ARE WE GOING?}
\end{center}

\noindent
\textbf{\boldmath
{\color{red} The mystery: why does each molecule of hydrogen have $\sim\!23$ bits of entropy at
standard temperature and pressure?}
\vskip 1em \noindent
{\color{brickred}
The goal: derive and understand the formula
for the entropy of a classical ideal monatomic gas:
\[   S = kN\left( \frac{3}{2} \ln k T + \ln\frac{V}{N} + \gamma
\right) \]
including the mysterious constant $\gamma$.}
\vskip 1em \noindent
{\color{darkred}
The subgoal: compute the entropy of a single classical particle in a 1-dimensional box.}
\vskip 1em \noindent
The sub-subgoal:  explain entropy from the ground up, and compute the entropy of a classical harmonic
oscillator.}
 
 \end{minipage}} \end{center} \vskip 1em 
 
 To understand something deeply, it can be good to set yourself a concrete goal.
 To avoid getting lost in the theory of entropy, let's try to understand the entropy of hydrogen gas.
 This is a `diatomic' gas since a hydrogen molecule has two atoms.  At standard temperature
 and pressure it's close to `ideal', meaning the molecules don't interact much.   It's also close
 to `classical', meaning we don't need to know quantum mechanics to do this calculation.  
 Also, when the hydrogen is not extremely hot, its molecules don't vibrate much---but they do tumble around.   
 
 Given all this, we can derive a formula for the entropy $S$ of some hydrogen gas as a function of its
 temperature $T$,
 the number $N$  of molecules, the volume $V$, and a physical constant $k$ called `Boltzmann's
 constant'.  This formula also involves a rather surprising constant which I'm calling $\gamma$. 
 We'll figure that out too.  It's so weird I don't want to give it away!
 
As a warmup, we will derive the formula for the entropy of
an ideal `monatomic' gas---a gas made of individual atoms, like helium or neon or argon.
Sackur and Tetrode worked this out in 1912.   Their result, called the Sackur--Tetrode
equation, is similar to the one for a diatomic gas.  

But before doing a monatomic gas, we'll figure out the entropy of a \emph{single atom} 
of gas in a box.  That turns
out to be a good start, since in an ideal monatomic gas the atoms don't interact, and the
entropy of $N$ atoms---as we'll see---is just $N$ times the entropy of a single atom.

But before we can do any of this, we need to understand what entropy is, and how
to compute it.     It will take quite a bit of time to compute the entropy of a classical harmonic oscillator!
But from then on, the rest is surprisingly quick.
 
 \vfill \eject \begin{center}\noindent\fbox{\begin{minipage}{30em}
 
\begin{center}
\entry{FIVE KINDS OF ENTROPY}
\end{center} \vskip 1em

\noindent
\textbf{\boldmath Entropy in thermodynamics}: the change in entropy as we change a system's internal energy by an infinitesimal amount $dE$ while keeping it in thermal equilibrium is $dS = dE/T$, where $T$ is the temperature.

\vskip 1em \noindent
\textbf{\boldmath Entropy in classical statistical mechanics:} $S = -k \int_X p(x) \ln (p(x)) d\mu(x)$
where $p$ is a probability distribution on the measure space $(X,\mu)$ of states and $k$ is Boltzmann's constant.

\vskip 1em \noindent
\textbf{\boldmath Entropy in quantum statistical mechanics:} $S = -  k\,\mathrm{tr}(\rho \ln \rho) $  where $\rho$ is a density matrix.

\vskip 1em \noindent
\textbf{\boldmath Entropy in information theory:} $H = -\sum_{i \in X} p_i \log p_i $  where
$p$ is a probability distribution on the set $X$.

\vskip 1em \noindent
\textbf{\boldmath Algorithmic entropy:} the entropy of a string of symbols is the length of the
shortest computer program that prints it out.

 \end{minipage}} \end{center} \vskip 1em 
 
Before I actually start explaining entropy, a warning: it can be hard
at first to learn about entropy 
 because there are many
kinds---and people often don't say which kind they're talking about.
Here are 5 kinds.  Luckily, they are closely related!

In thermodynamics we primarily have a formula for the \textit{change} in
entropy: if you change the internal energy of a system by an infinitesimal 
amount $dE$ while keeping it in thermal equilibrium, the infinitesimal
change in entropy is $dS = dE/T$ where $T$ is the temperature.   

Later, in classical statistical mechanics, Gibbs explained entropy in terms
of a probability distribution $p$ on the space of states of a
classical system.  In this framework, entropy is the integral of $-p
\ln p $ times a constant $k$ called Boltzmann's constant.  

Later von Neumann generalized Gibbs' formula for entropy from
classical to \textit{quantum} statistical mechanics!  He replaced the
probability distribution $p$ by a so-called density matrix $\rho$,
and the integral by a trace. 

Later Shannon invented information theory, and a formula for the entropy
of a probability distribution on a set (often a finite set).   This is often called 
`Shannon entropy'.  It's just a special case of Gibbs' formula for entropy in 
classical statistical mechanics,  but without the Boltzmann's constant.   

Later still, Kolmogorov invented a formula for the entropy of a
\textit{specific} string of symbols.  It's just the length of the shortest
program, written in bits, that prints out this string.  It depends on
the computer language, but not too much.  

There's a network of results connecting all these 5 concepts of
entropy.    I will first explain Shannon entropy, then entropy in 
classical statistical mechanics, and then entropy in thermodynamics.
While this is the reverse of the historical order, it's the easiest way to go.

I will not explain entropy in quantum statistical mechanics: for that
I would feel compelled to teach you quantum mechanics first.  Nor will I explain 
algorithmic entropy.

 \vskip 1em \vfill \eject \begin{center}\noindent\fbox{\begin{minipage}{30em}

\begin{center}
\entry{FROM PROBABILITY TO INFORMATION}
\end{center} \vskip 1em

\vskip 1em

\noindent
\textbf{\boldmath How much information do you get when you learn an event of probability $p$ has
happened?    It's 
\[          { \color{darkred}   - \log p } \]
where we can use any base for the logarithm, usually $e$ or $2$. 
\vskip 1em 
\noindent
\textbf{Example:} Suppose I flip 3 coins that you know are fair.   I tell you the outcome: ``heads, tails, heads''.
That's an event of probability $1/2^3$, so the information you get is
\[              - \log \left( \frac{1}{2^3} \right) = 3 \log 2  \]
or ``3 bits'' for short, since $\log 2$ of information is called a \define{bit}.   }

\end{minipage}} \end{center} \vskip 1em

Here is the simplest link between probability and information:
when you learn that an event of probability $p$ has happened, how much
information do you get?  We say it's $-\log p$.   We take a logarithm so that when you multiply probabilities, information adds.  The minus sign makes information come out positive.

\define{Beware: when I write `$\log$' I don't necessarily mean the logarithm base 10.}   I mean
that you can use whatever base for the logarithm you want; this choice is like a choice of
units.   Whatever base $b$ you decide to use, I'll call $\log_b 2$ a `bit'.
For example, if I flip a single coin that you know is fair, and you see that it comes up
heads, you learn of an event that's of probability $1/2$, so the amount of information 
you learn is
\[         - \log_b \frac{1}{2} = \log_b 2. \]
That's one bit!   Of course if you use base $b = 2$ then this logarithm actually equals 1, which is nice.

To understand the concept of information it helps to do some puzzles.

\begin{puzzle}  First I flip 2 fair coins and tell you the outcome.
Then I flip 3 more and tell you the outcome.  How much information did
you get?
\end{puzzle}

\begin{puzzle}  I roll a fair 6-sided die and tell you the outcome.
Approximately how much information do you get, using logarithms base
2?
\end{puzzle}

\begin{puzzle} When you flip 7 fair coins and tell me the outcome, 
how much information do I get?
\end{puzzle}

\begin{puzzle}  Every day I eat either a cheese sandwich, a salad,
or some fried rice for lunch---each with equal probability.  I
tell you what I had for lunch today.  Approximately how many bits of
information do you get?
\end{puzzle}

\begin{puzzle} I have a trick coin that always lands heads up.  You
know this.  I flip it 5 times and tell you the outcome.  How much
information do you receive?
\end{puzzle}

\begin{puzzle} I have a trick coin that always lands heads up.  You
believe it's a fair coin.  I flip it 5 times and tell you the outcome.
How much information do you receive?
\end{puzzle}

\begin{puzzle} I have a trick coin that always lands with the same
face up.  You know this, but you don't know which face always comes
up.  I flip it 5 times and tell you the outcome.  How much information
do you receive?  
\end{puzzle}

These puzzles raise some questions about the nature of probability,
like: is it subjective or objective?  People like to argue about those
questions.  But once we get a probability $p$, we can convert it to
information by computing $-\log p$.

 \vfill \eject \begin{center}\noindent\fbox{\begin{minipage}{30em}
 
\begin{center}
\entry{UNITS OF INFORMATION} 
\vskip 1.5em

\textbf{\boldmath{
An event of probability 1/2 carries one \define{bit} of information. \\ \vskip 0.5em
An event of probability 1/$e$ carries one \define{nat} of information. \\ \vskip 0.5em
An event of probability 1/3 carries one \define{trit} of information. \\ \vskip 0.5em
An event of probability 1/4 carries one \define{crumb} of information. \\ \vskip 0.5em
An event of probability 1/10 carries one \define{hartley} of information. \\ \vskip 0.5em
An event of probability 1/16 carries one \define{nibble} of information. \\ \vskip 0.5em
An event of probability 1/256 carries one \define{byte} of information. \\ \vskip 0.5em
An event of probability 1/$2^{8192}$ carries one \define{kilobyte} of information. \\ \vskip 0.5em}}
\end{center} \vskip 1em

\end{minipage}} \end{center} \vskip 1em

There are many units of information.  Using $ \textrm{information} =
- \log p $ we can relate these to probabilities.  For example if you
see a number in base 10, and each digit shows up with probability
1/10, the amount of information you get from each digit is one
`hartley'.   

How many bits are in a hartley?   Remember: no matter what base you use, I call $\log 10$ a
hartley and $\log 2$ a bit.  There are $\log 10 / \log 2$ bits in a hartley.  This number
has the same value no matter what base you use for your logarithms!   If you use base 2, it's
\[          \log_2 10 / \log_2 2 = \log_2 10 \approx 3.32  .\]
So a hartley is about 3.32 bits.   

If you flip 8 fair coins and tell me what answers you got, I've learned of an event that has
probability $1/2^8 = 1/256$.   We say I've received a `byte' of information.  This equals 8 bits of information.   Similarly, if you flip $1024 \times 8$ fair coins and tell me the outcome, I receive a kilobyte
of information.   

Or at least that's the old definition.  Now many people define a kilobyte to be
$1000$ bytes rather than $1024$ bytes, in keeping with the usual meaning of the prefix.
If you want $1024$ bytes you're supposed to ask for a `kibibyte'.
When we get to a terabyte, the new definition based on powers of $10$ is about $10\%$ 
less than the old one based on powers of $2$: $10^{12}$ bytes rather than $2^{40} \approx
1.0995 \times 10^{12}$.   If you want the old larger amount of information you should
ask for a `tebibyte'.

Wikipedia has an article that lists many strange 
\href{https://en.wikipedia.org/wiki/Units_of_information}{units of information}.
Did you know that 2 bits is a `crumb'?  Did you even need to know?
No, but now you do.

Feel free to dispose of this unnecessary information!  All this is just for fun---but I want you to
get used to the formula
\[ \textrm{information} = - \log p \]

\vfill \eject \begin{center}\noindent\fbox{\begin{minipage}{30em}
  
\begin{center}
\entry{THE INFORMATION IN A LICENSE PLATE NUMBER}

\vskip 1em
\includegraphics[width = 25 em]{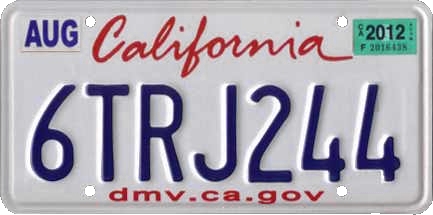}
\vskip 1em

\textbf{\boldmath If there are $N$ different possible license plate numbers, all equally likely, how many
bits of information do you learn when you see one?}

\end{center} 
\end{minipage}} \end{center} \vskip 1em

If you think $N$ alternatives are equally likely, when you see which one actually
occurs, you gain an amount of information equal to $\log_b N$.  Here the choice of base
$b$ is up to you: it's a choice of units.    But what is this in bits?  No matter what base you use, 
\[   \log_b N = \log_2 N  \times \log_b 2 .\]
Since we call $\log_b 2$ a `bit',  this means you've learned $\log_2 N$ bits of information.

Let's try it out!

\vskip 1em

\begin{puzzle} Suppose a license plate has 7 numbers and/or letters
on it.  If there are 10+26 choices of number and/or letter, there are
$36^7$ possible license plate numbers.  If all license plates are
equally likely, what's the information in a license plate number in bits---approximately?
\end{puzzle}

\begin{center}
\includegraphics[width = 25 em]{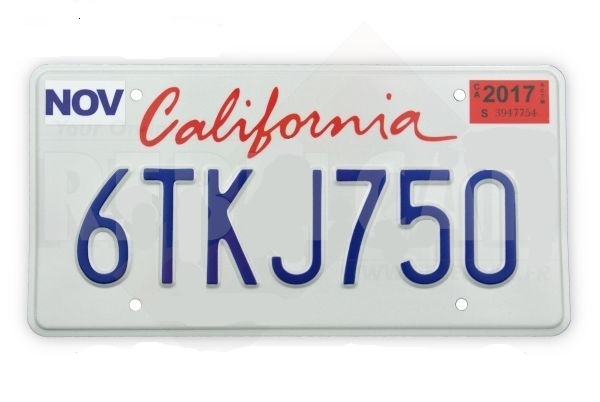}
\end{center}

But wait!  Suppose I tell you that all license
plate numbers have a number, then 3 letters, then 3 numbers!  
You have just learned a lot of information.  So the remaining
information content of each license plate is presumably less.  Let's work it out.

\vskip 1em

\begin{puzzle}  How much information is there in a license
plate number if they all have a number, then 3 letters, then 3
numbers?  (Assume they're all equally probable and there are 10
choices of each number and 26 choices of each letter.)
\end{puzzle}
\vskip 1em
The moral: when you learn more about the possible choices, the
information it takes to describe a choice drops.  

\vfill \eject \begin{center}\noindent\fbox{\begin{minipage}{30em}

\begin{center}
\entry{THE INFORMATION IN A LICENSE PLATE}
\end{center} \vskip 2em

\textbf{\boldmath How much unknown information do the atoms in a license plate contain?
\vskip 1em
Aluminum has an entropy of about 28 joules/kelvin per mole at standard temperature
and pressure.  A mole of aluminum weighs about 27 grams.   A typical license plate might
weigh 150 grams, and thus have
\[     150 \; \textrm{g} \times \frac{28 \; \textrm{J/K} \cdot \textrm{mole}}{27 \; \textrm{g}/\textrm{mole}} \approx  160 \; \textrm{J/K} \]
of entropy.   But a joule/kelvin of entropy is about $10^{23}$ bits of unknown information.
Thus, the atoms in such a license plate contain about
\[   160 \times 10^{23} \textrm{\; bits} \approx \color{darkred} 1.6 \cdot 10^{25} \; \textrm{bits} \]
of unknown information.
}

\textbf{\boldmath }

\end{minipage}} \end{center} \vskip 1em 

Last time we talked about the information in a license plate number.   A license plate number
made of 7 numbers and/or letters contains
\[    \log_2 (36^7)  \approx 36.189 \]  
bits of information if all combinations  are equally likely.    How does this compare to the information in the actual metal of the license plate?

These days most license plates are made of aluminum, and they weigh roughly between 100 and 200
grams.    Let's say 150 grams.   If we work out the entropy of this much aluminum, and express it
in bits of unknown information, we get an enormous number: roughly
{\boldmath \[    \color{darkred}      16,000,000,000,000,000,000,000,000   \; \textbf{bits}!  \]}
\vskip -0.8em
Here is the point.  While the information \emph{on} the license plate and the information
\emph{in} the license plate can be studied using similar mathematics, the latter dwarfs the
former.   Thus, when we are doing chemistry and want to know, for example, how much the entropy
of the license plate increases when we dissolve it in hydrochloric acid, the information in the
writing on the license plate is irrelevant for all practical purposes.

Some people get fooled by this, in my opinion, and claim that ``information'' and ``entropy'' are
fundamentally unrelated.   I disagree.   

\vfill \eject \begin{center}\noindent\fbox{\begin{minipage}{30em}

\begin{center}
\entry{JUSTIFYING THE FORMULA FOR INFORMATION} 
\end{center} \vskip 1em
\vskip 1.5em

\textbf{\boldmath
\noindent
Why do we say the information of an event of probability $p$ is
\[       I(p) = -\log_b p \]
for some base $b > 1$?   Here's why:
\vskip 1em \noindent
\define{Theorem}.   Suppose $I \maps (0,1] \to \R$ is a function that is:
\vskip 1em \noindent
\define{1.\  Decreasing:} $p < q$ implies $I(p) > I(q)$.    This says less probable events 
have more information.
\vskip 1em \noindent
\define{2.\  Additive:} $I(p q) = I(p) + I(q)$.  This says the information of the combination of 
two independent events is the sum of their separate informations.
\vskip 1em \noindent
Then for some $b > 1$ we have $I(p) = - \log_b p$.
}

\end{minipage}} \end{center} \vskip 1em 
 
 The information of an event of probability $p$ is $-\log p$, where
you get to choose the base of the logarithm.  
But why?   This is the only option if we want less probable events to have more
information, and information to add for independent events.  

Proving this will take some math---but don't worry, you won't need to know this stuff
for the rest of this `course'.

Since we're trying to prove $I(p)$ is a logarithm function, let's write
\[ I(p) = f(\ln(p)) \]
and prove $f$ has to be linear:
\[ f(x) = cx. \]
As we'll see, this gets the job done.

Writing $I(p) = f(x)$ where $x = \ln p$, we can check that
Condition 1 above is equivalent to
\[ x < y \textrm{ implies } f(x) > f(y) \textrm{ for all } x,y \le 0.  \]
Similarly, we can check that Condition 2 is equivalent to
\[ f(x+y) = f(x) + f(y) \textrm{ for all } x,y \le 0 . \]

Now what functions $f$ have 
\[  f(x+y) = f(x) + f(y) \]
for all $x,y \le 0$?

If we define $f(-x) = -f(x)$, $f$ will become a function from the
whole real line to the real numbers, and it will still obey $f(x+y) =
f(x) + f(y)$.   So what functions obey this equation?
The obvious solutions are
\[ f(x) = cx \]
for any real constant c.  But are there any other solutions?

\emph{Yes}, if you use the \href{https://en.wikipedia.org/wiki/Axiom_of_choice}{axiom of choice}! 
Treat the reals as a vector space over the rationals.  Using the axiom
of choice, pick a basis.  To get $f \colon \R \to \R$
that's linear over the rational numbers, just let $f$ send each
basis element to whatever real number you want and extend it to a linear
function defined on all of $\R$.   
This gives a function $f$ that obeys $f(x+y) = f(x) + f(y)$.

However, no solutions of $f(x+y) = f(x) + f(y)$ meet our other condition
\[ x < y \textrm{ implies } f(x) > f(y) \textrm{ for all } x,y \le 0 \]
except for the familiar ones $f(x) = cx$.  For a proof see
Wikipedia: they show all solutions except the familiar ones are so discontinuous
their graphs are \emph{dense in the plane!}

\begin{itemize}
\item
Wikipedia, \href{https://en.wikipedia.org/wiki/Cauchy\%27s_functional_equation}{Cauchy's
functional equation}.
\end{itemize}

\noindent So, our conditions imply $f(x) = cx$ for some $c$, and since $f$ is
decreasing we need $c < 0$.  So our formula $I(p) = f(\ln p)$
says
\[ I(p) = c \ln p \]
but this equals $-\log_b p$ if we take $b = \exp(-1/c)$.  And this number
$b$ can be any number $> 1$.    QED.

\vskip 1em
Thus, if we want a more general concept of the information associated to
a probability, we need to drop Condition 1 or 2.   For example, we
could replace additivity by some other rule.  People have tried this!
Indeed, there is a world of generalized entropy concepts including 
\href{https://en.wikipedia.org/wiki/Tsallis_entropy}{Tsallis entropies},
\href{https://en.wikipedia.org/wiki/Renyi_entropy}{R\'enyi entropies} and others.   

 \vfill \eject \begin{center}\noindent\fbox{\begin{minipage}{30em}
 
 \hypertarget{sec:WHAT_IS_PROBABILITY}{}
 
\begin{center}
\entry{WHAT IS PROBABILITY?} 
\vskip 1em

\bf{
\emph{The theory of probabilities is at bottom nothing but common sense reduced to calculus; it enables us to appreciate with exactness that which accurate minds feel with a sort of instinct for which of times they are unable to account.}  --- Pierre-Simon Laplace 
\vskip 1em
\emph{In no other branch of mathematics is it so easy for experts to blunder as in probability theory.} ---
Martin Gardner 
}
\end{center}
 
 \end{minipage}} \end{center} \vskip 1em
  
Since I've defined information in terms of probability,
you may naturally wonder ``what is probability?''   I won't seriously try to answer this. 
This question has stirred up many debates over the centuries, and even today there's 
not a fully accepted answer.   It deserves a whole book---and this is not that book.   
Luckily, we don't really \emph{need} to know
exactly what probability is to do calculations with it: we mainly need to set up some rules for working with it.   This may seem like a cop-out.  But it's a strange and
wonderful feature of 
science that we can achieve great reliability in our results by sidestepping certain difficult questions, 
like someone who can make their way safely through a jungle by avoiding the quicksand and snakes.

One approach to probability goes like this.  Suppose you repeat
some experiment $N$ times, doing your best to make the conditions the same each time.
Suppose that $M$ of these times some event $E$ occurs.   You may then say that the probability 
of event $E$ happening under these conditions is $M/N$.    This approach is called `finite frequentism'.
Unfortunately, this approach can lead you to say a coin has probability $1$ of landing heads up
if it does so the first time, or first $3$ times, you flip it.   
 
Another approach goes like this.  You may say that some event $E$ has probability $p$ under some conditions if when you set up these conditions $N$ times, and the event $E$ happens $M$ times, the fraction $M/N$ approaches $p$ in  the limit $N \to \infty$.     This approach is called `hypothetical frequentism', because in real experiments you can't take the limit $N \to \infty$.     But you can hope that when $N$ becomes large enough, the fraction $M/N$ usually becomes close to the limiting probability $p$---whatever that means.

Another approach, called `Bayesianism', treats a probability of an event $E$  under some specified conditions as a measure of your degree of belief that $E$ will happen under these conditions.     But what is `degree of belief'?   One answer involves bets.  For example, perhaps to believe an event has probability $1/2$ means you're willing to take a bet where you win more when the event happens than you lose if it does not.  

Bayesians tend to focus on the rules  for \emph{updating} your probabilities as you learn new things, the most famous being `\href{https://en.wikipedia.org/wiki/Bayes\%27_theorem}{Bayes' rule}'.  Even if agents start by assigning different probabilities to an event, if they follow the same rules for changing these probabilities as they learn new things, under certain circumstances we can prove their probabilities will converge to the same value.   

For a passionate and intelligent discussion of these issues, I recommend E.\ T.\ Jaynes' book \textsl{Probability Theory: the Logic of Science}.    Later we'll meet his `principle of maximum entropy', another important approach to working with probabilities.

\vfill \eject \begin{center}\noindent\fbox{\begin{minipage}{30em}

\hypertarget{sec:PROBABILITY_MEASURES}{}
 
 \begin{center}
\entry{PROBABILITY MEASURES} 
\end{center}
\vskip 1em

\noindent\textbf{\boldmath{A \define{measure} on a set $X$ is a function that assigns to certain so-called
\define{measurable} subsets $S \subseteq X$ a number $m(S) \in [0,\infty]$, obeying these rules:
\begin{itemize}
\item $\emptyset, X \subseteq X$ are measurable and 
\[           m(\emptyset) = 0\]
\item If $S,T \subseteq X$ are measurable and $S \subseteq T$, then $T - S$ is 
measurable and
\[       m(T) = m(S) + m(T - S)  \]
\item  If a countable collection of subsets $S_i \subseteq X$ are measurable, then their union is measurable, and if $S_i$ are disjoint then 
\[           m\left( \bigcup_{i = 1}^\infty S_i \right) = \sum_{i = 1}^\infty m(S_i)  \]
\end{itemize}
\noindent We say $m$ is a \define{probability measure} if $m(X) = 1$.
}}
 
 \end{minipage}} \end{center} \vskip 1em
 
 It is easier to do calculations with probabilities than say exactly what they mean!  I will take a rough-and-ready approach to working with them, but first let's take a peek at how mathematicians do it.   If you don't care, it's safe to move right on to the next tweet.
 
 We start with any set.  We call elements of $X$ `outcomes' and subsets of $X$ `events'.    We can sometimes  get into trouble trying to assign a probability to \emph{every} subset of $X$.   So, we'll only try to assign probabilites to events in some collection $\mathcal{M}$ with these properties:
 \begin{itemize}
 \item $\emptyset \in \mathcal{M}$ and $X \in \mathcal{M}$.
 \item If $S, T \in \mathcal{M}$ and $S \subseteq T$ then the set of elements of $T$ that are not in
 $S$, called $T - S$, is in $\mathcal{M}$.
 \item If $S_i \in \mathcal{M}$ for $i = 1, 2, \dots$ then the union $\bigcup_{i = 1}^\infty S_i $ is
 in $\mathcal{M}$.
 \end{itemize}
 We call elements of  $\mathcal{M}$ \define{measurable} subsets of $X$.  A \define{measure} is then a function $m \maps \mathcal{M} \to [0,\infty]$ obeying these rules:
 \begin{itemize}
 \item $m(\emptyset) = 0$ 
 \item If $S, T \in \mathcal{M}$ and $S \subseteq T$ then $ m(T) = m(S) + m(T - S) $.
 \item If the sets $S_i \in \mathcal{M}$ are disjoint then $  m\left( \bigcup_{i = 1}^\infty S_i \right) = \sum_{i = 1}^\infty m(S_i) $.
 \end{itemize}
 If $m$ also obeys $m(X) = 1$ then we say $m$ is a \define{probability measure}, and for any
 $S \in \mathcal{M}$ we say $m(S)$ is the \define{probability} of the \define{event} $S$.  But we will also
 be interested in other measures, like the measure on the real line called `\href{https://en.wikipedia.org/wiki/Lebesgue_measure}{Lebesgue measure}'.   This is closely connected to the symbol `$dx$' that shows up in integrals, because for any measurable set $S \subseteq \R$, its Lebesgue measure is
 \[       \int_{-\infty}^{\infty} \chi_S(x) \, dx \]
 where $\chi_S(x)$ is $1$ for $x \in S$ and $0$ for $x \notin S$.   Indeed, people often get sloppy 
 and say $dx$ `is' Lebesgue measure, and I may do that too.   By the way, Lebesgue measure is one where we cannot take $\mathcal{M}$ to be the collection of all subsets of $\R$.  
 
 There is an extensive theory of measures.    We will not need it here, but if you're interested, you can try a book like Halsey Royden's \textsl{Real Analysis}, where
 I learned the basics myself, or Terry Tao's \textsl{\href{https://terrytao.files.wordpress.com/2012/12/gsm-126-tao5-measure-book.pdf}{An Introduction to Measure Theory}}, which has a legal
 free version online.
 
Here are some puzzles about measures.  

\begin{puzzle}
Let $X$ be any set and define $\mathcal{M}$ to be the collection of \emph{all}
subsets of $X$.  Show that there is a measure $m \maps \mathcal{M} \to [0,\infty]$ called \define{counting measure}
such that for any $S \subseteq X$, $m(S)$ is the number of elements of $S$, or $\infty$ if
$S$ is infinite.
\end{puzzle}

\begin{puzzle}
Let $X$ be any set and define $\mathcal{M}$ as before.  Suppose $p$ is 
a \define{probability distribution} on $X$, meaning a function $p \maps X \to [0,\infty)$ with $\sum_{i \in X} p(i) = 1$.   Show that there is a probability measure $m \maps \mathcal{M} \to [0,\infty]$ such that for any $S \subseteq X$, 
\[          m(S) = \sum_{i \in S} p(i)  .\]
In this situation we usually write $p(i)$ as $p_i$ and call it the \define{probability} of the
\define{outcome} $i \in X$.   For any $S \subseteq M$ we call $m(S)$ the probability of the
event $S$.
\end{puzzle}

In the next puzzles $X$ is any set, $\mathcal{M}$ obeys the three rules for 
a collection of measurable subsets of $X$, and $m \maps \mathcal{M} \to [0,\infty]$ is a measure.

\begin{puzzle}
Show that if $S, T \in \mathcal{M}$ then the union $S \cup T$ is in $\mathcal{M}$.
\end{puzzle}

\begin{puzzle}
Show that if $S, T \in \mathcal{M}$ then the intersection $S \cap T$ is in $\mathcal{M}$.
\end{puzzle}

\begin{puzzle}
Show that if $S_i \in \mathcal{M}$ for $i = 1, 2, \dots$ then the intersection $\bigcap_{i=1}^\infty S_i$ is in $\mathcal{M}$.
\end{puzzle}

\begin{puzzle}
Show that if $S, T \in \mathcal{M}$ and $S \subseteq T$ then $m(S) \le m(T)$.
\end{puzzle}

\begin{puzzle}
Show that if $S_i \in \mathcal{M}$ for $i = 1, 2, \dots$ then 
\[          m\left( \bigcup_{i=1}^\infty S_i \right) \le \sum_{i = 1}^\infty m(S_i)  .\]
\end{puzzle}

 \begin{puzzle}
Show that if $m$ is a probability measure and $S \in \mathcal{M}$ then $0 \le m(S) \le 1$.
\end{puzzle}

One of the main uses of a measure $m$ on a space $X$ is that it lets us integrate certain
functions $f \maps X \to \R$.   Alas, not all functions!    It's only reasonable to try to integrate \define{measurable} functions $f \maps X \to \R$, which have the property that if $S \subseteq \R$
is measurable, its inverse image $f^{-1}(S) \subseteq X$ is measurable.   And even measurable
functions can cause trouble, because when we try to integrate them we might get $+\infty$,
$-\infty$, or something even worse.  For example, what's
\[             \int_{-\infty}^\infty  x^2 \sin x \, dx ?  \]
There's no good answer.    We say a function $f \maps X \to \R$ is \define{integrable} if it is 
measurable and its integral over $X$, defined in a certain way I won't explain here, gives a well-defined real number.

\vfill \eject \begin{center}\noindent\fbox{\begin{minipage}{30em} 

\begin{center}
\entry{SHANNON ENTROPY: A FIRST TASTE} 
\end{center} \vskip 1em

\textbf{\boldmath
\noindent
When you learn an event of probability $p$ has
happened, the amount of information you get is $- \log p$.   
\vskip 1em \noindent
\define{Question.}  Suppose you know a coin lands heads up $\frac{2}{3}$ of the time and 
tails up $\frac{1}{3}$ of the time.   What is the average or `expected' amount of information 
you get when you learn which side landed up?
\vskip 1em
\noindent
\define{Answer.}  $\frac{2}{3}$ of the time you get $- \log \frac{2}{3}$ of information, and
$\frac{1}{3}$ of the time you get $-\log \frac{1}{3}$.  So, the expected amount of information you get is
\[ \textstyle{  -\frac{2}{3} \log\frac{2}{3} - \frac{1}{3} \log \frac{1}{3} }  \]
\vskip 1em
\noindent
You can do the same thing whenever you have any number of 
probabilities that add to 1.  The expected information is called the \define{Shannon entropy}.  
}

 \end{minipage}} \end{center} \vskip 1em 
 
 You flip a coin.  You know the probability that it lands heads up.
How much information do you get, on average, when you discover which
side lands up?   It's not hard to work this out.  It's a simple example
of `Shannon entropy'.   Roughly speaking, entropy is information that you 
\textsl{don't know}, that you could get if you did enough experiments.  Here the experiment
is simply flipping the coin and looking at it.

\vskip 1em

\begin{puzzle} Suppose you know a coin lands heads up
$\frac{1}{2}$ of the time and tails up $\frac{1}{2}$ of the time.  What is the
expected amount of information you get from a coin flip?
If you use base 2 for the logarithm, you get the expected information
measured in bits.  What do you get?
\end{puzzle}

\begin{puzzle} Suppose you know a coin lands heads up
$\frac{1}{3}$ of the time and tails up $\frac{2}{3}$ of the time.  What is the
expected amount of information you get from a coin flip?
\end{puzzle}

\begin{puzzle} Suppose you know a coin lands heads up
$\frac{1}{4}$ of the time and tails up $\frac{3}{4}$ of the time.  What is the
expected amount of information you get from a coin flip, in bits?
\end{puzzle}

If you solve these you'll see a pattern: the Shannon entropy is biggest when the coin is
fair.  As it becomes more and more likely for one side to land up than
the other, the entropy drops.  You're more sure about what will
happen... so you learn less, on average, from seeing what happens!

We've been doing examples where your experiment has just two possible
outcomes: heads up or
down.  But you can do Shannon entropy for any number of 
outcomes.  It measures how ignorant you are of what will
happen.  That is: how much you learn on average when it does!
 
\vfill \eject \begin{center}\noindent\fbox{\begin{minipage}{30em}
 
\begin{center}
\entry{SHANNON ENTROPY: A SECOND TASTE} 
\end{center} \vskip 1em

\textbf{\boldmath \noindent According to the weather report there's a $\frac{1}{4}$ chance that it will rain 1 centimeter, a $\frac{1}{2}$ chance it will rain 2 centimeters, and a $\frac{1}{4}$ chance it will rain 3 centimeters.
\vskip 1em \noindent
\textbf{Question.}  What is the `expected' amount of rainfall?
\vskip 1em
\noindent
\textbf{Answer.} $\frac{1}{4} \cdot 1 + \frac{1}{2} \cdot 2 + \frac{1}{4} \cdot 3 = 2$ centimeters.
\vskip 1em
\noindent
\textbf{Question.}  What is the `expected' amount of information you learn
when you find out how much it rains?
\vskip 1em
\noindent
\textbf{Answer.} $- \frac{1}{4} \log \frac{1}{4} - \frac{1}{2} \log \frac{1}{2} - 
\frac{1}{4} \log \frac{1}{4} =  \frac{3}{2} \log 2 $, or in other words, $\frac{3}{2}$ bits.   This is the \define{Shannon entropy} of the weather report.}

\end{minipage}} \end{center} \vskip 1em 
 
 If the weather report tells you it'll rain different amounts with
different probabilities, you can figure out the `expected' amount of
rain.  You can also figure out the expected amount of information
you'll learn when it rains.  This is called the  `Shannon entropy'.

Shannon entropy  is closely connected to information, but we can also
think of it as a measure of ignorance.  This may seem paradoxical.   But it's not.
Shannon entropy is the 
expected amount of information that you \emph{don't know} when all
you know is a probability distribution, which you \emph{will know}
when you see a specific outcome chosen according to this probability
distribution.

For example, consider a weather report that says
it will rain 1 centimeter with probability 0, 2 centimeters
with probability 1, and 3 centimeters with probability 0.
The Shannon entropy of this weather report is
\[   - 0 \log 0 - 1 \log 1 - 0 \log 0 = 0 \]
since by convention we set $p \log p = 0$ when $p = 0$, this being the limit
of $p \ln p$ as $p$ approaches $0$ from above.   

What does it mean that this weather report has zero Shannon entropy?  
It means that when we see a specific outcome chosen according to this 
probability distribution, we learn nothing!   The weather report says it
will rain 2 centimeters with probability 1.   When this
happens, we learn nothing that the weather report didn't already tell us.

The Shannon entropy doesn't depend on the amounts of rain, or even whether
the forecast is about centimeters of rain or dollars of income.  It
only depends on the probabilities of the various outcomes.
So Shannon entropy is a universal, abstract concept.

Shannon entropy is closely connected to Gibbs entropy, which
was already known in physics.  But by lifting entropy to a
more general level and connecting it to digital information, Shannon
helped jump-start the information age.  In fact a paper of his was the
first to use the word `bit'!

\vfill \eject \begin{center}\noindent\fbox{\begin{minipage}{30em}

\begin{center}
\entry{THE DEFINITION OF SHANNON ENTROPY} 
\vskip 2em

\textbf{\boldmath
\noindent
Suppose you believe there are $n$ possible outcomes with probabilities 
$p_1, \dots, p_n \ge 0$ that sum to $1$.
\vskip 1em
\noindent
The average amount of information you learn when one of these outcomes happens, chosen
according to this probability distribution, is the \define{Shannon entropy}:
{\boldmath
\[  \color{darkred} \scalebox{1.2}{$\displaystyle{ H = - \sum_{i=1}^n p_i \log p_i }$} \]}
\vskip 0.2em
Shannon entropy is larger for probability distributions that are more spread out, and 
smaller for probability distributions that are more sharply peaked.
} 
\end{center} 

\end{minipage}} \end{center} \vskip 1em 
  
I've been leading up to it with examples, but here it is in general: Shannon entropy!   Gibbs had already used a similar formula in physics---but with base $e$ for the logarithm, an integral instead of a sum, and multiplying the answer by Boltzmann's constant.  Shannon applied it to digital information.

Here's where the formula for Shannon entropy comes from.  We have some set of outcomes, say $X$.   We have a \define{probability distribution} on this set, meaning a function $p \maps X \to [0,1]$
such that
\[             \sum_{i \in X} p_i = 1  . \]
If we have any function $A \maps X \to \R$, we define its  \define{expected value} to be
\[           \langle A \rangle =     \sum_{i \in X}  p_i A_i  .\]
It's a kind of average of $A$ where each value $A(i)$ is `weighted', i.e.\ multiplied,
by the probability of the $i$th outcome.   We saw an example in the last tweet: the expected amount of rainfall.

We've seen that if you believe the $i$th outcome has probability $p_i$, the amount
of information you learn if the $i$th outcome actually occurs is $-\log p_i$.   Thus, the \emph{expected}
amount of information you learn is
\[            \langle - \log p \rangle = - \sum_{i \in X} p_i \log p_i    .\]
And this is the \define{Shannon entropy}!  We denote it by $H$, or more precisely $H(p)$, so
\[         H(p) = - \sum_{i \in X} p_i \log p_i  . \]
In the box above I was taking $X$ to be the set $\{1, \dots, n\}$.  This is often a good thing to do when there are finitely many outcomes.

Let's get to know the Shannon entropy a little better.

\begin{puzzle}
Let $X = \{1,2\}$ so that we know a probability distribution $p$ on $X$ if we know $p_1$, since
$p_2 = 1 - p_1$.   Graph  the Shannon entropy $H(p)$ as a function of $p_1$.  Show that it has a 
maximum at $p_1 = \frac{1}{2}$ and minima at $p_1 = 0$ and $p_1 = 1$.
\end{puzzle}

\noindent
This makes sense: if you believe $p_1 = 1$ then you learn nothing when an outcome happens chosen according to the probability distribution $p$: you are sure outcome $1$ will occur, and it does (with probability $1$).   Similarly, if you believe $p_1 = 0$ you learn nothing when an outcome happens
according to this probability distribution, since you are sure outcome $2$ will occur.  On the other hand, if
$p_1 = \frac{1}{2}$ you are maximally undecided about what will happens, and you learn $1$ bit of
information when it does.

\begin{puzzle}
Let $X = \{1,2,3\}$.  Draw the set of probability distributions on $X$ as an equilateral triangle
whose corners are the probability distributions $(1,0,0)$, $(0,1,0)$, and $(0,0,1)$.  Sketch contour
lines of $H(p)$ as a function on this triangle.  Show it has a maximum at $p = (\frac{1}{3}, \frac{1}{3}, \frac{1}{3})$ and minima at the corners of the triangle.
\end{puzzle}

\noindent
Again this should make intuitive sense.      Here is a harder puzzle along the same lines:

\begin{puzzle}
Let $X = \{1,\dots, n\}$.   Show that $H(p)$ has a maximum at $p = (\frac{1}{n}, \dots, \frac{1}{n})$
and minima at the probability distributions where $p_i = 1$ for some particular $i \in X$.
\end{puzzle}

Here is one of the big lessons from all this:

\begin{center}
\define{Shannon entropy is larger for probability distributions that are more spread out, and 
smaller for probability distributions that are more sharply peaked.}
\end{center}

Indeed, you can think of Shannon entropy as a measure of how spread out a probability distribution is!
The more spread out it is, the more you learn when an outcome occurs, drawn from that distribution.

Another important way to think about Shannon entropy is that it sets a limit on how much we can 
compress messages that are drawn from a given probability distribution.   This is made precise by a theorem 
Shannon proved in his original 1948 paper.  I won't explain it here, but this result is fundamental for 
understanding the role of entropy in communication and data storage:

\begin{itemize}
\item Wikipedia, \href{https://en.wikipedia.org/wiki/Shannon\%27s_source_coding_theorem}{Shannon's source coding theorem}.
\item Claude E.\ Shannon, \href{https://web.archive.org/web/20090216231139/http://plan9.bell-labs.com//cm//ms//what//shannonday//shannon1948.pdf}{A mathematical theory of communication}, 
\textsl{Bell System Technical Journal} \textbf{27} (1948), 379--423, 623--656.
\end{itemize}

 \vfill \eject \begin{center}\noindent\fbox{\begin{minipage}{30em} 

\begin{center}
\entry{THE PRINCIPLE OF MAXIMUM ENTROPY} 
\vskip 1.5em
\textbf{\boldmath
\noindent Suppose there are $n$ possible outcomes.  At first you have no reason to think any is more probable than any other.  \vskip 1em
Then you learn some facts about the correct probability
distribution---but not enough to determine it uniquely! Which probability distribution $p_1, \dots, p_n$ should you choose?   
\vskip 1em
The \define{principle of maximum entropy} says:  
\vskip 2em
\define{Of all the probability distributions consistent with the facts \break you've learned, choose the one with the largest Shannon entropy}
\[  \scalebox{1.2}{$\displaystyle{ \color{darkred} H = - \sum_{i=1}^n p_i \log p_i }$} \]}
\end{center} \vskip 1em

 \end{minipage}} \end{center} \vskip 1em

What's Shannon entropy good for?  For starters, it gives a principle for
choosing the `best' probability distribution consistent with what you
know.   \emph{Choose the one that maximizes the Shannon entropy!}

This is 
called the `principle of maximum entropy'.  This
principle first arose in statistical mechanics, which is the application of
probability theory to physics---but we can use it elsewhere too.

For example: suppose you have a die with faces numbered 1,2,3,4,5,6.  At first you think it's fair.  But then you somehow learn that the average of the numbers that comes up when you roll it is 5.   Given this, what's the probability that if you roll it, a 6 comes up?   

Sounds like an unfair question!   But you can figure out the probability distribution on $\{1,2,3,4,5,6\}$
that maximizes Shannon entropy subject to the constraint that the mean
is 5.  According to the principle of maximum entropy, you should use
this to answer my question!

But is this correct?

The problem is figuring out what `correct' means!
But in statistical mechanics we use the principle of maximum entropy
all the time, and it seems to work well.  The brilliance of
E.\ T.\ Jaynes was to realize it's a general principle of reasoning, not
just for physics.

The principle of maximum entropy is widely used outside physics,
though still controversial.  But I think we should use it to figure out some basic properties of a
gas---like its energy or entropy per molecule, as a function of
pressure and temperature.

To do this, we should generalize Shannon entropy to `Gibbs entropy',
replacing the sum by an integral.  Or else we should `discretize' the
gas, assuming each molecule has a finite set of states.  It sort of
depends on whether you prefer calculus or programming.  Either
approach is okay if we study our gas using classical statistical
mechanics.

Quantum statistical mechanics gives a more accurate answer.  It uses a more
general definition of entropy---but the principle of maximum
entropy still applies!

I won't dive into any calculations just yet.  Before doing a gas, we
should do some simpler examples---like the die whose average
roll is 5. But I can't resisting mentioning one philosophical point.
In the box above I was hinting that maximum entropy works when your `prior' is
uniform:

\begin{quote}
\textbf{\boldmath
Suppose there are $n$ possible outcomes.  \define{At first you have no reason to think any is more probable than any other.}}
\end{quote}

\noindent
This is an important assumption: when it's not true, the principle of maximum
entropy as we've stated it does not apply.
But what if our set of events is something like a line?  There's no
obvious best probability measure on the line!  And even good old Lebesgue
measure $dx$ depends on our choice of coordinates.  To handle this, we
need a generalization of the principle of maximum entropy, called the principle
of maximum \emph{relative} entropy.

In short, a deeper treatment of the principle of maximum entropy pays
more attention to our choice of `prior': what we believe \emph{before} we
learn new facts.  And it brings in the concept of `\href{https://en.wikipedia.org/wiki/Kullback%E2%80%93Leibler_divergence}{relative entropy}':
entropy relative to that prior.   But we won't get into this here, because we will
always be using a very bland prior, like assuming that each of finitely many outcomes
is equally likely.

\vfill \eject \begin{center}\noindent\fbox{\begin{minipage}{32em}

\begin{center}
\entry{ADMITTING YOUR IGNORANCE}
\end{center}
\vskip 1.5em

\begin{center}
\textbf{\boldmath
Suppose you describe your knowledge of a system with $n$ states \\
using a probability distribution $p_1, \dots, p_n$.  \\
Then the 
Shannon information
\[  \scalebox{1}{$\displaystyle{ H = - \sum_{i=1}^n p_i \log p_i }$} \]
measures your \textit{ignorance} of the system's state.  
\vskip 1em
So, choosing the maximum-entropy probability distribution \\ consistent with 
the facts you know \\ amounts to \\
{\color{darkred} \textit{not pretending to know more than you do}}.
}
\end{center}

\end{minipage}} \end{center} \vskip 1em

Remember: if we describe our knowledge using a probability
distribution, its Shannon entropy says how much we expect to learn
when we find out what's really going on.  We can roughly say it measures our `ignorance'---though ordinary
language can be misleading here.

\begin{center}
\includegraphics[width = 10em]{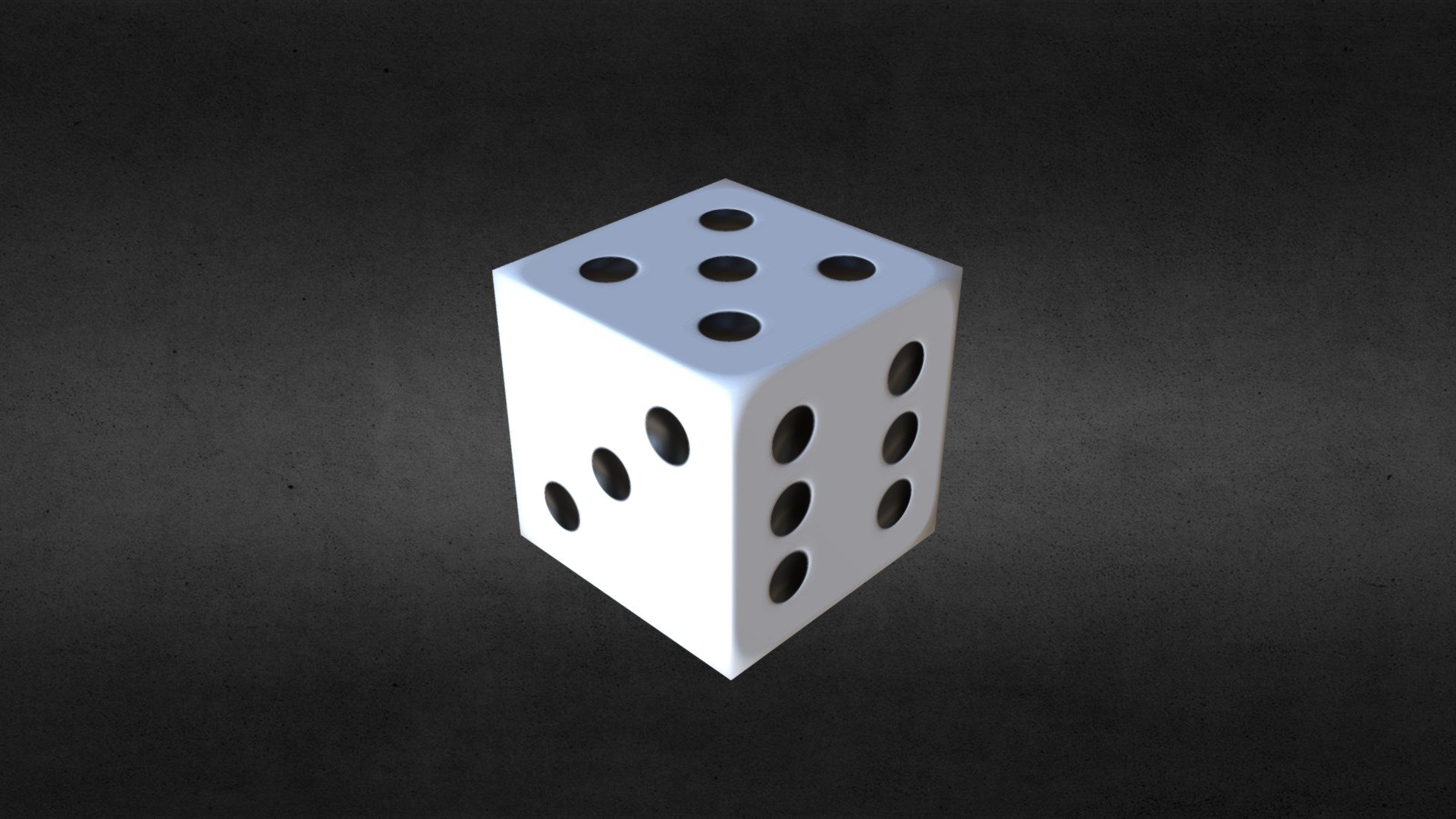} 
\end{center} \vskip 1em

At first you think this ordinary 6-sided die is fair.  But then you
learn that no, the average of the numbers that come up is 5.  What are
the probabilities $p_1, \dots, p_n$ for the different faces to come up?

This is tricky: you can imagine different answers!

You could guess the die lands with 5 up every time.  In other words, 
$p_5 = 1$.  This indeed gives the correct average.  But the entropy of this
probability distribution is 0.  So you're claiming to have no
ignorance at all of what happens when you roll the die!

Next you might guess that it lands with 4 up half the time and 6 up half the time. 
 In other words, $p_4 = p_6 = \frac{1}{2}$.
This probability distribution has 1 bit of entropy.  Now you are
admitting more ignorance.   But how can you be so sure that 5
never comes up?

Next you might guess that $p_4 = p_6 = \frac{1}{4}$ and $p_5 = \frac{1}{2}$.
We can compute the entropy of this probability distribution.  
It's higher: 1.5 bits.   
Good, you're being more honest now!  But how can you be sure that 1,
2, or 3 never come up?  You are still pretending to know stuff!

Keep improving your guess, finding probability distributions with mean
5 with bigger and bigger entropy.  The bigger the entropy gets, the
more you're admitting your ignorance!    If you do it right, your guess
will converge to the unique maximum-entropy solution.   

But there's a more systematic way to solve this problem.

 \vfill \eject \begin{center}\noindent\fbox{\begin{minipage}{34em}

\begin{center}
\entry{THE BOLTZMANN DISTRIBUTION} 

\end{center} \vskip 1em

\noindent \textbf{\boldmath Suppose you want to maximize the Shannon entropy
\[  \scalebox{1}{$\displaystyle{ - \sum_{i=1}^n p_i \log p_i }$} \]
of a probability distribution $p_1, \dots, p_n$, 
subject to the constraint that  the expected
value of some quantity $A_i$ is some number $c$:
\[    \qquad   \sum_{i=1}^n   p_i A_i = c    \qquad  (\ast) \]
Then try the \define{Boltzmann distribution:}
\[         p_i = \frac{\exp(-\beta A_i)}{\displaystyle{\sum_{i=1}^n \exp(-\beta A_i)}}  \]
If you can find $\beta$ that makes ($\ast$) hold, this is the answer you seek!
}

\end{minipage}} \end{center} \vskip 1em 
 
How do you actually \emph{use} the principle of maximum entropy?

If you know the expected value of some quantity and want to maximize
entropy given this, there's a great formula for the probability
distribution that usually does the job!   It's called the `Boltzmann distribution'.
In physics it also goes by the names `Gibbs distribution' or
`canonical ensemble', and in statistics it's called an
`exponential family'.

In the Boltzmann distribution, the probability $p_i$ is proportional to
$\exp(-\beta A_i)$ where $A$ is the quantity whose expected value
you know.  Since probabilities must sum to one, we must have
\[         p_i = \frac{\exp(-\beta A_i)}{\displaystyle{\sum_{i=1}^n \exp(-\beta A_i)}}  .\]
It is then easy to find the expected value of $A$ as a function of the
number $\beta$: just plug these probabilities into the formula
\[ \langle A \rangle =  \sum_{i=1}^n   A_i p_i  \]
The hard part is inverting this process and finding $\beta$ if you know what you
want $\langle A \rangle$ to be.

When and why does the Boltzmann distribution actually work?  That's a bit of a long story, so I'll 
explain it later.  First, let's use the Boltzmann distribution to solve the puzzle I mentioned 
last  time:

\begin{quote}
At first you think this ordinary 6-sided die is fair.  But then you
learn that no, the average of the numbers that come up is 5.  What are
the probabilities $p_1, \dots, p_n$ for the different faces to come up?
You can use the Boltzmann distribution to solve this puzzle!
\end{quote}

To do it, take $1 \le i \le 6$ and $A_i = i$.  Stick the Boltzmann
distribution $p_i$ into the formula $\sum_i A_i p_i = 5$ and get 
a polynomial equation for $\exp(-\beta)$.  You can solve this with a 
computer and get $\exp(-\beta) \approx 1.877$.

So, the probability of rolling the die and getting the number $1 \le
i \le 6$ is proportional to $\exp(-\beta i) \approx 1.877^i$.  You
can figure out the constant of proportionality by demanding that the
probabilities sum to $1$---or just look at the formula for the Boltzmann
distribution.  You should get these probabilities:
\[ p_1 \approx  0.02053, \; p_2 \approx 0.03854, \;
p_3 \approx 0.07232, \; p_4 \approx 0.1357, \;
p_5 \approx 0.2548, \; p_6 \approx 0.4781. \]
You can compute the entropy of this probability distribution, and you
get roughly $1.97$ bits.  You'll remember that last time 
we got entropies up to 1.5 bits just by
making some rather silly guesses.

So, using the Boltzmann distribution, you can find the maximum-entropy die that rolls 5
on average.   Later, we'll see how the same math lets us find the maximum-entropy state
of a box of gas that has some expected value of energy!

\vfill \eject \begin{center}\noindent\fbox{\begin{minipage}{30em}

\begin{center}
\entry{MAXIMIZATION SUBJECT TO A CONSTRAINT}

\vskip 1em
\noindent \textbf{\boldmath To maximize a smooth function $f$ of several variables \\
subject to a constraint on some smooth function $g$, \\ look for a point where
\define{
\[  \nabla f = \lambda \nabla g \]}
for some number $\lambda$.
\[
\begin{tikzpicture}
\node [] at (0, 0) {\includegraphics[scale = 1]{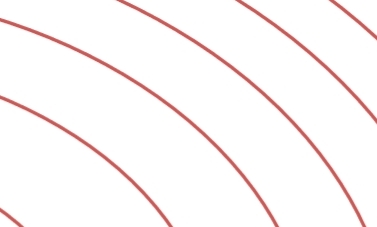}};
\node [] at (0, 0.8) {\textbf{ \boldmath{$\nabla f$}}};
\node [circle,fill=black,draw=black, scale=.3] at (-0.1,-0.1) (A) {};
\node [] at (0.6, 0.7) (B) {};
\node [] at (-1.2, -2) {};
\node [] at (1.5, -2.2) {$\scriptstyle{g = \mathrm{constant}}$};
\node [] at (3.8, -2) {};
\draw[->, -Triangle, draw=black,line width=0.5] (A) to (B);
\end{tikzpicture} 
\]
}

\end{center} \end{minipage}} \end{center} \vskip 1em 
  
When we're trying to maximize entropy subject to a constraint, we're doing a problem of the above sort.   If you don't know how to do problems like this, it's time to learn about Lagrange multipliers.  You can
find this in any book on calculus of several variables.    But the idea is in the picture above.  Say we've
got two smooth functions $f, g \maps \R^n \to \R$ and we have a point on the surface
$ g = \mathrm{constant}$ where $f$ is as big as it gets on this surface.    The
gradient of $f$ must be perpendicular to the surface at this point.   Otherwise we
could move along the surface in a way that made $f$ bigger!  For the same reason, the gradient of $g$ is \emph{always} perpendicular to the surface $g = \mathrm{constant}$.   So $\nabla f $ and $\nabla g$ must
point in the same or opposite directions at this point.  Thus, as long as the gradient of $g $ is nonzero, we must have 
\[    \nabla f = \lambda \, \nabla g \]
for some number $\lambda$, called a \define{Lagrange multiplier}.   So, solving this equation along with
\[        g = \mathrm{constant} \]
is a way to find the point we're looking for---though we still need to check we've found
a maximum, not a minimum or something else.

We can write a formula that means the exact same thing as $\nabla f = \lambda \nabla g$ using differentials:
\[    df = \lambda dg \]
This is what we'll do from now on.   Gradients are vector fields while differentials
are 1-forms.  If you don't know what this means, you can probably ignore this for now: 
the difference, while ultimately quite important, will not be significant for anything
we're doing.

\vfill \eject \begin{center}\noindent\fbox{\begin{minipage}{30em}

\hypertarget{sec:MAXIMIZING_ENTROPY_SUBJECT_TO_A_CONSTRAINT}{}

\begin{center}
\entry{MAXIMIZING ENTROPY SUBJECT TO A CONSTRAINT}
\vskip 1em
\noindent \textbf{\boldmath To maximize the entropy 
\[ \scalebox{1}{$\displaystyle{H = - \sum_{i=1}^n p_i \ln p_i }$} \]
subject to a constraint on the expected value 
\[ \scalebox{1}{$\displaystyle{\langle A \rangle =  \sum_{i=1}^n p_i A_i }$} ,\]
it's good to look for a probability distribution such that 
\define{
\[  d H = \lambda\, d \langle A \rangle  \]}
for some number $\lambda$.   This will then be a  \\ \define{Boltzmann distribution}:
\[ \color{darkred}       p_i = \frac{\exp(-\lambda A_i)}{\displaystyle{\sum_{i=1}^n \exp(-\lambda A_i)}}  \]
\[
\scalebox{0.5}{
\begin{tikzpicture}
\node [] at (0, 0) {\includegraphics[scale = 1]{contours_2.jpg}};
\node [] at (0.1, 0.8) {\textbf{ \boldmath{$d H$}}};
\node [circle,fill=black,draw=black, scale=.3] at (-0.1,-0.1) (A) {};
\node [] at (0.6, 0.7) (B) {};
\node [] at (-1.2, -2) {};
\node [] at (1.5, -2.1) {${\langle A \rangle =}$};
\node [] at (1.5, -2.5) {${\mathrm{constant}}$};
\node [] at (3.8, -2) {};
\draw[->, -Triangle, draw=black,line width=0.5] (A) to (B);
\end{tikzpicture} 
}
\]
}
\end{center} \end{minipage}} \end{center} \vskip 1em 

We've seen how to maximize a function subject to a constraint.  
Now let's do the case we're interested in: maximizing entropy subject to 
a constraint on the expected value of some quantity. 

Suppose we have a finite set of outcomes, say $1, \dots, n$.
Our `quantity' $A$ is just a number $A_1, \dots, A_n$ depending on the outcome.
For any probability distribution $p$ on the set of outcomes,
we can define its Shannon entropy and the expected value of $A$:
\[       H = - \sum_{i=1}^n p_i \ln p_i  , \qquad  \langle A \rangle = \sum_{i=1}^n p_i A_i    .\]
Here we are using base $e$ for the Shannon entropy, to simplify the calculations.
Let's try to find the probability distribution that maximizes $H$ on the surface $\langle A \rangle = c$.
We'll show that if such a probability distribution $p$ exists, and none of the $p_i$ are
zero, then $p$ must be a Boltzmann  distribution 
\[            p_i = \frac{\exp(-\lambda A_i)}{\displaystyle{\sum_{i = 1}^n  \exp(-\lambda A_i) }}\]
for some $\lambda \in \R$.   If you're willing to trust me on this, you can skip this calculation.

To use the method from last time---the Lagrange multiplier method---we'd like to use 
the probabilities $p_i$ as coordinates on the
space of probability distributions.   But they aren't independent, since
\[        \sum_{i=1}^n p_i = 1  .\]
To get around this, let's use all but one of the $p_i$ as coordinates, 
and remember that the remaining one is a function of those.   Let's use $p_2, p_3, \dots , p_n$
as coordinates, so that $p_1 = 1 - (p_2 + \cdots + p_n)$.
Furthermore, the space of all probability distributions on our finite set is
\[    \left\{p \in \R^n  \vert \; 0 \le p_i \le 1 , \; \sum_{i =1}^n p_i  = 1\right\} .\]
It looks like a closed interval when $n = 2$, or a triangle when $n = 3$, or a tetrahedron when $n = 4$, or some higher-dimensional version of a tetrahedron when $n$ is larger.  
In its interior this space looks locally like $\R^{n-1}$, so we can use the Lagrange multiplier
method, but it also has a boundary where one or more of the $p_i$ vanish, and
then this method no longer applies.   (We'll see an example of that later.)   

So, let's assume $p$ is a probability distribution maximizing 
the Shannon entropy $H$ on the surface $\langle A \rangle = c$, and also
suppose $p$ has $p_1, \dots, p_n > 0$.  Suppose that not all the values $A_i$ are equal,
since that makes the problem too easy---see why?  Then $d \langle A \rangle$ is never
zero, so from what I said last time, we must have
\[          dH = \lambda \, d \langle A \rangle \]
at the point $p$.   So let's see what this equation actually says.

Since 
\[    H = - \sum_{i =1}^n p_i \ln p_i  \]
we have
\[  
 dH = \displaystyle{ - \sum_{i=1}^n d(p_i \ln p_i)  } 
= \displaystyle{ - \sum_{i=1}^n   (1 + \ln p_i) dp_i. }
 \]
Similarly, since
\[    \langle A \rangle = \sum_{i =1}^n p_i A_i  \]
we have
\[    d \langle A \rangle = \sum_{i=1}^n A_i dp_i .\]
So, our equation $d H = \lambda \, d \langle A \rangle$ says
\[      - \sum_{i=1}^n   (1 + \ln p_i) dp_i =  \lambda \sum_{i=1}^n A_i dp_i  .\]
For these to be equal, the coefficients of $dp_i$ must agree for each of our coordinates
$p_2, \dots, p_n$.   But we have to remember that $p_1 = 1 - (p_2 + \cdots + p_n)$
and thus  $dp_1 = - (dp_2 + \cdots + dp_n)$.
Thus, for each $i = 2, \dots n$ we have
\[    (1 + \ln p_1)  - (1 + \ln p_i) = \lambda (-A_1 + A_i  )\]
and fiddling around we get
\[    \frac{p_i}{p_1} = \frac{\exp(-\lambda A_i)}{\exp(-\lambda A_1)}. \]
This says something cool: the probabilities $p_i$ are proportional to the exponentials $\exp(-\lambda A_i)$.  
And since the probabilities must sum to $1$, it's obvious what the constant of proportionality
must be:
\[            p_i = \frac{\exp(-\lambda A_i)}{\displaystyle{\sum_{i = 1}^n  \exp(-\lambda A_i) }}.  \]
So yes: $p_i$ must be given by the Boltzmann distribution!  

In summary, we've seen that \emph{if} there exists a probability distribution $p$
that maximizes the Shannon entropy among probability distributions with  $\langle A \rangle = c$,
and \emph{if} all the $p_i$ are positive, 
then $p$ must be a Boltzmann distribution.   But this raises other
questions.   When does such a probability distribution 
exist?    If it exists, is it unique?  And what if not all the $p_i$ are positive?

In what follows we'll dive down this rabbit hole and get to the bottom of it.   I'll just
state some facts---you may enjoy trying to see if you can prove them.
First,  there exists a probability distribution $p_1, \dots, p_n$ with $\langle A \rangle = c$ if
and only if 
\[    A_{\textrm{min}} \le  c \le A_{\textrm{max}}  \]
where $A_{\textrm{min}}$ is the minimum value and $A_{\textrm{max}}$ is the maximum value of the numbers $A_1, \dots, A_n$.   
Second, whenever 
\[          A_{\textrm{min}} \le c \le A_{\textrm{max}} , \]
 there exists a unique probability distribution $p_1, \dots, p_n$ maximizing Shannon entropy
 subject to the constraint $\langle A \rangle = c$.
Third, this unique maximizer $p$ has $p_i > 0$ for all $i$, and is thus a Boltzmann distribution,
whenever
\[          A_{\textrm{min}} < c < A_{\textrm{max}} . \]
When $c = A_{\textrm{min}}$, the unique maximizer assigns the same probability $p_i$
to each outcome $i$ with $A_i =  A_{\textrm{min}}$, while $p_i = 0$ for all other outcomes.
Similarly, when $c = A_{\textrm{max}}$,  the unique maximizer assigns the same probability $p_i$
to each outcome $i$ with $A_i =  A_{\textrm{max}}$, while $p_i = 0$ for all other outcomes.

It's good to look at an example:

\begin{puzzle} 
\label{puzzle:limiting_case}
Suppose there are only two outcomes, with $A_1 = -1$ and $A_2 = 1$.   Work out the Boltzmann distribution $p$ maximizing Shannon entropy subject to the constraint $\langle A \rangle = c$
for $-1 < c < 1$.   Show that as $\lambda \to +\infty$ this Boltzmann distribution has
\[        p_1 \to 1, p_2 \to 0 \]
while as $\lambda \to -\infty$ it has
\[      p_1 \to 0, p_2 \to 1 .\]
Show the probability distribution $p_1 = 1, p_2 = 0$ maximizes Shannon entropy subject
to the constraint $\langle A \rangle = -1$, while $p_1 = 0, p_2 = 1$ maximizes it subject
to the constraint $\langle A \rangle = 1.$  Show these two probability distributions are not Boltzmann distributions.
\end{puzzle}

\vfill \eject \begin{center}\noindent\fbox{\begin{minipage}{30em}

\begin{center}
\entry{THERMAL EQUILIBRIUM} 
\end{center} \vskip 1em

\noindent \textbf{\boldmath 
Suppose a system has finitely many states $i = 1, \dots, n$ with \\ energies $E_i$.
If the probability $p_i$ that it's in the $i$th state maximizes entropy subject to a
constraint on its expected energy:
\[              \langle E \rangle =  \sum_{i = 1}^n p_i E_i \]
we say it is in \define{thermal equilibrium}.
In this case $p_i$ is given by a \define{Boltzmann distribution} 
\[         p_i = \frac{\exp(-\beta E_i)}{\displaystyle{\sum_{i=1}^n \exp(-\beta E_i)}}  \]
at least if all the probabilities $p_i$ are positive.
}
 
 \end{minipage}} \end{center} \vskip 1em
 
Don't worry: the substance of what I'm saying here is almost the same as in the last box.  I'm merely attaching new words to the concepts, to make them sound more like physics:
\begin{itemize}
\item Before I said we had a set of $n$ `outcomes' numbered $1, 2, \dots, n$.   Now I'm talking
about `states'.    If we have a system with $n$ states, it means there are $n$ outcomes
when we do a measurement to completely determine which state it's in.   A `state' is some way 
for a physical system to be---that's vague but it's all we can say until we consider some specific
kind of system.   In classical physics the states form a set, usually infinite but sometimes finite.  
\item Before I said we had a `quantity' $A$ that depends on the outcome, taking the value $A_i$ 
in the $i$th outcome.  Now I'm calling this quantity the `energy' $E$.   Energy is a particularly interesting
quantity in physics, so we'll focus on that, without demanding that you know anything about it: for our present purposes, we can take any quantity and dub it `energy'.
\item Before I called the Lagrange multiplier $\lambda$.  Now I'm calling it $\beta$, because that's what physicists do in this particular context.
\end{itemize}

When a system maximizes entropy subject to a constraint on the expected value of its energy, and perhaps
also some other quantities, we say the system is in \define{thermal equilibrium}.  This is meant to suggest that an object just sitting there, not heating up or cooling down, is often best modeled this way.

You may have noticed the annoying clause ``at least if all the probabilities $p_i$ are positive''.
I only said that because I cannot tell a lie.   In Puzzle \ref{puzzle:limiting_case} we saw that as $\beta \to \pm \infty$, the Boltzmann distribution can converge to a non-Boltzmann 
probability distribution where some of the probabilities $p_i$ vanish.   This still counts as
thermal equilibrium, because it's still maximizing entropy subject to a constraint on expected
energy.  We'll learn more about this when we study the concept of `absolute zero'.

 \vfill \eject \begin{center}\noindent\fbox{\begin{minipage}{25em}

\begin{center}
\entry{COOLNESS}
\vskip 1.5em
\end{center} \vskip 1em

\noindent \textbf{\boldmath 
If a probability distribution $p_i$ 
maximizes entropy subject to a
constraint on the expected value of the energy $E_i$, then
\vskip 0.5em
\[      \scalebox{1.2}{$ p_i \propto e^{-\beta E_i} $} \]
\vskip 0.7em \noindent
where $\beta$ is the {\textbf {\color{blue}coolness,}}  inversely proportional to temperature.  So:
\vskip 1em
\begin{center}
{\color{blue}The cooler a system is, the less likely it is to be in a high-energy state!}
\end{center}
}

\end{minipage}} \end{center} \vskip 1em 

Say a system with finitely many states maximizes entropy subject to a
constraint on the expected value of some quantity $E$ that we choose to call `energy'.  Then its
probability of being in the $i$th state is proportional to $\exp(-\beta E_i)$ for some number
$\beta$.

When $\beta$ is big and positive, the probability of being in a state of high
energy is tiny, since $\exp(-\beta E_i)$ gets very small for large
energies $E_i$.  This means our system is \emph{cold}.

Conversely when $\beta$ is small and positive, $\exp(-\beta E_i)$ drops off very slowly
as the energy $E_i$ gets bigger.  So, high-energy states become quite likely when $\beta$
is small and positive.   This means our system is \emph{hot}.

It turns out $\beta$ is inversely proportional to the temperature---more about that later.  But in modern physics $\beta$ is just as important as
temperature.  It comes straight from the principle of maximum entropy!

So $\beta$ deserves a name.   And its name is `\textbf{\color{blue} coolness}'.

By the way, the formula
\[      p_i \propto e^{-\beta E_i} \]
is only strictly true when $\beta$ is finite.  There's also a limiting case $\beta \to +\infty$, 
when $p_i = 0$ except for states of the very lowest energy.   And there's a limiting case
$\beta \to -\infty$, where $p_i = 0$ where except for states of the very \emph{highest}
energy.   I'll say a bit about these oddities later.   First I'll say more about what coolness has to do
with temperature.

\vfill \eject \begin{center}\noindent\fbox{\begin{minipage}{30em}

\begin{center}
\entry{COOLNESS VERSUS TEMPERATURE}
\vskip 1.5em

\noindent \textbf{\boldmath 
{\textbf {\color{blue} Coolness $\beta$ }} is inversely proportional to
\define{temperature $T$:}
\[      {\color{blue}    \beta} = \frac{1}{k  \color{darkred}T} \]
where $k$ is \textbf{Boltzmann's constant}.  \vskip 1em  
{\color{blue} Coolness} is measured in joules${}^{-1}$, \\
{\color{darkred} temperature }is measured in kelvin, and \\
Boltzmann's constant is a conversion 
factor:
\[            k = 1.380649 \cdot 10^{-23}\; \frac{\textrm{joules}}{\textrm{kelvin}} \]
}
\end{center}
 \end{minipage}} \end{center} \vskip 1em

In statistical mechanics, coolness is inversely proportional to
temperature.  But coolness has units of energy${}^{-1}$, not
temperature${}^{-1}$.  So we need a constant to convert between
coolness and inverse temperature!  And this constant is very
interesting.

Remember: when a system maximizes entropy with a constraint on its
expected energy, the probability of it having energy $E$ is
proportional to $\exp(-\beta E)$ where $\beta$ is its coolness.
But we can only exponentiate dimensionless quantities!   (Why?)  
So $\beta$ has dimensions of 1/energy.

Since coolness is inversely proportional to temperature, we must have
$\beta = 1/kT$ where $k$ is some constant with dimensions of
energy/temperature.   This constant $k$ is called `Boltzmann's constant'.   It's tiny:
\[   k = 1.380649 \cdot 10^{-23} \; \textrm{joules/kelvin}.   \]
This is mainly because we use units of
energy, joules, suited to macroscopic objects like a cup of hot water.
Boltzmann's constant being tiny reveals that such things have
enormously many microscopic states!

Later we'll see that a single classical point particle, in empty
space, has energy $3kT/2$ when it's maximizing entropy at
temperature $T$.  The 3 here is because the atom can move in 3
directions, the $1/2$ because we integrate $x^2$ to get this
result.  The important part is $kT$.  The $kT$ says: if an ideal
gas is made of atoms, each atom contributes just a tiny bit of energy
per kelvin, or degree Celsius: roughly $10^{-23}$ joules.  So a little bit of
gas, like a gram of hydrogen, must have roughly $10^{23}$ atoms in
it.  This is a very rough estimate, but it's a big deal.

Indeed, the number of atoms in a gram of hydrogen is about $6 \cdot
10^{23}$.  You may have heard of Avogadro's number---this is
quite close to that.  So Boltzmann's constant gives a hint that matter is made of atoms---and 
even better, a nice rough estimate of how many per gram!

Later we will see that Boltzmann's constant has another important meaning:
it's a fundamental unit of entropy, a nat, expressed in joules/kelvin.

 \vfill \eject \begin{center}\noindent\fbox{\begin{minipage}{30em}

\begin{center}
\entry{TEMPERATURE}
\end{center} \vskip 1em

\begin{center}
\textbf{\boldmath 
If a system has finitely many states with energies $E_i$, \\
in thermal equilibrium at temperature $T$  \\ the probability that it's in the $i$th state is 
\[   \color{darkred}   p_i \propto  \exp(-E_i/kT) \]
where $k$ is Boltzmann's constant and \\
$T$ can be positive, negative, or even infinite:}
\[
\begin{tikzpicture}
\node [] at (0, 0) {\includegraphics[width = 10 em]{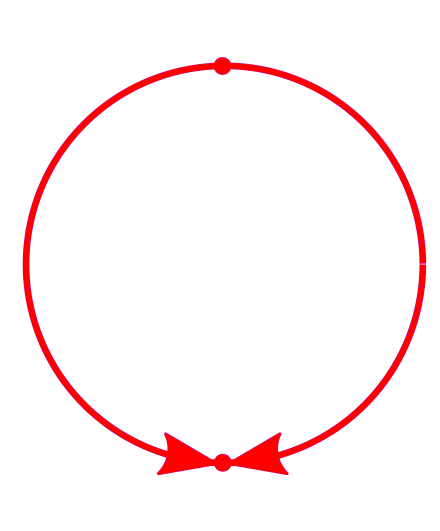}};
\node [] at (0, 2.2) {\textbf{ \boldmath{$T = \infty$}}};
\node [] at (-2, -1.5) {\textbf{ \boldmath{$T < 0$}}};
\node [] at (2, -1.5) {\textbf{ \boldmath{$T > 0$}}};
\end{tikzpicture} 
\]
\end{center} 

 \end{minipage}} \end{center} \vskip 1em

A system with finitely many states can be pretty weird.  It can have negative temperature!   Even weirder: as you heat it up, its temperature becomes large and positive, then it reaches infinity, and then it `wraps around' and become large and negative.

The reason: coolness is more fundamental than temperature.   The coolness $\beta$ is inversely proportional to the temperature $T$:
\[           \beta = 1/kT. \]
 When the temperature goes up to infinity and then suddenly becomes a large negative number, it's really just the coolness going down to zero and becoming negative.   Temperatures `wrap around' infinity, as shown in the picture.

A system with finitely many states can have negative or infinite temperature
because in thermal equilibrium, its probability of being in the $i$th state is 
\[         p_i = \frac{\exp(-\beta E_i)}{\displaystyle{\sum_{i=1}^n \exp(-\beta E_i)}}  \]
where $E_i$ is the energy of the $i$th state, and this makes sense for any $\beta \in \R$.   Moreover, the probability \(p_i\) depends continously on $\beta$, even as $\beta$ passes through zero.   This means a large positive temperature is almost like a large negative temperature!   

But the circle of temperature can be misleading.  
Temperatures wrap around $T = \infty$ but not $T = 0$.  A system with a small
positive temperature is very different from one with a small negative
temperature!  That's because $p_i$ for $\beta \gg 0$ is very different than it is for $\beta \ll 0$. 

For a system with finitely many states we can take
the limit of the Boltzmann distribution when $\beta \to +\infty$; then the system will only
occupy its lowest-energy state or states.  We can also take the limit when
$\beta \to -\infty$; then the system will only occupy its highest-energy
state or states.  In terms of temperature, this means that the limit where $T$ approaches
zero from above is very different than the limit where $T$ approaches zero from below.

So, for a system with finitely many states, the best picture of
possible thermal equilibria is not a circle of temperatures but a closed interval
of coolness: the coolness $\beta$ can be anything in $[-\infty, +\infty]$, which
topologically is a closed interval.   In terms of coolness, $+\infty$
is different from $-\infty$, but approaching $0$ from above is the
same as approaching it from below.   But in terms of temperature, approaching $0$ 
from above is different from approaching $0$ from below, while a temperature of $+\infty$ 
is the same as a temperature of $-\infty$.

Now, if all this seems very weird, here's why: we often describe
physical systems using infinitely many states, with a lowest possible
energy but no highest possible energy.  In this case the sum in the
Boltzmann distribution can't converge for $\beta < 0$, so negative
temperatures are ruled out.

However, some physical systems can be \emph{approximately} described using a finite set
of states (or in quantum theory, a finite-dimensional Hilbert space of states).  
Then the things I just said hold true!   And people enjoy studying these systems, and their
strange properties, in the lab.

It's good to look at a simple example, and work everything out explicitly:

\begin{puzzle}
\label{puzzle:two-state_system}
Suppose a system has two states with energies $E_1 < E_2$.   Compute
the probabilities $p_i$ that it is in either of these states in thermal equilibrium 
as a function of the coolness $\beta$.    Then express these probabilities as a function of 
the temperature $T$.    Using these functions $p_i(T)$: 
\begin{itemize}
\item Show that when $0 < T < +\infty$ the system is more likely to be in the lower-energy
state: $p_1(T) > p_2(T)$.
\item Show that when $-\infty < T < 0$ the system is more likely to be in the higher-energy
state: $p_1(T) < p_2(T)$.  
\item Show that 
\[        \lim_{T \to +\infty} p_i(T) = \lim_{T \to -\infty} p_i(T)  \]
so we can speak unambiguously of the probabilities $p_i$ at infinite temperature.
\item Show that at infinite temperature the system has an equal probability of being
in either state.
\item Show that as $T$ approaches zero from above, the probability of the system being
in the lower energy state approaches $1$.
\item Show that as $T$ approaches zero from below, the probability of the system being
in the higher energy state approaches $1$.
\end{itemize}
\end{puzzle}

 \vfill \eject \begin{center}\noindent\fbox{\begin{minipage}{30em}

\begin{center}
\entry{INFINITE TEMPERATURE}
\vskip 1.5em

\noindent \textbf{\boldmath 
If a system has finitely many states with energies $E_i$, \\
in thermal equilibrium at temperature $T$ \\
the probability that it's in the $i$th state is 
\vskip 0.5em
\[   \color{darkred}    \scalebox{1.2}{$ p_i \propto e^{-\beta E_i} $} \]
\vskip 0.5em
where $\beta = 1/kT$ and $k$ is Boltzmann's constant.
\vskip 1em \noindent
{\color{red}
When $\beta = 0$ the system's temperature becomes infinite, \\ and all states become
equally probable!
}}
\end{center} \vskip 1em

 \end{minipage}} \end{center} \vskip 1em 
 
The probability of finding a system in a particular state decays
exponentially with energy when the coolness $\beta$ is positive.
But for a system with finitely many states, $\beta$ can be zero.
Then it becomes equally probable for the system to be in any state!

Zero coolness means `utter randomness': that is, maximum entropy.

Here's why.  The probability distribution with the largest entropy is
the one where all probabilities $p_i$ are all equal.  This happens at
zero coolness!  When $\beta = 0$ we get $\exp(-\beta E_i) = 1$ 
for all $i$.  The probabilities $p_i$ are proportional to these
numbers $\exp(-\beta E_i) = 1$, so they're all equal.

It seems zero coolness is impossible for a system with infinitely many
states.  With infinitely many states, all equally probable, the
probability of being in any state would be zero.  In other words,
there's no uniform probability distribution on an infinite set.

One way out: replace sums with integrals.  For the usual measure
on $[0,1]$,  called the Lebesgue measure $dx$, we have $\int_0^1 dx = 1$.  
So this is a `probability measure'
that we could use to describe a system at zero coolness, whose space
of states is $[0,1]$.

But replacing sums by integrals raises all sorts of interesting
issues.  For example, there's a unique way to sum over a finite set of
states, but an integral over an infinite set of states depends on a choice
of measure.    So a choice of measure is a significant extra structure 
we're slapping on our set of states.

We'll need to think about these issues later, since to compute the entropy of a
classical ideal gas we'll need integrals.  But we'll
encounter difficulties, which are ultimately resolved using quantum
mechanics.

Anyway: infinite temperature is really zero coolness, and at least for
systems with finitely many states, the entropy becomes as large as
possible at zero coolness.

 \vfill \eject \begin{center}\noindent\fbox{\begin{minipage}{30em}

\begin{center}
\entry{NEGATIVE TEMPERATURE}
\vskip 1em

\textbf{\boldmath 
If a system has finitely many states with energies $E_i$, \\
in thermal equilibrium at temperature $T$ \\
the probability that it's in the $i$th state is 
\vskip 0.5em
\[   \color{darkred}    \scalebox{1.2}{$ p_i \propto e^{-\beta E_i} $} \]
\vskip 0.5em
where $\beta = 1/kT$ and $k$ is Boltzmann's constant.
\vskip 1em \noindent
{\color{red} When $\beta < 0$ the system becomes `hotter than infinitely hot'. \\ 
Its temperature is negative---but the higher the
energy of a state, the more probable it is!}
}
\end{center} 
\end{minipage}} \end{center} \vskip 1em

A system with finitely many states can reach infinite
temperature.  It can get even hotter, but then its temperature `wraps
around' and become negative!    

The possibility of negative temperatures was first discussed by the physicist 
Lars Onsager in 1949, and they have been created in the lab with a variety of
systems that---within some approximation---can be described as having
finitely many states.  In quantum theory, this happens for systems that have
finite basis of `energy eigenstates': states with well-defined energies $E_i$.
For example, the nucleus of an atom may have just two spin states, and if we put it in
a magnetic field these will have different energies.  The result is the system we
studied in Puzzle \ref{puzzle:two-state_system}.

Here is a generalization with more energy states, all equally spaced:

\begin{puzzle}
Consider a system with $2N+1$ states labeled by an integer $n$ with $-N \le n \le N$, where the $n$th state has energy $E_n  =  \alpha n$ for some energy $\alpha > 0$.   Compute the Boltzmann distribution for this system at coolness $\beta$ for all $\beta \in \R$.   Compute the expected energy $\langle E \rangle$ as a function of $\beta$.  What is the qualitative difference in your result between the case
of positive temperature ($\beta > 0$) and negative temperature ($\beta < 0$)?
\end{puzzle}

For more, try this:

\begin{itemize}
\item Wikipedia, \href{https://en.wikipedia.org/wiki/Negative_temperature}{Negative
temperature}.
\end{itemize}

\vfill \eject \begin{center}\noindent\fbox{\begin{minipage}{30em}

\hypertarget{sec:ABSOLUTE_ZERO}{}

\begin{center}
\entry{ABSOLUTE ZERO: THE LIMIT OF INFINITE COOLNESS}
\end{center} \vskip 1em

\noindent \textbf{\boldmath 
If a system
with finitely many states having energies $E_i$ is in thermal equilibrium, the probability  $p_i$ that it's in the $i$th state is proportional to $\exp(-\beta E_i)$
where $\beta$ is the coolness.
\vskip 1em \noindent
In the limit of infinite coolness, $\beta \to +\infty$, these probabilities go to zero
except for the states of lowest energy, which all become equally probable.
\vskip 1em \noindent
The limit $\beta \to + \infty$ is also the limit where $T$ approaches zero from above,
commonly called {\color{blue} absolute zero}.
}

 \end{minipage}} \end{center} \vskip 1em

The limit where $T$ approaches zero from above is often called \define{absolute zero}.  Why?
First people made up
various temperature scales like Celsius, where zero was the freezing point of
water, and Fahrenheit, where zero is the freezing point of a 
mixture of water, ice, and ammonium chloride.   But researchers discovered
that nature had a more fundamental concept of zero temperature:
the limit of infinite coolness!    This happens as the temperature approaches  $-273.15 \; {}^\circ$C, or
roughly $-459.67 \; {}^\circ$F.    This discovery led Kelvin to propose a shifted version of
Celsius where zero is absolute zero.    This was originally called `absolute Celsius',
but now it is called the \define{Kelvin scale}.    This is the scale of temperature I'll always use here.
The size of the degrees is a somewhat arbitrary convention, but the zero is not: it's 
absolute zero.

\hypertarget{sec:HAGEDORN_TEMPERATURE}{}

\vfill \eject \begin{center}\noindent\fbox{\begin{minipage}{30em}

\begin{center}
\entry{THE HAGEDORN TEMPERATURE}
\end{center} \vskip 1em

\noindent \textbf{\boldmath 
If a system has a countable infinity of states $n = 1, 2, 3, \dots $ \\ with energies 
$E_n$, the Boltzmann distribution
 \[         p_n = \frac{\exp(-E_n/kT)}{\displaystyle{\sum_{n=1}^\infty \exp(-E_n/kT)}}  \]
is either: 
\vskip 1em
1) defined for all $0 < T < +\infty$ \\ \\
2) undefined for all $0 < T < +\infty$ \\ \\
3) defined for all $0 < T < T_{\mathrm{H}}$ but not for $ T_{\mathrm{H}}< T < +\infty$,  
where $T_{\mathrm{H}}$ is some temperature called the \define{Hagedorn temperature}.
}

 \end{minipage}} \end{center} \vskip 1em
 
 We've been discussing systems with finitely many states, but many physical systems have a countable infinity of states.   So let's think a bit about those.   We can copy everything we've done so far, but we have to be careful.    For thermal equilibrium to be possible at some temperature $T$, we need the 
Boltzmann distribution 
 \[         p_n = \frac{\exp(-E_n/kT)}{\displaystyle{\sum_{n=1}^\infty \exp(-E_n/kT)}}  \]
 to make sense.   But it might not.   Sometimes the sum fails to converge!    This happens when
 the terms $\exp(-E_n/kT)$ don't go to zero fast enough as $n \to +\infty$.

Let's investigate this issue.   We'll assume that
\[   \sum_{n=1}^\infty \exp(-E_n/kT)  \]
converges for some $T > 0$.  Then the energies $E_n$ must be bounded below: otherwise the terms
$\exp(-E_n/kT)$ will get bigger and bigger.     Furthermore for any $E \in \R$ there can be at most finitely many $E_n$ less than $E$: otherwise we'd be adding up infinitely many terms greater than $\exp(-E/kT)$.   As a result, we can reorder the states so their energies are nondecreasing:
\[    E_1 \le E_2 \le E_3 \le \cdots  \]
and $E_n \to +\infty$.   

Reordering a sum can't change its convergence or value if it's a sum of nonnegative numbers, like the sum we have here.  So we might as well assume we've listed the energies in nondecreasing order as above.    Then there are two cases:
\begin{enumerate}
\item The energies $E_n$ approach $+\infty$ so fast that $\sum_{n=1}^\infty \exp(-E_n/kT)$ converges for all  $0 < T < +\infty$.   Then our system can be in thermal equilibrium at any finite positive
temperature.   This is the nicest situation, and the one we typically expect..
\item  The energies $E_n$ approach $+\infty$ slowly enough that  $ \sum_{n=1}^\infty \exp(-E_n/kT)$ converges when $T$ is small enough, but not otherwise.    In this case there is some temperature $T_{\mathrm{H}}$, called the \define{Hagedorn temperature}, such that our system can be in thermal
equilibrium at positive temperatures $T$ below $T_{\mathrm{H}}$, but not above $T_{\mathrm{H}}$.    
\end{enumerate}
\noindent In both cases,  $\sum_{n=1}^\infty \exp(-E_n/kT)$ diverges for all $-\infty \le T < 0$ and $T = +\infty$.  So, for a system with a countable infinity of states, if thermal equilibrium exists for some positive temperature, it cannot exist for negative or infinite temperatures.  

The second case is weird and interesting.   It's named after Rolf Hagedorn, who in 1964 noticed that this was a possibility in nuclear physics.   He considered a model of nuclear matter where the energies $E_n$ approach $+\infty$ in a roughly logarithmic way.   As you heat it, its expected energy keeps increasing, but its temperature can never exceed $T_{\mathrm{H}}$.   This model turned out to be
incorrect, but it's interesting anyway.

Now let's solve some puzzles on systems with a countable infinity of states.  Some of these show up
in quantum mechanics, but you don't need to know quantum mechanics to do these puzzles.

\begin{puzzle}
Show that for a system with a countable infinity of states, if thermal equilibrium is possible for some
negative temperature, it is impossible for positive or infinite temperatures.
\end{puzzle}

\begin{puzzle}  
Work out the Boltzmann distribution when $E_n = n E$ for some energy $E$, and show that it is well-defined for all temperatures $0 < T < +\infty$.
\end{puzzle}

The next puzzle is a lot like the previous one---a bit more messy, but worthwhile because of its great importance in physics.

\begin{puzzle}
For a system called the \href{https://en.wikipedia.org/wiki/Quantum_harmonic_oscillator}{quantum harmonic oscillator} of frequency $\omega$ we have $E_n = (n + \frac{1}{2}) \hbar \omega$, where $\hbar$ is the reduced Planck's constant.   Work out the Boltzmann distribution in this case, and 
show it is well-defined for all temperatures $0 < T < +\infty$.   
\end{puzzle}

\begin{puzzle}
For a system called the \href{https://en.wikipedia.org/wiki/Primon_gas}{primon gas} we have $E_n = E \ln n$ for some energy $E$.  Show that the Boltzmann distribution is well-defined for 
small enough positive temperatures, but there is a Hagedorn temperature.    Give a formula for the
Boltzmann distribution in terms of the \href{https://en.wikipedia.org/wiki/Riemann_zeta_function}{Riemann zeta function}:
\[        \zeta(s) = \sum_{n = 1}^\infty n^{-s}   .\]
\end{puzzle}

You can show that for the primon gas the sum $\sum_{n = 1}^\infty \exp(-E_n/kT)$ diverges at the Hagedorn temperature.   But it can go the other way, too:

\begin{puzzle}
Find energies $E_n$ with a Hagedorn temperature such that $\sum_{n = 1}^\infty \exp(-E_n/kT)$ converges at the Hagedorn temperature.
\end{puzzle}

Various other strange things can happen, as you should expect when dealing with infinite series.  For example, it's possible that the Boltzmann distribution is well-defined at some temperature but the expected value of the energy is infinite!   But I'll resist the lure of these rabbit holes and turn to something much more important: systems with a \emph{continuum} of states.   We will need to get
good at these to compute the entropy of hydrogen.   Now our sums become integrals, and various new things happen.

 \vfill \eject \begin{center}\noindent\fbox{\begin{minipage}{30em}
 
 \hypertarget{sec:THE_FINITE_VERSUS_THE_CONTINUOUS}{}

\begin{center}
\entry{THE FINITE VERSUS THE CONTINUOUS }
\vskip 1em
\begin{tabular}{cc}  
\textbf{THE FINITE}   & \textbf{THE CONTINUOUS}  \\   \\           
\textbf{\boldmath $p$ a probability distribution }  &
\textbf{\boldmath $p$ a probability distribution  }
\\
\textbf{\boldmath  on  $\{1,\dots,n\}$}  &
\textbf{\boldmath on  $\R$ }
\\  \\  
\textbf{\boldmath Gibbs entropy  } &
\textbf{\boldmath Gibbs entropy } 
\\  
\textbf{\boldmath $\displaystyle{S(p) = - k\sum_{i =1}^n p_i \ln p_i}$ } &
\textbf{\boldmath  $\displaystyle{S(p) = - k\int_{-\infty}^\infty p(x) \ln p(x) \, d x}$ } 
\\  \\
\textbf{\boldmath $S(p)$ always $\ge 0$ } & \textbf{\boldmath $S(p)$ not always $\ge 0$ } 
\\ \\
\textbf{\boldmath $S(p)$ always finite }  & \textbf{\boldmath $S(p)$ not always finite } 
\\ \\
\textbf{\boldmath $S(p)$ invariant under } & 
\textbf{\boldmath $S(p)$ not invariant under } \\
\textbf{\boldmath  permutations of $\{1,\dots, n\}$ } & 
\textbf{\boldmath  reparametrizations of $\R$ } \\
\end{tabular} 
\\
\end{center} \vskip 1em

\end{minipage}} \end{center} \vskip 1em

You can switch from finite sums to integrals in the definition of
entropy, and we'll need to do this to compute the entropy of hydrogen.
But be careful: a bunch of things change!   

We need to switch from finite sums to integrals when we switch from a finite set
of states to a \hyperlink{sec:PROBABILITY_MEASURES}{measure space} of states.
I'll illustrate the ideas with the real line, $\R$.   We
define a \define{probability distribution} on $\R$ to be an integrable 
function $p \maps \R \to [0,\infty)$ with
\[              \int_{-\infty}^\infty p(x) \, dx = 1 .\]
Such a probability distribution has a \define{Gibbs entropy}
given by
\[           S(p) = - k \int_{-\infty}^\infty p(x) \ln p(x) \, dx. \]
We can also define \define{Shannon entropy}, where we leave out Boltzmann's constant
$k$ and use whatever base we want for the logarithm:
\[          H(p) = - \int_{-\infty}^\infty p(x) \log p(x) \, dx. \]
I should warn you that many writers reserve the term `Shannon entropy' only for a sum
\[           H(p) = -\sum_{i \in X} p_i \log p_i  .\]
While that convention has advantages,  I want to use the term `Shannon entropy' to signal
that I'm leaving out the factor of $k$.

Unlike the entropy for a probability distribution on a finite set,
the entropy of a probability distribution on $\R$ can be negative!   This is 
disturbing.   Earlier I said that the Shannon entropy of a probability distribution 
is the expected amount of information you learn when an outcome is
chosen according to that distribution.    How can this be negative?   

The answer is that this interpretation of entropy, valid for probability distributions
on a finite or even a countably infinite set, is \emph{not true} in the continuous case!
We have to adapt our intuitions.

Look at an example.  Let $p_\epsilon$ be the probability distribution on $\R$ given by
\[     p_{\epsilon}(x) =
\left\{   \begin{array}{ccl}
 \displaystyle{\frac{1}{\epsilon}}  & \textrm{if } 0 \le x \le \epsilon  \\ \\
 0                              & \textrm{otherwise.} 
 \end{array} \right. 
 \]
 For small $\epsilon$ it's a tall thin spike near $0$.   Let's work out its Shannon entropy:
 \[     \begin{array}{ccl}         H(p) &=& \displaystyle{ - \int_{-\infty}^\infty p(x) \log p(x) \, dx }  \\ \\
 &=& \displaystyle{ -\int_0^\epsilon \frac{1}{\epsilon} \log \frac{1}{\epsilon} \, dx } \\ \\
 &=& \log \epsilon .
 \end{array}
 \]
 We're just integrating a constant here, so it's easy.   When $\epsilon = 1$ the entropy is zero,
 and when $\epsilon$ becomes smaller than $1$ the entropy becomes negative!
 
 Why?   We need a distance scale to define the entropy of a 
 probability distribution on the real line.   If I measure distance in centimeters, I'll think the entropy of a probability distribution is bigger than you, who measures it in meters.  And if I measure it in kilometers,
 I'll think the entropy is smaller---and possibly even negative.
 
Let's see how this works.  If I measure distance in different units from you, my coordinate
 $y$ on the real line will not equal your coordinate $x$: instead we'll have
 \[                y = c x \]
 for some $c > 0$.   Then my probability distribution, say $q$, will have
 \[        \int_{-\infty}^\infty q(y) \, dy = \int_{-\infty}^\infty q(cx) \, d(cx) =  c \int_{-\infty}^\infty q(cx) \, dx \]
so we must have
 \[        q(cx) = \frac{1}{c} p(x)   \]
 to make this integral equal $1$.   In other words, stretching out a probability distribution must also flatten it out, making it less `tall'---and its entropy increases.   Indeed:
 
 \begin{puzzle}
 Show that $H(q) = H(p) + \ln c $.
 \end{puzzle}
 
 Thanks to this formula choosing $0 < c < 1$ compresses a probability distribution and makes it
 taller, reducing its entropy.  Inevitably, this can make the entropy negative if $c$ is small enough.
  
 In summary: in the continuous case, entropy is not invariant under reparametrizations: our choice of coordinates matters!  And this can make entropy negative.  This applies not only to $\R$ but many other
 measure spaces we'll be considering, like $\R^n$.  This issue will be very important.
 
After learning this, it should be less of a shock that the entropy of a probability distribution on $\R$
can be infinite, or even undefined:

\begin{puzzle}
Find three probability distributions $p$ on the real line that have entropy $+\infty$, $-\infty$, and
undefined because it's of the form $+\infty - \infty$.  
\end{puzzle}

\vfill \eject \begin{center}\noindent\fbox{\begin{minipage}{30em}

\begin{center}
\entry{ENTROPY, ENERGY AND TEMPERATURE}
\vskip 1.5em
\noindent \textbf{\boldmath 
Suppose a system has some measure space $X$ of states with energy $E \maps X \to \R$.
In thermal equilibrium the probability distribution on states, $p \maps X \to \R$, maximizes the
Gibbs entropy
\[ \color{darkred}  S = - k \int_X p(x) \ln p(x) \, dx  \]
subject to a constraint on the expected value of energy:
\[    \color{darkred} \langle E \rangle = \int_X p(x) E(x) \, dx \]
Typically when this happens $p$ is the Boltzmann distribution
\[    \color{darkred}    p(x) = \frac{e^{-E(x)/kT}}{\displaystyle{\int_X e^{-E(x)/kT} \, dx}} \]
where $T$ is the temperature and $k$ is Boltzmann's constant.
 \vskip 1em   
Then as we vary $ \langle E \rangle$ we have
\vskip 0.3em
\[   \color{darkred}    \scalebox{1.4}{$ d\langle E\rangle = T d S $} \]
}

\end{center}  \end{minipage}} \end{center} \vskip 1em 

We can now generalize a lot of our work from a finite set of states to a general 
measure space.  I won't redo all the arguments, just state the results and point out a
couple of caveats. 

For any measure space $X$ we say a function $p \maps X \to [0,\infty)$ is a
\define{probability distribution} if it's measurable and
\[     \int_X p(x) \, dx = 1 .\]
We can define a version of \define{Shannon entropy} for $p$ by
\[   H = - \int_X p(x) \log p(x) \, dx, \]
but physicists mainly use the \define{Gibbs entropy}, defined by
\[   S = - k \int_X p(x) \ln p(x) \, dx. \]
As I warned you last time, this can take values in $[-\infty,\infty]$, though we are mainly
interested in cases when it's finite.   If we think of $X$ as the space of states of some system,
we can pick any measurable function $E \maps X \to \R$ and call it the `energy'.    
Its \define{expected value} is then
\[    \langle E \rangle = \int_X E(x) p(x) \, dx \]
at least when this integral converges.   

We say the probability distribution $p$ describes \define{thermal equilibrium}
if it maximizes $S$ subject to a constraint $\langle E \rangle = c$.    Typically when this happens
 $p$ is a \define{Boltzmann distribution}
\[     
  p(x) = \frac{\displaystyle{e^{-\beta E(x)}}}{\displaystyle{\int_X e^{-\beta E(x)} \, dx}}  
 \]  
 where $\beta$ is called the \define{coolness}.      I say `typically' because even
 when $X$ is a finite set, we saw in Puzzle \ref{puzzle:limiting_case} that there can 
 be thermal equilibria that are not Boltzmann distributions, but only
 \emph{limits} of Boltzmann distributions as $\beta \to +\infty$
 or $\beta \to -\infty$.  This can also happen for other measure spaces $X$.   I will
 not delve into this, because my goal now is to get to some physics.
 
As before, we can write $\beta = 1/kT$, at least if $\beta \ne 0$,
 and then write the Boltzmann distribution as
\[    p(x) =  \frac{\displaystyle{e^{-E(x)/kT}}}{\displaystyle{\int_X e^{-E(x)/kT} \, dx}}  . \]
Also as before, the Boltzmann distributions obey the crucial relation
 \[    d H = \beta d \langle E \rangle .\]
Rewriting this in terms of Gibbs entropy $S = k H$ and temperature $T = 1/k \beta$, it
becomes this famous relation between temperature, entropy and the expected energy:
\[      T d S = d \langle E \rangle .\]
Notice that the units match here.  The Shannon entropy $H$ is dimensionless, but since $k$ has units of energy/temperature, the Gibbs entropy $S = kH$ has units of energy/temperature.
Thus $T dS$ has units of energy, as does $d \langle E \rangle $.
 
\vfill \eject \begin{center}\noindent\fbox{\begin{minipage}{28em}
 
 \begin{center}
\entry{THE CHANGE IN ENTROPY}
\vskip 1.5em
\textbf{\boldmath
As we change the temperature of a system from $T_0$ to $T_1$ while keeping it in 
thermal equilibrium, the change in its entropy is
\vskip 0.5em
\[ \color{darkred}
\scalebox{1.2}{$\displaystyle{  S(T_1) - S(T_0) = \int_{T_0}^{T_1} \frac{d\langle E \rangle}{T} }$ } \]
\vskip 0.5em
where $\langle E \rangle$ is its expected energy at temperature $T$.
}
\end{center}
\end{minipage}} \end{center} \vskip 1em

Last time we saw that as we change the expected energy $\langle E \rangle$ of a system while keeping it in thermal equilibrium, this fundamental relation holds:
\[  T dS = d \langle E \rangle .\]
We can rewrite this as 
\[    dS = \frac{d \langle E \rangle}{T}  \]
and then integrate this from one temperature to another---remember, as the expected energy
changes, so does the temperature.  We get
\[ \displaystyle{ \int_{T_0}^{T_1} \frac{d\langle E \rangle}{T} = S(T_1) - S(T_0) } .\]

This is the main way people do experiments to `measure entropy'.  Slowly heat something up,
keeping track of how much energy it takes to increase its temperature each little bit.  Using
this data you can approximately calculate the integral at left---and that gives the change in entropy!

But so far we're just measuring
\emph{changes} in entropy.   How can you figure out the actual value of the entropy?
One way is to assume the Third Law of Thermodynamics, which says that in thermal
equilibrium the entropy approaches zero
as the temperature approaches zero from above.  This gives 
\[ \displaystyle{ \int_{0}^{T_1} \frac{d\langle E \rangle}{T} = S(T_1) } .\]
This is how people often `measure the entropy' of a system in thermal 
equibrium.   They heat it up starting from absolute zero, very slowly so---they hope---it
is close to thermal equilibrium at every moment---and they take data on how much
energy is used, and approximately calculate the integral at left!

But this relies on the Third Law of Thermodynamics.  So where does that come from?

\vfill \eject \begin{center}\noindent\fbox{\begin{minipage}{30em}

\hypertarget{sec:THIRD_LAW}{}

\begin{center}
\entry{THE THIRD LAW OF THERMODYNAMICS} 
\vskip 1em
\textbf{\boldmath
If a system has countably many states, \\ with just one of lowest energy,  \\
and thermal equilibrium is possible for this system \\ 
for some temperature $T > 0$, \\
then its entropy in thermal equilibrium approaches zero \\
as $T$ approaches zero from above:
\[          \color{darkred}    \lim_{T \to 0^+} S(T) = 0 \]
}

\end{center} 

 \end{minipage}} \end{center} \vskip 1em
 
Some people say the Third Law of Thermodynamics this way: ``entropy is zero at
absolute zero''.   But it's not really that simple---indeed, other people
say it's impossible to reach absolute zero.   Above I've stated a version
of the Third Law that's actually a theorem.    Let's prove it!

Actually, let's prove it now for systems with only finitely many states.  It'll be 
easier to handle systems with countably infinite number of states 
\hyperlink{sec:THIRD_LAW_REVISITED}{later},
when we've developed more tools.  And by the way, we'll see the Third Law \emph{isn't 
always true} for systems with a continuum of states.  It will fail for all three of the 
problems on our big to-do list: the classical harmonic oscillator, the classical particle
in a box and the classical ideal gas.  This is often taken as a failure of classical mechanics,
since switching to quantum mechanics makes the Third Law hold for these systems.

Let's show that for a system with finitely many states $i = 1, \dots, n$ with energies $E_i$,
as the temperature $T$ approaches zero from above, the entropy of the system in thermal equilibrium approaches $k \ln N$ where $N$  is the number of lowest-energy states.  
In thermal equilibrium
\[      p_i \propto e^{- E_i/kT} . \]
Thus, all states with the lowest energy have the same probability, while as the temperature approaches
zero from above, any higher-energy states have $p_i \to 0$.  So, as the temperature approaches
zero from above, the probability of the system being in any one of its $N$ lowest-energy states approaches $1/N$, and we get
\[    \lim_{T \to 0^+} S(T) = \lim_{T \to 0^+} -k \sum_{i=1}^n p_i \ln p_i  = 
- k \sum_{i=1}^N \frac{1}{N} \ln\left(\frac{1}{N}\right) = k \ln N .\]
In particular, if the system has just one lowest-energy state, we get  the \define{Third Law of Thermodynamics}:
\[     \lim_{T \to 0^+} S(T) = 0 . \]
Here $T \to 0^+$ means that $T$ is approaching zero from above.

But beware: for systems with lots of lowest-energy states, their entropy in thermal equilibrium can be large even near absolute zero!  Also, a related problem: systems may take a ridiculously long time to reach equilibrium!  This is especially true for systems that have many states whose energies are \emph{very near} the lowest energy, like a piece of glass.  You can put a piece of glass in a fancy refrigerator and 
try to cool it to near absolute zero.  If it has one lowest-energy state, its entropy should approach
zero.   If this happened, the glass  would change from glass to a crystal, which has less entropy.  But 
nobody has seen glass turn into a crystal when they cool it down.    If this happens, it probably does so only after an absurdly long time, much longer than the age of the Universe.  This phenomenon is called
`\href{https://en.wikipedia.org/wiki/Geometrical_frustration}{frustration}'.    People like to argue about frustration and the Third Law, so I am not trying to give you the final word here, just alert you to the issue.   You can learn a bit more here:
\begin{itemize}
\item
Wikipedia, \href{https://en.wikipedia.org/wiki/Third_law_of_thermodynamics}{Third law of 
thermodynamics}.
\end{itemize}

By the way: for systems with finitely many states, it's possible to have negative
temperatures, and the Third Law has a counterpart saying what happens when the temperature approaches zero \emph{from below}:

\begin{puzzle}
Show that for a system with finitely many states,
\[    \lim_{T \to 0^-} S(T) =  k \ln M \]
where $M$ is the number of states of highest energy.   
\end{puzzle}

However, most systems we'll be studying won't have a state of highest energy.

\vfill \eject \begin{center}\noindent\fbox{\begin{minipage}{30em}

\begin{center}
\entry{MEASURING ENTROPY}
\end{center} \vskip 1em

\begin{center}
\textbf{\boldmath
If we assume the entropy of a system approaches zero as $T$ \\
approaches zero from above, we have
\vskip 0.2em
\[ \color{darkred} \scalebox{1.2}{$\displaystyle{ \int_{0}^{T_1} \frac{d\langle E \rangle}{T} = S(T_1)  }$} \]
\vskip 0.5em \noindent
Using this assumption, we can
do experiments to measure \\
the entropy of different substances \\ at standard temperature and pressure:
\begin{itemize}
\item
iron: $\sim$5 bits per atom 
\item
water: $\sim$12 bits per molecule
\item
hydrogen: $\sim$23 bits per molecule 
\end{itemize}
}
\end{center}

 \end{minipage}} \end{center} \vskip 1em 
 
People actually do experiments and use the above formula to figure out the entropy of many
substances in thermal equilibrium assuming their entropy vanishes as the temperature approaches
absolute zero.   They slowly heat up a substance and keep track of how much energy is needed
to raise its temperature as they go, so they can approximately calculate the integral shown.
They usually report the answers in joules/kelvin per mole, but I enjoy `bits per molecule'. 
 
As you can see, liquids tend to have more entropy than solids, and gases tend to have even more.
My goal in this course is to teach you how to approximately compute
some of these entropies from first principles.  Unfortunately the only substances that
are simple enough for us to handle are gases.

This is a good opportunity to explain some conventions.    A \href{https://en.wikipedia.org/wiki/Mole_(unit)}{mole} is defined to be exactly $6.02214076 \cdot 10^{23}$---this is called \href{https://en.wikipedia.org/wiki/Avogadro_constant}{Avogadro's number}, and it's close to the number of hydrogen atoms in a gram.  A joule/kelvin of Gibbs entropy corresponds to about $7.242297 \cdot 10^{22}$ nats of Shannon entropy: the number here is the reciprocal of \href{https://en.wikipedia.org/wiki/Boltzmann_constant}{Boltzmann's constant}, which is defined to be exactly $1.380649 \cdot 10^{-23}$ joules per kelvin.  A bit is $\ln 2$ nats.    From these three facts, we see 1 joule/kelvin of Gibbs entropy per mole corresponds to about $0.173516$ bits/molecule of Shannon entropy.

By the way, what is `\href{https://en.wikipedia.org/wiki/Standard_temperature_and_pressure}{standard temperature and pressure}'?  Annoyingly, this phrase means
different things to different organizations.   I will try to always use it to mean a temperature
of 298.15 K and a pressure of 1 bar.   The temperature here equals 25 ${}^\circ$C, which seems a bit random compared to 0 ${}^\circ$C---but convenient, because it's close to room temperature.
A pressure of 1 \href{https://en.wikipedia.org/wiki/Bar_(unit)}{bar}, or more officially 100 kilopascals, is slightly less than a `\href{https://en.wikipedia.org/wiki/Standard_atmosphere_(unit)}{standard atmosphere}', which is a unit of pressure intended to equal the average air pressure at sea level.    A \href{https://en.wikipedia.org/wiki/Pascal_(unit)}{pascal} is an official
SI unit: it's a pressure of one newton per square meter.
 
 \vfill \eject \begin{center}\noindent\fbox{\begin{minipage}{30em}

\begin{center}
\entry{THE EQUIPARTITION THEOREM} 
\vskip 2em
\textbf{\boldmath
Suppose the energy of a system with $n$ degrees of freedom is a positive 
definite quadratic form $E \maps \R^n \to \R$, for example
\[     \phantom{ c_i > 0}   \qquad \qquad     E(x) = \sum_{i=1}^n \frac{c_i x_i^2}{2}  \qquad \qquad c_i > 0  \]
Then in thermal equilibrium at temperature $T$, \\ the expected value of the energy
is
\[    \color{darkred}     \langle E \rangle  = \frac{1}{2} n k T   \]
where $k$ is Boltzmann's constant.   
}
\end{center} \vskip 1em

 \end{minipage}} \end{center} \vskip 1em 
 
Temperature is very different from energy.  But sometimes---not very often,
but sometimes---the expected energy of a system in thermal
equilibrium is proportional to its temperature.  The equipartition
theorem says this happens when the
energy depends quadratically on several real variables, defining
a \href{https://en.wikipedia.org/wiki/Definite_quadratic_form}{positive
definite quadratic form} on $\R^n$.  For example, it
happens for a classical harmonic oscillator.

Some people get confused and try to apply the equipartition theorem
where it doesn't apply.   They foolishly conclude that temperature is always proportional 
to energy.

This theorem does \emph{not} apply to quantum systems.
Indeed, when people tried to apply the equipartition theorem to a mirrored box
of light they ran into a problem called the \href{https://en.wikipedia.org/wiki/Ultraviolet_catastrophe}{ultraviolet catastrophe}.  Classically the box of light is a system where the energy
is a positive definite quadratic form, but $n = \infty$, so they got an \emph{infinite} expected
value of the energy!   Quantum mechanics saves the day and makes the answer finite.

\vskip 1em
\begin{center}
\includegraphics[width = 25 em]{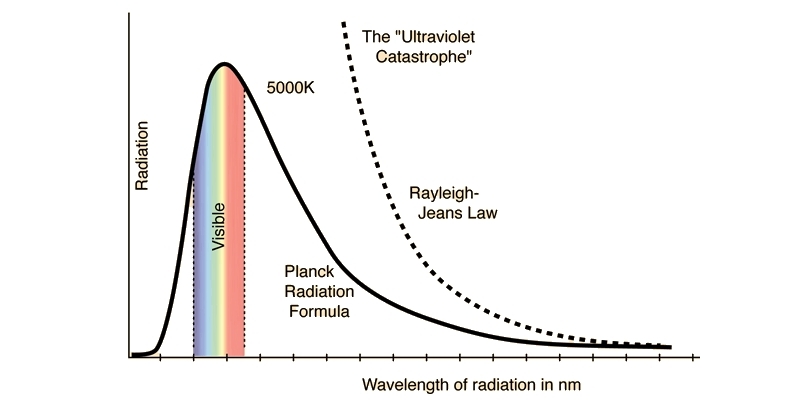}
\end{center}

The equipartition theorem also doesn't apply to \emph{classical} systems unless
the energy is quadratic.   So it's very limited in its applicability, but still useful.

Let's prove this result!   To prove a theorem, you have to understand the definitions.
We'll start with some background.
 
 \vfill \eject \begin{center}\noindent\fbox{\begin{minipage}{30em}

\begin{center}
\entry{THE EQUIPARTITION THEOREM---BACKGROUND} 
\vskip 2em
\textbf{\boldmath
Suppose the energy of a system with $n$ degrees of freedom is some function
\[  E \maps \R^n \to \R \]   
Let $k$ be Boltzmann's constant. \\
Suppose $p \maps \R^n \to \R$ is a probability distribution maximizing the entropy
\[        S = -k \int_{\R^n} p(x) \ln p(x) \, d^n x \]
subject to a constraint on the expected energy
\[       \langle E \rangle = \int_{\R^n} E(x) p(x) \, d^n x \]
Then $p$ must be a \define{Boltzmann distribution}:
\[        p(x) = \frac{e^{-\beta E(x)}}{\displaystyle \int_{\R^n} e^{-\beta E(x) }d^n x} \]
for some number $\beta > 0$.  \\
\vskip 1em 
The temperature $T$ is defined so that $\beta = 1/kT$.
}
\end{center} \vskip 1em

 \end{minipage}} \end{center} \vskip 1em 
 
 We're defining entropy with an integral now, unlike a sum as before,
and sticking Boltzmann's constant into the definition of entropy, as
physicists do, so that entropy has units of energy over temperature.

Given the formula for the energy $E$ as a function on $\R^n$,
we'll have to find the Boltzmann distribution and then compute $\langle E
\rangle$ as a function of $T$.
 
 \vfill \eject \begin{center}\noindent\fbox{\begin{minipage}{35em}

 \hypertarget{sec:PROOF_OF_EQUIPARTITION_THEOREM_2}{}

\begin{center}
\entry{PROOF OF THE EQUIPARTITION THEOREM: 1} 
\vskip 2em
\textbf{\boldmath
Special case: a system with 1 degree of freedom where 
the energy $E \maps \R \to \R$ is $E(x) = x^2/2$. 
\vskip 1em
The Boltzmann distribution is
\[       p(x) \; = \;
 \frac{e^{-\beta E(x)}}{\displaystyle \int_{-\infty}^\infty e^{-\beta E(x) }\, dx} 
 \; = \; \frac{e^{-\beta x^2/2}}{\displaystyle \int_{-\infty}^\infty e^{-\beta x^2/2}dx} 
\]
so the expected energy is
\[   \langle E \rangle = \int_{-\infty}^\infty E(x) p(x) \, dx \; = \;
 \frac{\displaystyle \int_{-\infty}^\infty \frac{x^2}{2} e^{-\beta x^2/2} \, dx}{\displaystyle \int_{-\infty}^\infty e^{-\beta x^2/2}dx}
 \; = \;
 \frac{\displaystyle \frac{1}{2\beta} \int_{-\infty}^\infty \beta x^2 e^{-\beta x^2/2} \, dx}{\displaystyle \int_{-\infty}^\infty e^{-\beta x^2/2}dx}
\]
so doing a substitution with  $u^2 = \beta x^2$:
\[  {\color{darkred} \langle E \rangle } \;=  \;
 \frac{\displaystyle \frac{1}{2\beta} \int_{-\infty}^\infty u^2 e^{-u^2/2} \, du}{\displaystyle \int_{-\infty}^\infty e^{-u^2/2}du} 
  \;= \;\frac{1}{2\beta}  \;= \;  {\color{darkred} \frac{1}{2} kT}
 \]
since
\\
\[   \int_{-\infty}^\infty e^{-u^2/2}du = 
\int_{-\infty}^\infty u^2 e^{-u^2/2}du = \sqrt{2 \pi} \]
}
\end{center} \vskip 1em

 \end{minipage}} \end{center} \vskip 1em
 
 We'll do two special cases before proving the general result.  First
let's do a system with 1 degree of freedom where the energy is $E(x)
= x^2/2$.  In this case, after a change of variables, the Gibbs
distribution becomes a Gaussian with mean 0 and variance 1, and that 
gives the desired result.  Or just do the integrals and see what you get!
The expected energy $\langle E \rangle $ is $\frac{1}{2}kT$.
 
 \vfill \eject \begin{center}\noindent\fbox{\begin{minipage}{35em}

\begin{center}
\entry{PROOF OF THE EQUIPARTITION THEOREM: 2} 
\vskip 2em
\textbf{\boldmath
More general case: a system with $n$ degrees of freedom where 
the energy $E \maps \R^n \to \R$ is
\[   E(x) = \frac{1}{2} \|x\|^2 = \frac{1}{2} \sum_{i=1}^n x_i^2   \]
We can reduce this to the case with 1 degree of freedom:
\[ { \color{darkred}  \langle E \rangle } \; = \;
 \frac{\displaystyle \int_{\R^n} \tfrac{1}{2} \|x\|^2 \, e^{-\beta \|x\|^2 /2} \, d^nx}{\displaystyle \int_{\R^n} e^{-\beta \|x\|^2 /2} d^nx} \; = \;
 \scalebox{1.5}{$\displaystyle{\sum_{i=1}^n}$}\; \frac{\displaystyle \int_{\R^n} \tfrac{1}{2} x_i^2 \, e^{-\beta \|x\|^2 /2} \, d^nx}{\displaystyle \int_{\R^n} e^{-\beta \|x\|^2 /2} d^nx} \; = \; { \color{darkred} \frac{n}{2} k T}
\]
}
\end{center} \vskip 1em

 \end{minipage}} \end{center} \vskip 1em 
 
Next we do a system with $n$ degrees of freedom where the energy
is a sum of $n$ terms of the form $x_i^2/2$.  It's no surprise that each
degree of freedom contributes $\frac{1}{2}kT$ to the expected energy,
giving
\[ \langle E \rangle = \frac{1}{2} nkT \]
But make sure you follow my calculation above.  I skipped a couple of steps!

 \vfill \eject \begin{center}\noindent\fbox{\begin{minipage}{35em}
 
\hypertarget{sec:PROOF_OF_EQUIPARTITION_THEOREM_3}{}

\begin{center}
\entry{PROOF OF THE EQUIPARTITION THEOREM: 3} 
\vskip 2em
\textbf{\boldmath
General case: a system with $n$ degrees of freedom where \\
the energy $E \maps \R^n \to \R$ is any positive definite quadratic form.   Then
\[       E(x) = \frac{1}{2} \|A x \|^2  \]
for some invertible $n \times n$ matrix $A$.     When $A$ is a diagonal matrix this
gives
\[      \phantom{ c_i > 0}   \qquad \qquad     E(x) = \sum_{i=1}^n \frac{c_i x_i^2}{2}  \qquad \qquad c_i > 0  \]
We can reduce the general case to the previous case
by a change of variables $y = Ax$:
\[ { \color{darkred}  \langle E \rangle } \; = \;
 \frac{\displaystyle \int_{\R^n} \tfrac{1}{2} \|Ax\|^2 \, e^{-\beta \|A x\|^2 /2} \, d^nx}{\displaystyle \int_{\R^n} e^{-\beta \|Ax\|^2 /2} \, d^nx} \; = \;
 \frac{\displaystyle \int_{\R^n} \tfrac{1}{2} \|y\|^2 \, e^{-\beta \|y\|^2 /2} \, d^n y}{\displaystyle \int_{\R^n} e^{-\beta \|y\|^2 /2}\, d^n y}
 \; = \;  { \color{darkred} \frac{n}{2} kT }
\]
}
\end{center} \vskip 1em

 \end{minipage}} \end{center} \vskip 1em
 
Finally let's do the general case.   A \define{quadratic form} on $\R^n$ is a map
$Q \maps \R^n \to \R$ such that
\[          Q(x) = \sum_{i,j = 1}^n   q_{ij} x_i x_j  \]
for some numbers $q_{ij} \in \R$.    We say it's \define{positive definite} if
\[        x \ne 0 \implies Q(x) > 0  . \]
One can prove that a quadratic form $Q \maps \R^n \to \R$ is positive definite if and only if
\[           Q(x) = \frac{1}{2} \| A x \|^2 \]
for some invertible $n \times n$ matrix $A$.  The factor of $1/2$ here is just to make our
calculations easier.

Thanks to this, if we have a system whose space of states is $\R^n$ and its energy
function $E \maps \R^n \to \R$ is a positive definite quadratic form, we can compute
\[      \langle E \rangle =
 \frac{ \displaystyle{\int_{\R^n} E(x) \exp(-\beta E(x)) \, dx}}{\displaystyle{\int_{\R^n} \exp(-\beta E(x)) \, dx}} \]
by reducing it to the \hyperlink{sec:PROOF_OF_EQUIPARTITION_THEOREM_3}{previous case} using a change of variables.  We get
\[   \langle E \rangle = \frac{1}{2} nkT \]
So, each degree of freedom still contributes $\frac{1}{2}kT$ to the
expected energy.  That's the equipartition theorem!

But be careful.  The equipartition theorem doesn't apply when the energy is an arbitrary
function of $n$ variables.  It also fails when we switch from
classical to quantum statistical mechanics.    

People sometimes memorize the conclusion of the equipartition theorem, $E = \frac{1}{2}n k T$, without learning that it holds only for classical systems whose energy is a positive definite quadratic form.    These people sometimes get fooled into thinking $\langle E \rangle$ is \emph{always} proportional to $T$.    Some of these poor benighted souls go around saying that temperature is just a measure of energy per degree of freedom.   This completely ignores the subtlety of the concept of temperature. 

As we've seen, the \emph{truly general} relation between temperature and energy, for systems in thermal equilibrium, also involves  entropy:
\define{
  \[              T dS = d \langle E \rangle .\]
  }

 \vfill \eject \begin{center}\noindent\fbox{\begin{minipage}{30em}
 
\begin{center}
\entry{THE AVERAGE ENERGY OF AN ATOM} 
\vskip 2em
\textbf{\boldmath
Since an atom of helium gas can move in 3 directions, and its energy 
depends quadratically on its velocity and not on position, the equipartition theorem
says that \textit{classically} its expected energy should be 
\[    \color{darkred}     \langle E \rangle  = \frac{3}{2} k T   \]
where $T$ is temperature and $k$ is Boltzmann's constant, about $1.38 \cdot 10^{-23}$ joules/kelvin. 
\vskip 1em
So, heating an atom of helium gas 1 ${}^\circ$C should take
\[    \frac{3}{2} \times 1.38 \cdot 10^{-23} \; \textrm{joules}
= 2.07 \cdot 10^{-23} \; \textrm{joules} \]
This is very close to the truth.
}
\end{center} \vskip 1em

 \end{minipage}} \end{center} \vskip 1em 
 
 We can finally start reaping the rewards of all our thoughts about
entropy!  The equipartition theorem lets us estimate how much energy
it takes to heat up one atom of helium one degree Celsius.
And it works!

Of course we don't heat up an individual atom: we heat up a bunch.  A
mole of helium is about $6.02 \cdot 10^{23}$ atoms, so heating up a mole of helium
one degree Celsius (= 1 kelvin) should take about
\[ 6.02 \cdot 10^{23} \times 2.07 \cdot 10^{-23} \approx 12.46
\textrm{ joules} \]
And this is very close to correct!  It seems the experimentally
measured answer is $12.6$ joules.

What are the sources of error?  Most importantly, our calculation
neglects the interaction between helium atoms.  Luckily this is very
small at standard temperature and pressure.  We're also neglecting quantum
mechanics.  Luckily for helium this too gives only small corrections
at standard temperature and pressure.
 
 It's important here that helium is a \href{https://en.wikipedia.org/wiki/Monatomic_gas}{monatomic gas}.  
 In hydrogen, which is a \href{https://en.wikipedia.org/wiki/Diatomic_molecule}{diatomic gas}, we
get extra energy because this molecule can tumble around, not just
move along.  We'll try that next.
 
 \vfill \eject \begin{center}\noindent\fbox{\begin{minipage}{30em}

\begin{center}
\entry{THE ENERGY OF HYDROGEN} 

\vskip 1em
\includegraphics[scale = 0.024]{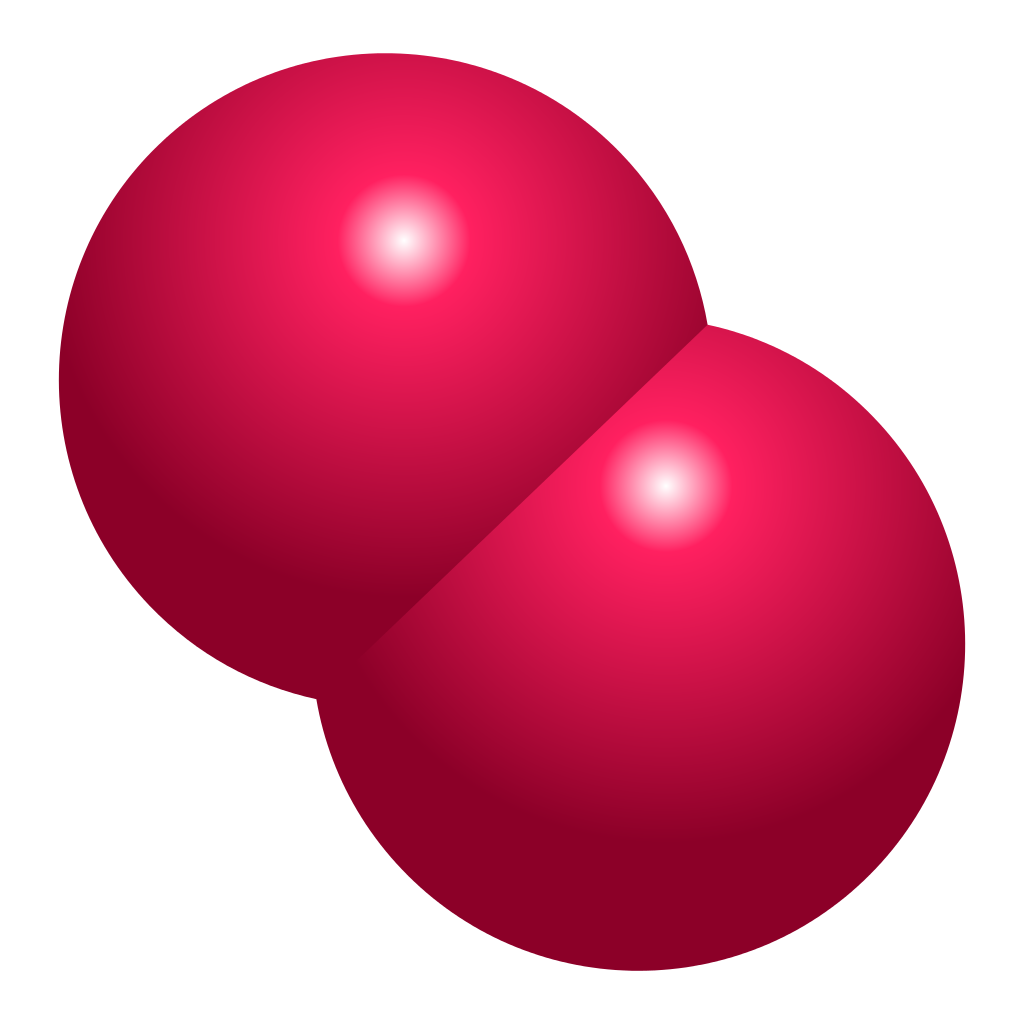}

\vskip 1em
\textbf{\boldmath
If we treat a molecule of hydrogen as a dumbbell 
whose position takes 3 numbers and whose axis takes 
2 numbers to describe, we can \textit{try} to use the equipartition theorem
to estimate its expected energy as
\[    \color{darkred}     \langle E \rangle  = \frac{5}{2} k T   \]
where $T$ is temperature and $k = 1.38 \cdot 10^{-23}$ joules/kelvin. 
\vskip 1em
In this approximation, heating a molecule of hydrogen gas \\ 1 kelvin takes
\[    \frac{5}{2} \times 1.38 \cdot 10^{-23} \; \textrm{joules}
= 3.45 \cdot 10^{-23} \; \textrm{joules} \]
In reality it takes $3.39 \cdot 10^{-23}$ joules \\ at standard temperature
and pressure.  Not bad!
}
\end{center} \vskip 1em

 \end{minipage}} \end{center} \vskip 1em
 
 A molecule of hydrogen gas is a blurry quantum thing, but let's
pretend it's a classical solid dumbbell that can move and tumble but
not spin around its axis.  Then it has 3+2 = 5 degrees of freedom, and
we can use the equipartition theorem to estimate its energy.

For $T$ significantly less than 6000 kelvin, hydrogen molecules
don't vibrate with the two atoms moving toward and away from each 
other.  They don't spin around their axis until
even higher temperatures.  But they tumble like a dumbbell as soon as
$T$ exceeds about 90 kelvin.  

We need quantum mechanics to compute these things.  But at room
temperature and pressure, we can pretend a hydrogen gas is made of
classical solid dumbbells that can move around and tumble but not spin
around their axes!  In this approximation the equipartition theorem
tells us $\langle E \rangle = \frac{5}{2} kT$.

This is fine as far as it goes---but our goal in this course is to compute the \emph{entropy} of hydrogen.
We'll start with a useful warmup: the classical harmonic oscillator.

 \vfill \eject \begin{center}\noindent\fbox{\begin{minipage}{36em}
 
\begin{center}
\entry{ENTROPY OF THE HARMONIC OSCILLATOR: 1} 

\vskip 2em
\noindent
\textbf{\boldmath
A classical harmonic oscillator has energy
\[           E = \frac{p^2}{2m} + \frac{\kappa q^2}{2} \]
where $q$ is its position, $p$ its momentum, $m$ its mass and $\kappa$ 
its spring constant.  
\\ $\;$ \\
By the equipartition theorem, in thermal equilibrium at temperature $T$ \\
it has expected energy $\langle E \rangle  =  k T$
where $k$ is Boltzmann's constant.   \\ $\;$ \\
So, using $d\langle E \rangle = T dS$, its entropy is
\[   S = \int dS = \int \displaystyle{ \frac{d\langle E \rangle}{T} = k \int
\frac{ d T}{T}} = k (\ln T \; +\; C) \]
Since this does not approach $0$ as $T \to 0$ from above,
the Third Law of Thermodynamics \emph{doesn't hold} for the
classical harmonic oscillator.
\vskip 1em 
But what is this constant $C$?  \\ For that we must think harder.
}
\end{center} \vskip 1em

 \end{minipage}} \end{center} \vskip 1em 
 
 You may have already studied the classical harmonic oscillator.  We can use it to model a rock of mass $m$ hanging on a spring with spring constant $\kappa$.  But the harmonic oscillator is not just about springs!  Almost any classical system that vibrates can be approximately modeled as a collection of classical harmonic oscillators.   Such systems include radiation in a box, the surface of a drum, a violin string, a vibrating molecule (when treated classically), and many more.   So, the harmonic oscillator is fundamental to physics.
 
 In classical mechanics we learn how to answer this question: if we know the position $q$ and momentum $p$ of an classical harmonic oscillator \emph{now}, what it will they be at some time in the future?   In classical statistical mechanics we can ask other questions, like: how much entropy does a classical harmonic oscillator have in thermal equilibrium?  Here we \emph{don't} know the position and momentum of the oscillator: instead, we know its temperature $T$.   This determines a probability distribution on the space of pairs $(p,q) \in \R^2$, which has some entropy $S$.  Using the equipartition theorem and the formula $d \langle E \rangle = T dS$, we can show 
\[        S = k(\ln T + C)  .\]
So, the entropy grows logarithmically with temperature.   And it does not go to
zero as $T$ approaches zero: instead, it goes to negative infinity.   So the \hyperlink{sec:THIRD_LAW}{Third Law of 
Thermodynamics} \emph{does not hold} for the classical harmonic oscillator! 

That may seem shocking, but it actually makes sense.  The Third Law holds only for \hyperlink{sec:THIRD_LAW_REVISITED}{certain special systems}.   Furthermore, \hyperlink{sec:THE_FINITE_VERSUS_THE_CONTINUOUS}{we've seen} that the entropy of
a sharply peaked probability distribution on a continuous state space is negative.   We'll see that the Boltzmann distribution for the classical oscillator gets more and more sharply peaked near $q = p = 0$ as the temperature approaches
zero from above.  So in fact, it makes perfect sense that the entropy approaches $-\infty$.

However, the classical harmonic oscillator is just an approximation to the 
quantum harmonic oscillator, which \emph{does} obey the Third Law.    It's a good approximation at high temperatures, but bad at low temperatures.   In fact, this 
business of negative entropies at low temperature is not something that happens in the real world.
It's just a defect of classical mechanics.   It's trying to tell us that quantum mechanics is better.

Another point: you'll have noticed that constant $C$ here.  What is it?
We can make progress with a bit of dimensional analysis.  The quantity
$\ln T$ is a funny thing: if we change our units of temperature, it doesn't get
multiplied by a constant factor, the way physical quantities usually do.  
It changes by \emph{adding} a constant!  So $k \ln T$ doesn't have
dimensions of entropy.  But
\[ S = k (\ln T + C) \]
must have dimensions of entropy.   The constant $C$ must somehow save the day!
How does that work?   Let's see.
 
 \vfill \eject \begin{center}\noindent\fbox{\begin{minipage}{35em}

\begin{center}
\entry{ENTROPY OF THE HARMONIC OSCILLATOR: 2} 

\vskip 2em
\noindent
\textbf{\boldmath
Classically, a harmonic oscillator at temperature $T$ has entropy
\[   S = k (\ln T \; +\; C) \]
Writing $C = -\ln(T_0)$ for some constant $T_0$, this
gives
\define{
\[   S = k \ln(T/T_0)\]
}
Dimensional analysis implies $T_0$ must have units of temperature!
\vskip 1em
But what is this temperature $T_0$?   For that we must think harder.
}
\end{center} \vskip 1em
 \end{minipage}} \end{center} \vskip 1em 
 
 The formula $S = k (\ln T + C)$ is a bit scary from the viewpoint of
 dimensional analysis.  We usually avoid working with the logarithm of a dimensionful
 quantity, like $\ln T$, because it transforms in a funny way when we change our 
 units.   But if we write $C = -\ln T_0$ then we get $S = k \ln(T/T_0)$, and we
 see the solution to our  problem!  If $T_0$ has units of temperature, then $T/T_0$ 
 is dimensionless, so $\ln(T/T_0)$ doesn't change at all when we change our units.   
 In other words: now $\ln(T/T_0)$ is dimensionless, so $S = k\ln(T/T_0)$
 has units of entropy as it should.
 
 So, the constant $C$ must equal $-\ln T_0$ for some temperature
$T_0$ that we can compute for any harmonic oscillator.  What is it?
This is a fascinating puzzle.

For starters, what could this temperature $T_0$ possibly depend on?  
Obviously the mass $m$, the spring constant 
$\kappa$ and Boltzmann's constant $k$.
But there's no way to form a quantity with units of temperature from
just $m, \kappa$ and $k$.   So we need an extra ingredient.
And it turns out, remarkably, that the extra ingredient is \href{https://en.wikipedia.org/wiki/Planck_constant}{Planck's constant} $\hbar$.

This should be absolutely shocking!   Planck's constant is associated to \emph{quantum}
mechanics, but we're trying to compute the entropy of a \emph{classical}
harmonic oscillator.  How does Planck's constant get into the game? 
We'll say more about this later.

We \emph{can} compute a quantity with units of temperature from $m,
\kappa, k$ and $\hbar$.  The frequency of our oscillator is
$\omega = \sqrt{k/m} $, and it's a famous fact that $\hbar \omega$
has units of energy.  $k$ has units of energy/temperature... so
$\hbar\omega/k$ has units of temperature.

Thus, our
temperature $T_0$ must be $\hbar \omega/k$ times some dimensionless
purely mathematical constant, which I'll call $1/\alpha$.
$\alpha$ must be something like $\pi$ or $2$, or if we're
really unlucky, $e^{666}$ ---though in physics our purely mathematical dimensionless
constants are usually numbers
fairly close to $1$, not huge or tiny numbers.

So, the entropy of a classical harmonic oscillator is
\[ S = k \ln(T/T_0) = k \ln(\alpha kT/\hbar \omega).  \]
This is far as I can get without breaking down and doing some real work.
Later we will compute $\alpha$.

 \vfill \eject \begin{center}\noindent\fbox{\begin{minipage}{30em}

\hypertarget{sec:ENTROPY_HARMONIC_OSCILLATOR_3}{}

\begin{center}
\entry{ENTROPY OF THE HARMONIC OSCILLATOR: 3} 
\noindent
\vskip 2em
\textbf{\boldmath
We've seen a classical harmonic oscillator \\ with frequency $\omega$ has entropy 
\[ \color{darkred}  S = k \ln(\alpha kT/\hbar \omega)\]
when it's in thermal equilibrium at temperature $T$. \vskip 1em  Here $k$ is Boltzmann's constant, \\
$\hbar$ is Planck's constant,  \\  and $\alpha$ is some dimensionless
mathematical constant. \\
We'll figure it out later.}
\end{center} \vskip 1em

\end{minipage}} \end{center} \vskip 1em

Even though we don't know $\alpha$, this formula is already very interesting!  $kT$ is known to be the typical energy scale of thermal fluctuations at temperature $T$.   $\hbar \omega$ is the spacing between energy levels of a \emph{quantum} harmonic oscillator with frequency $\omega$.
The ratio $kT/\hbar\omega$ is therefore roughly the number of energy eigenstates in
which we may find a \emph{quantum} harmonic oscillator with high
probability when it's at temperature $T$.    

Thus, $S$ is roughly $k$ times the logarithm of
the number of states that we're likely to find a  \emph{quantum} harmonic oscillator in,
when it's at temperature $T$.   This may seem mysterious.    After all, we weren't trying 
to do quantum mechanics, much less count quantum states.   

In 1912, Otto Sackur and Hugo Tetrode ran into the same issue when trying to solve the problem we're working up to now: computing the entropy of a classical  ideal gas.  They discovered---and so shall we---that Planck's constant appears in the answer.    For the fascinating story of how they did it, read this:

\begin{itemize}
\item Walter Grimus, \href{https://arxiv.org/abs/1112.3748}{On the 100th anniversary of the Sackur-Tetrode equation}, \textsl{Annalen der Physik} \textbf{525} (2013), A32--A35.
\end{itemize}

We'll learn more about this business of counting states 
\hyperlink{sec:HARMONIC_OSCILLATOR_ENTROPY}{later}, when we relate entropy to something called the `\hyperlink{sec:MEANING_PARTITION}{partition function}', which can be understood as the `number of accessible states'.    This viewpoint will also explain the constant $\alpha$.  But now let's calculate this constant.

\vfill \eject \begin{center}\noindent\fbox{\begin{minipage}{30em}

\begin{center}
\entry{ENTROPY OF THE HARMONIC OSCILLATOR: 4} 

\vskip 2em
\noindent
\textbf{\boldmath
A classical harmonic oscillator has \define{energy}
\[           E(p,q) = \frac{p^2}{2m} + \frac{\kappa q^2}{2} \]
where $p$ is its momentum, $q$ its position, $m$ its mass and $\kappa$ 
its spring constant.   
\vskip 1em
At temperature $T$, the probability density of its
momentum and position
 is the \define{Boltzmann distribution:}
\[   \rho(p,q) = \frac{e^{-\beta E(p,q)}}{\displaystyle{\int_{-\infty}^\infty \!\int_{-\infty}^\infty e^{-\beta E(p,q)} \, \frac{dp\, dq}{h}}} \]
where $\beta = 1/kT$, $k$ is Boltzmann's constant, \\
and $h = 2 \pi \hbar$ is the original `unreduced' Planck's constant.
\vskip 1em
The oscillator's \define{entropy} at temperature $T$ is thus
\[    S = - k \int_{-\infty}^\infty \! \int_{-\infty}^\infty \rho(p,q) \ln \rho(p,q) \;  \frac{dp \, dq}{h} \]
}
\end{center} \vskip 1em

 \end{minipage}} \end{center} \vskip 1em
 
Last time we found a formula for the
entropy of a classical harmonic oscillator... which includes a
mysterious purely mathematical dimensionless constant $\alpha$.
Now let's figure out $\alpha$.   To do this, we'll grit our teeth and actually do the integral needed 
to calculate the entropy---but only in one easy case!  This will be enough to determine $\alpha$.

First, recall the basics.  The energy $E(p,q)$ of our harmonic
oscillator at momentum $p$ and position $q$ determines its Boltzmann
distribution at temperature $T$, which I'll call $\rho(p,q)$ now since
the letter $p$ is already being used.  Integrating $-\rho \ln \rho$ over the space of 
states of the harmonic oscillator, which is the $pq$
plane, we get the Shannon entropy.   We then multiply this by Boltzmann's constant 
$k$ to get the Gibbs entropy.   

But here's the surprise: the Shannon entropy must be dimensionless, but the measure
$dp \, dq$ has units of momentum times position-- or in other words, action.
Thus the Shannon entropy \emph{cannot possibly} be
\[  \int_{-\infty}^\infty \! \int_{-\infty}^\infty \rho(p,q) \ln \rho(p,q) \;  dp \, dq .\]
To get an answer that makes sense, we must divide $dp \, dq$ by some quantity with units of action!  

To get the correct answer---that is, the one measured in experiments---we must divide $dp \, dq$
by Planck's original constant \(h\), not the so-called `reduced' Planck's constant \(\hbar = h/2\pi\).
Planck's constant \(h\) is defined to be exactly $6.62607015 \cdot 10^{-34}$ joule-seconds.  Thus, the 
Gibbs entropy of the classical harmonic oscillator is
\[    S = - k \int_{-\infty}^\infty \! \int_{-\infty}^\infty \rho(p,q) \ln \rho(p,q) \;  \frac{dp \, dq}{h}. \]
We just need to do this integral.

But what's so good about \(d p \, d q/ h\)?  Why do we divide by \(h\) and not, say \(\hbar\)?
This is related to Bohr and Sommerfeld's early approach to quantum physics,
the `\href{https://en.wikipedia.org/wiki/Old_quantum_theory}{old quantum theory}', which was later subsumed by the theory of `\href{https://en.wikipedia.org/wiki/Geometric_quantization}{geometric quantization}'.   In Bohr and Sommerfeld's approach, when we quantize a classical system with one position and one momentum degree of freedom, there is one quantum state for each region of area $h$ in the $pq$ plane.    More generally, when we quantize a classical system
with $n$ position and $n$ momentum degrees of freedom, there should be one quantum state for each region $R \subset \R^{2n}$ with
\[            \int_R \frac{d^n p \, d^n q}{h^n}  = 1. \]
So, delving into the whys of quantum mechanics and geometric quantization would shed more light on what we are doing now.  But when Sackur and Tetrode computed the entropy of an ideal gas and compared 
it to experiment, they just went ahead and did an integral using the measure \(d^n p \, d^n q/h^n\),
and discovered that this gives the correct answer!

\vfill \eject  \begin{center}\noindent\fbox{\begin{minipage}{30em}

\begin{center}
\entry{ENTROPY OF THE HARMONIC OSCILLATOR: 5} 

\vskip 2em
\noindent
\textbf{\boldmath
We can choose units of length, time, mass and temperature
\\ to make a classical harmonic oscillator's
mass $m$, its spring constant $\kappa$,  Boltzmann's constant $k$, and 
the reduced Planck's constant $\hbar$ all equal $1$. \vskip 1em
Then at $T = 1$ the Boltzmann distribution of the oscillator is
\[   \rho(p,q) \; = \; \frac{e^{- (p^2 + q^2)/2}}{\displaystyle{\int_{-\infty}^\infty \!\int_{-\infty}^\infty e^{- (p^2 + q^2)/2} \; \frac{dp dq}{2\pi}}} \; = \; e^{- (p^2 + q^2)/2} \]
so its entropy is 
\[    S = - \int_{-\infty}^\infty \! \int_{-\infty}^\infty \, e^{- (p^2 + q^2)/2} \ln\left( e^{-(p^2 + q^2)/2} \right) \;  \frac{dp dq}{2\pi} \]
\vskip 1em
Let's do this integral!
}
\end{center} \vskip 1em

 \end{minipage}} \end{center} \vskip 1em
 
Let's compute the Boltzmann distribution $\rho(p,q)$ and the entropy
$S$.  To keep the formulas clean, we'll work in units where $m =
\kappa = k = \hbar = 1$, and compute everything at one special
temperature: $T = 1$.

In this setup $h = 2\pi$, and 
\[   e^{-\beta E(p,q)}  = e^{-(p^2 + q^2)/2}  \]
is a beautiful Gaussian with integral
\[        \int_{-\infty}^\infty \! \int_{-\infty}^\infty \, e^{- (p^2 + q^2)/2}  = 2 \pi .\]
These two factors of $ 2\pi $ cancel when we compute the denominator of
the probability distribution $\rho(p,q)$:
\[    \int_{-\infty}^\infty \!\int_{-\infty}^\infty e^{- (p^2 + q^2)/2} \; \frac{dp dq}{2\pi} = \frac{2\pi}{2\pi} = 1.\]
 Thus, we get simply
 \[     \rho(p,q) = e^{-(p^2 + q^2)/2} .\]
The entropy of the harmonic oscillator is thus
\[    S = - \int_{-\infty}^\infty \! \int_{-\infty}^\infty \, e^{- (p^2 + q^2)/2} \ln\left( e^{-(p^2 + q^2)/2} \right) \;  \frac{dp dq}{2\pi} \]
when $\kappa = k = \hbar = T = 1$.   Next let's do this integral.
 
\vfill \eject \begin{center}\noindent\fbox{\begin{minipage}{30em}

\begin{center}
\entry{ENTROPY OF THE HARMONIC OSCILLATOR: 6} 

\vskip 2em
\noindent
\textbf{\boldmath
When $m = \kappa = k = \hbar = T = 1$ \\
 the entropy of a classical harmonic oscillator is
 \vskip 0.5em
\[ \begin{array}{ccl}   S &=& 
\displaystyle{ -\int_{-\infty}^\infty \! \int_{-\infty}^\infty \, e^{- (p^2 + q^2)/2} \ln \left( e^{-(p^2 + q^2)/2}\right) \;  \frac{dp \, dq}{2 \pi} }
 \\ \\
&=&
\displaystyle{\frac{1}{2 \pi} \int_0^{2\pi} \int_0^\infty \frac{r^2}{2}  e^{-r^2/2}  \;    r dr d\theta}
\qquad (\textrm{switching to polar}) 
\\ \\
&=&
\displaystyle{ \int_0^\infty \frac{r^2}{2}  e^{-r^2/2}  \;    r dr}
\; \qquad \qquad \qquad (\textrm{doing the $\theta$ integral}) 
\\ \\
&=& \displaystyle{ \int_0^\infty u e^{-u} du  }  
\qquad  \qquad \qquad \qquad \; (\textrm{substituting } u = r^2/2)
\\ \\
&=& 1  
\end{array}
\]
}
\end{center} \vskip 1em

 \end{minipage}} \end{center} \vskip 1em 
 
 Now let's do the integral to compute the entropy of the harmonic 
 oscillator.  We copy a famous
trick for computing the integral of a Gaussian.  First we switch to polar
coordinates in the $pq$ plane, where
\[    r^2 = p^2 + q^2 \textrm{\; and \; } dp \, dq = r  dr d\theta .\]
Then we integrate with respect to $\theta$, which cancels out the factor of
$1/2\pi$.    Then we do a substitution $u = r^2/2$.  But for us $r^2/2$ is
minus the logarithm of the Gaussian:
\[      - \ln (e^{-(p^2 + q^2)/2}) =  \frac{r^2}{2}  \]
so we're left with
\[    S = \int_0^\infty u e^{-u} \, d u\]
which we can do with integration by parts.

After all this work, we get an incredibly simple answer:
\[   S = 1 .  \]
So in the special case where $m = \kappa = k = \hbar = T = 1$, the entropy of a classical harmonic oscillator in thermal equilibrium is $1$.   

Now let's return to the general case, and finish the job.
 
 \vfill \eject \begin{center}\noindent\fbox{\begin{minipage}{30em}

\begin{center}
\entry{ENTROPY OF THE HARMONIC OSCILLATOR: 7} 
\noindent
\vskip 2em
\textbf{\boldmath
A classical harmonic oscillator with frequency $\omega$ has entropy 
\[  S = k \ln(\alpha kT/\hbar \omega) \]
for some dimensionless constant $\alpha$. \vskip 1em
But when $m = \kappa = k = \hbar = T = 1$ \\ 
we have $\omega = 1$ and $S = 1$, so we must have
\[  \alpha = e \]
and thus finally
\[ \color{darkred}  S = k \left( \ln\left(\frac{ kT}{\hbar\, \omega}\right)  + 1\right) \]
}
\end{center} \vskip 1em

 \end{minipage}} \end{center} \vskip 1em 
 
Knowing the entropy in one special case, we can figure out the constant $\alpha$ in our general formula for the entropy.   Our general formula says
 \[  S = k \ln(\alpha kT/\hbar \omega) .\]
 But when $m = k = T = \hbar = 1$ we get $\omega = k/m = 1$, and we saw last time that in this
 special case we get
 \[ S = 1. \]
 So $\alpha$ must equal $e$.  
 
Thus, the entropy of an oscillator with frequency $\omega$ at temperature $T$ is
\[ S = k \ln(ekT/\hbar\omega) =
 k \left( \ln\left(\frac{ kT}{\hbar\, \omega}\right)  + 1\right) .\]
The extra $1$ here is fascinating to me.  If we had slacked off,
ignored the possibility of a dimensionless constant $\alpha$, and crudely
used dimensional analysis to guess $S$ approximately the way people
often do, we might have gotten
\[ S = k \ln(kT/\hbar\omega)  \]
This would be off by 1 nat.  

\emph{What does the 1 extra nat mean?}  It seems pretty
mysterious now.  But later we'll understand it!    I \hyperlink{sec:ENTROPY_HARMONIC_OSCILLATOR_3}{already mentioned} that often entropy is
\emph{roughly} $k$ times the logarithm of something called the `number of accessible states'.   
But that formula is not exactly right: there's also an extra term related to energy, and that accounts for the 1 extra nat here.   Be patient, and you'll see what I mean.

 \vfill \eject \begin{center}\noindent\fbox{\begin{minipage}{30em}

\begin{center}
\entry{WHERE ARE WE NOW?}
\end{center}

\noindent
\textbf{\boldmath
{\color{red} The mystery: why does each molecule of hydrogen have $\sim\!23$ bits of entropy at
standard temperature and pressure?}
\vskip 1em \noindent
{\color{brickred}
The goal: derive and understand the formula
for the entropy of a classical ideal monatomic gas:
\[   S = kN\left( \ln\frac{V}{N} +  \frac{3}{2} \ln k T + \gamma
\right) \]
including the mysterious constant $\gamma$.}
\vskip 1em \noindent
{\color{darkred}
The subgoal: compute the entropy of a single classical particle in a 1-dimensional box.}
\vskip 1em \noindent
The sub-subgoal:  explain entropy from the ground up, and compute the entropy of a classical harmonic oscillator: \break
\[ \qquad \qquad S =  k \left( \ln \frac{kT}{\hbar \omega} + 1 \right)  \qquad \qquad  \color{darkgreen} \scalebox{1.5}{\checkmark}\]  }
 \end{minipage}} \end{center} \vskip 1em 
  
 Okay, so we've gotten somewhere!  By doing the right integral, we've figured out
 that the entropy $S$ of a classical harmonic oscillator of
 frequency $\omega$ in thermal equilibrium at temperature $T$ is 
 \[  S =  k \left( \ln \frac{kT}{\hbar \omega} + 1 \right).  \]  
 where $k$ is Boltzmann's constant and $\hbar$ is the reduced Planck's constant.  
 
We could compute the entropy of a single particle in a box the same way, and 
also the entropy of a classical ideal diatomic gas.   But the integrals get a bit hairy, so
people prefer to use a clever trick called the `partition function'.   It's definitely worth learning.
It's not merely a clever trick, it gives new insights on the relation between entropy, energy
and temperature.    So let's talk about it.
 
 \vfill \eject \begin{center}\noindent\fbox{\begin{minipage}{30em}
 
\begin{center}
\entry{THE PARTITION FUNCTION} 

\vskip 3em
\noindent
\textbf{\boldmath
If a system has a set of states $X$ \\ with measure $dx$ and \\ its energy is
a function $E \maps X \to \R$, \\ its \define{partition function} is
\[    Z(\beta) = \int_X e^{-\beta E(x)} \, dx \]
where $\beta$ is the coolness.   
}
\end{center} \vskip 1em

 \end{minipage}} \end{center} \vskip 1em
 
 I want to compute the entropy of a particle in a box, and ultimately the entropy of
 a box of hydrogen.  We could do it
directly, but that's a bit ugly.  It's better to use the `partition
function'.  This amazing function knows everything about statistical
mechanics.  From it you can get the entropy---and much more!
 
  \vfill \eject \begin{center}\noindent\fbox{\begin{minipage}{30em}

\begin{center}
\define{THE PARTITION FUNCTION AND THE BOLTZMANN DISTRIBUTION} 

\vskip 2em
\noindent
\textbf{\boldmath
If a system has a set of states $X$ with measure $dx$ \\ 
and its energy is
 $E \maps X \to \R$, \\ in thermal equilibrium at coolness $\beta$ its
probability distribution of states is the \define{Boltzmann distribution}:
\[ \displaystyle{ p(x) = \frac{  e^{-\beta E(x)} }{\int_X e^{-\beta E(x)} \, dx }= 
\frac{e^{-\beta E(x)}}{Z(\beta)}   } \]
where 
\[    Z(\beta) = \int_X e^{-\beta E(x)} \, dx \]
is its \define{partition function}.
}
\end{center} \vskip 1em

 \end{minipage}} \end{center} \vskip 1em 
 
In fact we've already seen the partition function: it's the thing you have to divide $e^{-\beta
E(x)}$ by to get a function whose integral is 1.   And that function whose integral is 1 is 
the Boltzmann distribution:
the probability distribution of states in thermal equilibrium
at coolness $\beta$.    So the partition function is a humble normalizing factor!  
And yet we'll see that it's incredibly powerful.  It's kind of surprising.

Like Lagrangians in classical mechanics, it's fairly easy to use partition functions, but it's harder
to understand what they `really mean'.   We will try.   But first let's see how to use them.
 
 \vfill \eject \begin{center}\noindent\fbox{\begin{minipage}{30em}

\begin{center}
\entry{THE PARTITION FUNCTION KNOWS ALL!} 

\vskip 2em
\noindent
\textbf{\boldmath
If a system has partition function
\[    Z(\beta) = \int_X e^{-\beta E(x)} \, dx \]
then in thermal equilibrium at coolness $\beta$ its expected energy is
\define{
\[  \langle E \rangle =  -  \frac{d }{d \beta} \ln Z \]
}
and its entropy is
\define{
\[     S = k\left(\ln Z - \beta \frac{d }{d \beta} \ln Z \right) \]
}
}
\end{center} \vskip 1em

 \end{minipage}} \end{center} \vskip 1em
 
 Here's how you can compute the expected energy $\langle E
\rangle$ and the entropy $S$ of any system starting from its partition
function $Z(\beta)$ as a function of the coolness $\beta$.    I'll show you why these 
formulas are true, and then we'll test them
out on the harmonic oscillator, where we have already computed the expected
energy and entropy by other methods.
 
  \vfill \eject \begin{center}\noindent\fbox{\begin{minipage}{30em}

\begin{center}
\entry{THE PARTITION FUNCTION  KNOWS  THE EXPECTED ENERGY} 

\vskip 1.5em
\noindent
\textbf{\boldmath
If a system has partition function
$ \displaystyle{  Z(\beta) = \int_X e^{-\beta E(x)} \, dx } $
then 
\[ \begin{array}{ccccl} \displaystyle{ -  \frac{d }{d \beta} \ln Z }
&=&  \displaystyle{- \frac{1}{Z} \frac{d}{d \beta} Z} &=&
\displaystyle{ - \frac{1}{Z} \frac{d}{d \beta}  \int_X e^{-\beta E(x)} \, dx}
 \\ \\
&&&=&  \displaystyle{ \frac{1}{Z} \int_X E(x) \, e^{-\beta E(x)} \, dx } \\ \\
&&&=&  \frac{ \displaystyle{\int_X E(x) \, e^{-\beta E(x)} \, dx }}{\displaystyle{\int_X e^{-\beta E(x)} \, dx}}  \\ \\&&&=&  \langle E \rangle 
\end{array}
\]
\vskip 1em
In short, the expected energy is \define{ $ \displaystyle \langle E \rangle =  -  \frac{d }{d \beta} \ln Z $}
}
\end{center} \vskip 1em

\end{minipage}} \end{center} \vskip 1em 
 
 The partition function is all-powerful!  For starters, if you know the partition function of a physical system, you can figure out its expected energy.   The expected energy $\langle E \rangle$ is minus the derivative of $\ln Z$ with respect to the coolness $\beta = 1/kT$. 

How do we show this?  Easy: just look at the calculation above!  We get a
fraction, which is the expected value of $E$ with respect to the Gibbs
distribution.

By the way, this trick of taking the derivative of the logarithm of a function is famous: it's
called a `logarithmic derivative'.  Notice that
\[    \frac{d}{dx} \ln f(x) = \frac{f'(x)}{f(x)} . \]
Thus the logarithmic derivative says how fast a function is growing compared to the
value of the function itself---like the interest rate in compound interest.

\vfill \eject \begin{center}\noindent\fbox{\begin{minipage}{35em}
 
\begin{center}
\entry{THE PARTITION FUNCTION KNOWS THE ENTROPY } 

\vskip 1.5em
\noindent
\textbf{\boldmath
If a system has Boltzmann distribution
\[ \displaystyle{ p(x) =  
\frac{e^{-\beta E(x)}}{Z} \qquad \textrm{ where } \qquad Z = \int_X e^{-\beta E(x)} \,dx  } \]
then its entropy in thermal equilibrium is 
\[ \begin{array}{ccccl} \displaystyle{ S } 
&=&  \displaystyle{-k \int_X p(x) \ln p(x) dx } \phantom{x}
&=&  \phantom{x} \displaystyle{-k \int_X   p(x) \ln \left( \frac{e^{-\beta E(x)}}{Z}\right) dx } \\ \\
&&&=& \phantom{x}  \displaystyle{ k \int_X   p(x) \Big(\! \ln Z + \beta E(x) \Big) dx } \\ \\
&&&=& \phantom{x} \displaystyle{k \Big(\!\ln Z + \beta \langle E \rangle \Big) }
\end{array}
\]
\vskip 0.5em
But since  $ \langle E \rangle =  -  \frac{d }{d \beta} \ln Z $, this gives
\define{
\[  \displaystyle{ S = k\left(\!\ln Z - \beta  \frac{d }{d \beta} \ln Z \right) } \]
}
}
\end{center} 

\end{minipage}} \end{center} \vskip 1em 
 
 The entropy is a bit more complicated.  But don't be scared!  The Boltzmann distribution $p(x)$ is a fraction, so the log of this fraction breaks into two parts:
\[   \ln p(x) = \ln \left( \frac{e^{-\beta E(x)}}{Z} \right) = - (\ln Z + \beta E(x))  .\]
Thus our integral for entropy breaks into two parts:
\[      S = -k \int p(x) \ln p(x) \, dx  
= k \int p(x) \ln Z \, dx +  k\beta \int p(x) E(x) \, dx.  \]
The first part is just $k \ln Z$ since the integral of $p(x)$ is $1$.  The second part
is $k\beta \langle E \rangle$.  If we  
use what we just learned about $\langle E 
\rangle $:
\[    \langle E \rangle = - \frac{d}{d \beta} \ln Z  \]
we get this formula for entropy in terms of the partition function:
\[  \displaystyle{ S = k\left(\!\ln Z - \beta  \frac{d }{d \beta} \ln Z \right).} \]
This formula seems hard to understand at first.  To extract its inner meaning, we need
a new concept: `free energy'.
 
 \vfill \eject \begin{center}\noindent\fbox{\begin{minipage}{33em}

\begin{center}
\entry{THE PARTITION FUNCTION KNOWS THE FREE ENERGY} 

\vskip 1.5em
\noindent
\textbf{\boldmath
To maximize entropy while holding expected energy constant, \\ you can just minimize
the  \define{free energy}
\[  \color{darkred}        F = \langle E \rangle - T S  \]
We've seen 
\[  \langle E \rangle =  -  \frac{d }{d \beta} \ln Z  \quad \textrm{ and } \quad
    S = k\left(\ln Z - \beta  \frac{d }{d \beta} \ln Z \right) \]
so with $\beta = 1/kT$ a little algebra shows
\vskip 0.5em
\[    \color{darkred}     F = - \frac{1}{\beta} \ln Z  \]
}
\end{center} 

\end{minipage}} \end{center} \vskip 1em 
 
We can understand the relation between entropy, energy and the partition function if we bring in a concept I haven't mentioned yet: the \define{free energy} 
\[  F = \langle E \rangle - TS .\]
Since we know formulas for $\langle E \rangle$ and $S$ in terms of the partition 
function, we can work out a formula for $F$.
And it's really simple!  Much simpler than $S$, for example.  It's 
just 
\[    F = -\frac{1}{\beta} \ln  Z .\]

But what's the \emph{meaning} of free energy?
Remember: to maximize the Shannon entropy $H$ subject to a constraint on expected energy,  we introduced the Lagrange multiplier $\beta = 1/kT$  
and maximized the quantity $H - \beta \langle E \rangle$.
But if you multiply this quantity by $-kT$, you get free energy:
\[     -kT(H - \beta \langle E \rangle) = \langle E \rangle - T S  = F.\]
So, as long as $T > 0$, maximizing entropy subject to a constraint on 
expected energy is equivalent to \emph{minimizing} free energy!

Thus, free energy turns a problem of maximizing
entropy subject to a constraint into a minimization problem without a constraint.   
The point is not that we've turned maximization
into minimization: that's just an arbitrary business with signs.   The point is
that free energy lets us stop thinking about the constraint.   

There's a huge amount to say about the free energy,
which is also called the `\href{https://en.wikipedia.org/wiki/Helmholtz_free_energy}{Helmholtz free energy}', since there are other
kinds.  You can think of $TS$ as the amount of
energy in useless random form, since it comes from entropy.  Since $\langle E \rangle$ is the total expected
energy, $F = \langle E \rangle - TS$ is the amount of `useful' energy.   More precisely, the free
energy is the maximum amount of work obtainable from
a system at a constant temperature.   But showing this would take us out of our way.
 
 \vfill \eject \begin{center}\noindent\fbox{\begin{minipage}{30em}
 
\hypertarget{sec:PARTITION_FUNCTION_KNOWS_ALL_REVISITED}{}

\begin{center}
\entry{THE PARTITION FUNCTION KNOWS ALL: REVISITED} 

\vskip 2em
\noindent
\textbf{\boldmath
If $Z(\beta)$ is the partition function of a system, in thermal equilibrium at coolness $\beta$ its expected energy is
\[ \color{darkred}  \langle E \rangle =  -  \frac{d }{d \beta} \ln Z \]
and its free energy is
\[    \color{darkred}     F = - \frac{1}{\beta} \ln Z  \]
We can compute its entropy from these using
\[   \color{darkred} F = \langle E \rangle -  TS \]
and we get   
\[   \color{darkred}  S = k \left(\ln Z - \beta \frac{d}{d \beta} \ln Z \right) 
\]
}
\end{center} 

 \end{minipage}} \end{center} \vskip 1em 
 
 Now we can tell a simpler story, which is easier to remember.  
Free energy, being the energy in useful form, is the expected energy minus the useless energy,
which is temperature times entropy.   Thus
\[             F = \langle E \rangle - TS \]
so
\[   \begin{array}{ccc}  S &=& \displaystyle{\frac{\langle E \rangle - F}{T}} \\  \\
&=&  k \beta (-F + \langle E \rangle) 
\end{array}
\]
and using our formulas for $F$ and $\langle E \rangle$ in terms of the partition function $Z$, we get
\[    S = k \left(\ln Z - \beta \frac{d}{d \beta} \ln Z \right) .
\]
 
The story here is more of a mnemonic than a true explanation, because I'm
not saying much what it means for energy to be `useful' or `useless'.   I've only given this hint: when a system is in thermal equilibrium, its free energy is minimized.
For more on the meaning of free energy, try a good book on thermodynamics, like this:
\begin{itemize}
\item
Frederick Reif, \textsl{Fundamentals of Statistical and Thermal Physics}, Waveland Press, Long Grove, Illinois, 2009. 
\end{itemize}
Right now I'd rather say a bit about the meaning of the partition function.
 
\vfill \eject \begin{center}\noindent\fbox{\begin{minipage}{30em}

\hypertarget{sec:MEANING_PARTITION}{}

\begin{center}
\entry{THE MEANING OF THE PARTITION FUNCTION} 

\vskip 1em
\noindent
\textbf{\boldmath
Say $X$ is a set where each point $i$ has an `energy' $E_i \in \R$. \\ 
Its \define{partition function} is 
\[    Z = \sum_{i \in X} e^{- \beta E_i}  \]
where $\beta \in \R$ is the coolness.   
\vskip 1em
The partition function counts the points of $X$---but it counts points with large energy
less, since they're less likely to be `occupied'.   
\vskip 1em 
If $\beta = 1/kT$, points with energy $E_i \gg kT$
count for very little. 
\vskip 1em
But as $T \to +\infty$, all points get fully counted and $Z \to |X|$.
\vskip 1em
In physics we call $Z$ the \define{number of accessible states}.
}
\end{center} 

 \end{minipage}} \end{center} \vskip 1em 
 
 Say we have a system with some countable set of states $X$.
 In thermal equilibrium at temperature $T$, the probability that the system
 is in its $i$th state is proportional to $\exp(-\beta E_i)$, where $E_i$ is the
 energy of that state and $\beta$ is the coolness.   Thus, physicists say the partition function
 \[        Z = \sum_{i \in X} e^{-\beta E_i} \]
 is the \define{number of accessible states}: roughly, the number of states the system 
 can easily be in at temperature $T$, where $\beta = 1/kT$.  
 
 This is a funny thing to say, because being `accessible' is not a yes-or-no matter.   A more precise statement is that the partition function counts states weighted by their \define{accessibility} $\exp(-\beta E_i)$.   States whose energy is low compared to $kT$ are highly accessible,  or probable, because $\exp(-\beta E_i)$ is close to $1$ if $E_i \ll kT$.   States of high energy are more inaccessible, or improbable, since $\exp(-\beta E_i)$ is close to $0$ if $E_i \gg kT$.   
 
Calling the partition function the `number of accessible states' emphasizes how
it generalizes the cardinality
$|X|$ of an ordinary set $X$, meaning its number of points.   Let's make this precise!
Let's call a set $X$ 
with a function $E \maps X \to \R$
an \define{energetic set}.  I will write it merely as $X$, so you need to
remember it comes with an energy function.  I will call its partition function $Z(X)$:
\[          Z(X) = \sum_{i \in X} e^{-\beta E_i}   .\]
If $X$ is finite we don't have to worry about the convergence of this sum.  My main message is this:

\begin{center}
\define{The partition function $Z(X)$ does for energetic sets \\
what the cardinality $|X|$ does for sets.}
\end{center}
\noindent For example, just like the cardinality, the partition function adds when you take disjoint unions, and 
multiplies when you take products!    Let's see why.

\begin{puzzle}
The disjoint union $X+X'$ of energetic sets $E \maps X \to
\R$ and $E' \maps X' \to \R$ is again an energetic
set: for points in $X$ we use the energy function $E$, while for points in
$X'$ we use the function $E'$.    Show that the partition function obeys
the law $Z(X+X') = Z(X) + Z(X')$, at least for finite energetic sets.
\end{puzzle}

\begin{puzzle}  
\label{puzzle:product_of_energetic_sets}
The cartesian product $X \times X'$ of energetic sets $E\maps X \to
\R$ and $E' \maps X' \to \R$ is again an energetic
set: define the energy of $(x,x') \in X \times X'$ to be
$E(x) + E(x')$.  This is how it really works in physics.   Show that the partition
 function obeys the law $Z(X \times X') = Z(X)\, Z(X')$, at least for finite
 energetic sets.
\end{puzzle}

\begin{puzzle}
Show that if $X$ is a finite energetic set, its partition function $Z(X)$ approaches
its cardinality $|X|$ as $T \to +\infty$.
\end{puzzle}

The key virtue of cardinality is that two sets are isomorphic---that is,
there exists a one-to-one and onto function between them---if and only if they
have the same cardinality.   This generalizes to energetic sets if we use
the partition function instead of the cardinality!   Let's say two energetic sets
with energy functions 
$E \maps X \to \R$ and $E' \maps X' \to \R$ are \define{isomorphic} if
there is a one-to-one and onto  $f \maps X \to X'$ which is compatible
with their energy functions, meaning
\[             E'(f(x)) = E(x) \]
for all $x \in X$.   

\begin{puzzle} 
Show that two finite energetic sets are isomorphic if and only if
they have the same partition function.  (Hint: the key is to show that the functions 
\(\exp(-E/kT)\) for various energies $E \in \R$ are linearly independent.
As a step toward this, show that a finite linear combination
\[         \sum_i c_i \exp(-E_i/kT)  \]
can only be zero if $c_i = 0$ for the smallest energy $E_i$.)
\end{puzzle}

If you're into category theory, here are some ways to go further.  If you're not,
please skip to the next page.

\begin{puzzle}  
Make up a category of energetic sets, where morphism are maps that are 
compatible with their energy functions.  Prove that it is a category.
\end{puzzle}

\begin{puzzle}  Show the disjoint union of energetic sets is the coproduct in this
category.
\end{puzzle}

\begin{puzzle}  Show that what I called the 
cartesian product of energetic sets is \emph{not} the
product in this category.
\end{puzzle}

\begin{puzzle}  Show that what I called the `cartesian product' of energetic sets
gives a symmetric monoidal structure on the category of energetic
sets.  So we should really write it as a tensor product $X \otimes
X'$, not $X\times X'$.
\end{puzzle}

\begin{puzzle}  Show this tensor product distributes over coproducts: $ X\otimes (Y+Z) 
\cong X\otimes Y + X\otimes Z$.
\end{puzzle}

We can go even further and define not only a partition function for energetic sets,
but also an expected energy, free energy, and entropy,
using the formulas we've seen 
\hyperlink{sec:PARTITION_FUNCTION_KNOWS_ALL_REVISITED}{earlier}.    
These obey a bunch of rules like this:

\begin{puzzle} Define the entropy of an energetic set by
\[         S(X) = k \left(\ln Z(X) - \beta \frac{d}{d\beta} \ln Z(X) \right)  .\]
Show that 
\[         S(X \otimes Y) = S(X) + S(Y)   .\]
\end{puzzle}
 
 \vfill \eject \begin{center}\noindent\fbox{\begin{minipage}{32em}

\hypertarget{sec:ENTROPY_COMES_IN_TWO_PARTS}{}

\begin{center}
\entry{ENTROPY COMES IN TWO PARTS} 

\vskip 1em
\noindent
\textbf{\boldmath
The entropy of a system in thermal equilibrium \\ is always the sum of two parts:
\begin{enumerate}
\item The \define{free energy part}:
\[            - \frac{F}{T} = k \ln Z  \]
This is Boltzmann's constant times the logarithm of the number of accessible states. 
\item The \define{expected energy part}:
\[            \frac{\langle E \rangle}{T}  \]
This equals $\frac{1}{2} n k T$ if the system has $n$ degrees of freedom and its energy
is a positive definite quadratic form.
\end{enumerate}
}
\end{center} 

 \end{minipage}} \end{center} \vskip 1em 

Before we dive into examples, it's good to think one last time about the 
entropy of a system in thermal equilibrium.   We've seen that this entropy is always the sum of two parts,
which we could call the \define{free energy part} $-F/T$ and the \define{expected energy part} $\langle E \rangle / T$.    But there are various ways to think about this.  One is simply that it follows from
$F = \langle E \rangle - TS$: the free energy is the expected energy minus the useless energy.   But here is 
another way to think about it.

In his early work, Boltzmann said the entropy of a system is $k$ times the logarithm of the number of states
it can occupy.   This is true if all these states are equally probable.   But typically some states are more probable
than others.   We could try to address this by replacing the number of states with the number of accessible
states
\[                    Z = \sum_{i \in X} e^{-\beta E_i}  . \]
Here we count states weighted by their accessibility $\exp(-\beta E_i)$.   If we try to follow Boltzmann's
prescription with this adjustment we get $k \ln Z = -F/T$.   This is the free energy part of the entropy.

In many situations this is close to the true entropy.   But this clearly can't be all there is to it.   After all, suppose
we add the same constant $c$ to the energy of each state.  Then the probability of each state in thermal equilibrium is unchanged, so the entropy must stay the same!    But the accessibility of each state gets multiplied
by $\exp(-\beta c)$, so we have to subtract $k \beta c$ from the free energy part of the entropy.   There 
must be some compensating term---and this is the expected energy part of the entropy, $\langle E \rangle/T$.    When we add $c$ to the energy of each state, this goes up by $c/T = k \beta c$.

Thus, in thermal equilibrium we can think of entropy as $k$ times the log of the number
of accessible states, `corrected' so that the result doesn't change when we add a constant
to the energy of every state.

\vfill \eject \begin{center}\noindent\fbox{\begin{minipage}{30em}
 
\begin{center}
\entry{THE POWER OF THE PARTITION FUNCTION}

\vskip 2em
\noindent
\textbf{\boldmath
A classical harmonic oscillator with mass $m$ and spring constant $\kappa$  
 has energy
\[           E(p,q) = \frac{p^2}{2m} + \frac{\kappa q^2}{2} \]
Its partition function is
\[ \color{red} Z(\beta) =
\displaystyle{ \int_{-\infty}^\infty \! \int_{-\infty}^\infty \, e^{-\beta E(p,q)} \;  \frac{dp \, dq}{h} }  \]
where $\beta$ is coolness and $h$ is Planck's constant.   \\ 
\vskip 0.5em
From this we can find its expected energy and free energy in 
thermal equilibrium:
\define{
\[  \langle E \rangle =  -  \frac{d }{d \beta} \ln Z  \qquad 
 F = - \frac{1}{\beta} \ln Z  \]
}
and then its entropy:
\define{
\[     S = \frac{\langle E \rangle -F}{T}  \]
}
where $T$ is temperature: $\beta = 1/kT$ where $k$
is Boltzmann's constant.
}
\end{center} \vskip 1em

 \end{minipage}} \end{center} \vskip 1em 
 
 To test the power of the partition function, let's use it to
figure out the entropy of a classical harmonic oscillator.  Here's the 
game plan.  First we'll compute the partition function by doing the integral in 
bright red.  Then we'll use it to compute the oscillator's expected energy and free
energy.  Then we'll subtract those and divide by temperature to get the entropy.

In fact, we've already worked out the answer to this problem:
\[ S = k \left( \ln \left( \frac{kT}{\hbar \omega}\right) + 1 \right).  \]
Our earlier approach led to some cool insights.  But it was `tricky', not
systematic.  The partition function method is systematic, so it's 
good for harder problems.   It will also give new insight into that pesky $+1$.

When we compute the entropy using a partition function, all the pain
is concentrated at one point: computing the partition function!  So let's get that
over with.  
 
 \vfill \eject \begin{center}\noindent\fbox{\begin{minipage}{33em}
\begin{center}
\entry{HARMONIC OSCILLATOR: PARTITION FUNCTION}

\vskip 2em
\noindent
\textbf{\boldmath
A classical harmonic oscillator has energy $E(p,q) = \frac{p^2}{2m} + \frac{\kappa q^2}{2}$ \\ and frequency $\omega = \sqrt{\kappa/m}$, so its partition function is
 \vskip 0.5em
\[ \begin{array}{ccl}   Z(\beta) &=& 
\displaystyle{ \int_{-\infty}^\infty \! \int_{-\infty}^\infty \, e^{-\beta \left(\frac{p^2}{2m} + \frac{\kappa q^2}{2}\right)} \;  \frac{dp \, dq}{h} }
 \\ \\
&=&
\displaystyle{\sqrt{\frac{m}{\kappa}}  \int_{-\infty}^\infty \! \int_{-\infty}^\infty \, e^{-\beta \left(\frac{x^2 + y^2}{2}\right)} \;  \frac{ dx \, dy}{h} }  \qquad (x = \frac{p}{\sqrt{m}}, \; y = \sqrt{\kappa}q) 
 \\ \\
&=&
\displaystyle{\frac{1}{h \omega}  \int_0^{2\pi} \int_0^\infty  e^{-\beta r^2/2}  \;    r dr \, d\theta}
\qquad \qquad (\textrm{switching to polar}) 
\\ \\
&=& \displaystyle{ \frac{2\pi}{h\omega }  \int_0^\infty e^{-\beta u} du  }  
\quad \qquad  \qquad \qquad \qquad ( u = r^2/2)
\\ \\
&=& \displaystyle{ \frac{1}{\hbar \omega} \cdot \frac{1}{\beta} }
\end{array}
\]
\[  \color{darkred} Z(\beta) = \frac{1}{\beta \hbar\omega} \]
}
\end{center} \vskip 1em

 \end{minipage}} \end{center} \vskip 1em 

 For the harmonic oscillator, the partition function is the integral of a Gaussian in two variables.  A
change of variables makes the Gaussian `round', and then we use polar
coordinates to do the integral.

The physicist Kelvin is said to have written 
\[       \int_{-\infty}^\infty e^{-x^2} d x = \sqrt{ \pi } \]
on the blackboard and said ``A mathematician is one to whom that is as obvious as that twice two makes four is to you.''   I find that rather obnoxious, but when I heard the story as a kid, I made damn sure I knew how to do this integral.  The usual trick is to compute the square of this integral using polar coordinates.

Now we're seeing something interesting.  The harmonic oscillator, whose energy depends quadratically on  two degrees of freedom, is physically more important than a system whose energy depends
quadratically on just one degree of freedom.  And when $\beta = h = \omega = 1$, the partition function of the harmonic oscillator is
\[     \int_{-\infty}^\infty \! \int_{-\infty}^\infty \, e^{-\left(\frac{x^2 + y^2}{2}\right)} \;   dx \, dy  =
\int_0^{2 \pi} \int_0^\infty e^{-r^2/2} r dr \; d \theta = 2 \pi \int_0^\infty e^{-u} d u = 2 \pi , \]
which is \emph{more fundamental} than the integral Kelvin wrote down.

 \vfill \eject \begin{center}\noindent\fbox{\begin{minipage}{30em}

\begin{center}
\entry{HARMONIC OSCILLATOR: EXPECTED ENERGY} 

\vskip 2em
\noindent
\textbf{\boldmath
A classical harmonic oscillator has partition function
\[   Z = \frac{1}{\beta \hbar \omega} \]
so its expected energy in thermal equilibrium is
\[   \langle E \rangle = -\frac{d}{d \beta} \ln Z = \frac{1}{\beta} \]
or
\[   \color{darkred}   \langle E \rangle =  kT \]
just as the equipartition theorem says it must be!
}
\end{center} \vskip 1em

 \end{minipage}} \end{center} \vskip 1em 
 
Once we know the partition function of the classical harmonic oscillator,
it's easy to compute its
expected energy: just use 
\[  \langle E \rangle = - \frac{d}{d \beta} \ln Z \]
and get
\[   \langle E \rangle = - \frac{d}{d\beta} \ln\left( \frac{1}{\beta \hbar \omega} \right) = 
\frac{1}{\beta}  .\]
We can also figure this out using the equipartition theorem.
Remember, the equipartition theorem applies to a classical system
whose energy is quadratic.  If it has $n$ degrees of freedom, then at
temperature $T$ it has
\[ \langle E \rangle = \frac{n}{2}  kT.  \]
Our harmonic oscillator has $n = 2$, so we get $\langle E \rangle = kT$.  
Good, this matches the partition function approach!
 
 \vfill \eject \begin{center}\noindent\fbox{\begin{minipage}{30em}

\begin{center}
\entry{HARMONIC OSCILLATOR: FREE ENERGY} 

\vskip 2em
\noindent
\textbf{\boldmath
A classical harmonic oscillator has partition function
\[   Z(\beta) = \frac{1}{\beta \hbar \omega} \]
so its free energy in thermal equilibrium is
\[   F = -\frac{1}{\beta} \ln Z =  -\frac{1}{\beta} \ln\left( \frac{1}{\beta \hbar \omega}\right) \]
\vskip 0.5em
or
\vskip -0.2em
\[   \color{darkred} F = -kT  \ln\left( \frac{kT}{\hbar \omega} \right)  \]
}
\end{center} \vskip 1em

 \end{minipage}} \end{center} \vskip 1em

The partition function lets us do more!  It lets us compute the
free energy, too, using
\[  F = - \frac{1}{\beta} \ln Z \]
Unlike the expected energy, the free energy involves Planck's constant:
\[   F = -kT  \ln\left( \frac{kT}{\hbar \omega} \right) . \]
Note $kT$ and $\hbar \omega$ both have units of energy, so
$kT/\hbar \omega$ is dimensionless, which is good because we're
taking its logarithm.  Also note that the free energy is negative at high
temperatures!  That may seem weird, but it turns out to be good
when we compute the entropy.

\vfill \eject \begin{center}\noindent\fbox{\begin{minipage}{34em}

\hypertarget{sec:HARMONIC_OSCILLATOR_ENTROPY}{}

\begin{center}
\entry{HARMONIC OSCILLATOR: ENTROPY} 

\vskip 2em
\noindent
\textbf{\boldmath
In thermal equilibrium at temperature $T$, \\ a classical harmonic oscillator  
has 
\[ \!\!\! \textrm{ expected energy } \langle E \rangle = k T 
\textrm{ and free energy }    F =-kT  \ln\left( \frac{kT}{\hbar \omega}\right) \]
so its entropy is
\vskip 0.1em
\[   S = \frac{\langle E \rangle - F}{T} = \frac{kT +  kT \ln {\displaystyle \left(\frac{kT}{\hbar \omega}\right) }}{T}  \]
\vskip 0.1em
or
\vskip 0em
\[   \color{darkred} S = k  \left( \ln \left( \frac{kT}{\hbar \omega} \right) + 1 \right) \]
}
\end{center} \vskip 1em

 \end{minipage}} \end{center} \vskip 1em
 
To compute the entropy of a classical harmonic oscillator, we just use
\[ S = \frac{ \langle E \rangle - F}{T} . \]
We get the answer we got before, of course:
\[   S = k  \left( \ln \left( \frac{kT}{\hbar \omega} \right) + 1 \right) .\]
\emph{But now we can finally understand the puzzling extra $+1$.}  

As \hyperlink{sec:ENTROPY_COMES_IN_TWO_PARTS}{we've seen}, the 
entropy of \emph{any} system in thermal equilibrium consists of two parts:  
\begin{enumerate}
\item
The free energy part, $-F/T$.   For the classical harmonic oscillator this is
\[    - \frac{F}{T} = k \ln \left(\frac{kT}{\hbar \omega} \right) .\]
\item
The expected energy part, $\langle E \rangle/T$.  For the classical harmonic oscillator this is
\[    \frac{\langle E \rangle}{T} = k .  \]
\end{enumerate}
The free energy part of the entropy is always 
$k$ times the logarithm of the \hyperlink{sec:MEANING_PARTITION}{number of accessible states}.    For the classical harmonic oscillator, the  expected energy part of the entropy must equal $k$ by the  \hyperlink{sec:PROOF_OF_EQUIPARTITION_THEOREM_3}{equipartition
theorem}, since the oscillator's energy depends on 2 degrees of freedom.  This is small compared to the 
free energy part when $\hbar \omega \ll kT$: that is,  when quantum effects are small compared to thermal effects.

\vfill \eject \begin{center}\noindent\fbox{\begin{minipage}{32em}

\begin{center}
\entry{PARTICLE IN A BOX: 
PARTITION FUNCTION} 

\vskip 3em
\noindent
\textbf{\boldmath
The energy of a classical free particle of mass $m$ \\
in a 1-dimensional box depends only on its momentum $p$:
\[        E(p,q) = \frac{p^2}{2m} \]
Its position $q$ is trapped in the interval $[0,L]$. 
\vskip 1em
 Its partition function is therefore
\[ 
 Z(\beta) =  \displaystyle{  \int_0^L \int_{-\infty}^\infty e^{-\beta E(p,q)} \, \frac{dp\, dq}{h}}
=  \displaystyle{\frac{L}{h}  \int_{-\infty}^\infty e^{-\beta p^2/2m} \, dp}  
= \displaystyle{   \frac{L}{h} \sqrt{\frac{2 \pi m}{\beta}} }
\]
}
\end{center} \vskip 1em

\end{minipage}} \end{center} \vskip 1em 

Now let's turn to our ultimate goal: computing the entropy of a box of gas.
As a warmup, let's figure
out the entropy of a \emph{single} particle in a box.   In fact, let's start with 
a free classical particle in a one-dimensional box: that is, in some interval $[0,L]$.

The first step is to compute its partition function.
As you can see, this is easy enough.  But the whole idea raises some questions.
Some people get freaked out by the concept of entropy for a single
particle---I guess because it involves probability theory for a single
particle, and they think probability only applies to large numbers of
things.

I sometimes ask these people ``how large counts as large?''
In fact the foundations of probability theory are just as mysterious
for large numbers of things as for just one thing.  What do
probabilities really mean?  We could argue about this all day:
Bayesian versus. frequentist interpretations of probability, etc.   I  
\hyperlink{sec:WHAT_IS_PROBABILITY}{said a tiny bit about this before},
and I won't say more now.

Large numbers of things tend to make large deviations less likely.
For example the chance of having all the gas atoms in a box all on the
left side is less if you have 1000 atoms than if you have just 2.
This makes us \emph{worry} less about using averages and probability.

But the math of probability works the same for small numbers of particles---even
one particle!
Even better, knowing the entropy of one particle in a box will help us
understand the entropy of a million particles in a box---at least if
they don't interact, as we assume for an `ideal gas'.

But why just a \emph{one-dimensional} box?
The answer is that particle in a 3-dimensional box is mathematically the same as 3
noninteracting distinguishable particles in a one-dimensional box!
The $x, y,$ and $z$ coordinates of the 3d particle act like positions of
three 1d particles.

 \vfill \eject \begin{center}\noindent\fbox{\begin{minipage}{30em}

\begin{center}
\entry{PARTICLE IN A BOX: 
EXPECTED ENERGY} 

\vskip 1.5em
\noindent
\textbf{\boldmath
A classical free particle of mass $m$ in a 1d box of length $L$ \\
has partition function
\[   Z = \frac{L}{h} \sqrt{\frac{2 \pi m}{\beta}} \]
The expected energy of any system in \\ thermal equilibrium is 
\[ \displaystyle \langle E \rangle = -\frac{d}{d \beta} \ln Z \]
So, by the miracle of basic calculus, we get
\[
 { \color{darkred}  \langle E \rangle  =
 \frac{1}{2\beta}  =  \frac{1}{2}\,kT   }\]
\vskip 0.5em
as we'd expect from the equipartition theorem!
}
\end{center} \vskip 1em

 \end{minipage}} \end{center} \vskip 1em 
 
We worked out the partition function of a classical free particle in a
1-dimensional box.  From this we can work out its expected energy.
Look how simple it is!  It's just $\frac{1}{2}kT$, where $k$ is
Boltzmann's constant and $T$ is the temperature!

Why is the final answer so simple?
We can use the chain rule
\[ \frac{d}{d \beta} \ln Z = \frac{1}{Z} \frac{d Z}{d \beta} \]
to see that only the power of $\beta$ in 
\[ Z = \frac{L}{h} \sqrt{\frac{2\pi m }{\beta}} \]
matters, not all the constants: these constants show up in $d Z/d
\beta$, but also in $1/Z$, and they cancel.  The length $L$, the
mass $m$, Planck's constant $h$, the factor of $2\pi$... none of
this junk matters!  Not for the expected energy, anyway.  Because
$Z$ is proportional to $\beta^{-1/2}$, we simply get $ \langle E
\rangle = \frac{1}{2}k T$.

More generally, if the partition function
of a system is proportional to $\beta^{-c}$, its expected energy will be
$c k T$:
\[    Z \propto \beta^{-c} \implies \langle E \rangle = c k T .\]
But when is the partition function of a system proportional to
$\beta^{-c}\,$?  It's enough for the system's energy to be a positive
definite quadratic form in $n$ real variables---which physicists call `degrees of
freedom'.  Then $c = n/2$.   We've already seen an example with 2 degrees of freedom: 
the classical harmonic oscillator.   We saw that in this example $Z \propto 1/\beta$.  This gives $\langle E \rangle = kT$.   But the result is quite general:

\begin{puzzle}
Suppose we have a system with state space $\R^n$ and energy function 
$E \maps \R^n \to \R$ that is a positive definite quadratic form, so that
\[        E(x) = \frac{1}{2} \|A x \|^2  \]
for some invertible $n \times n$ matrix $A$.   Show that its partition function is proportional
to $\beta^{-c}$ where $c = n/2$.   
\end{puzzle}

\noindent
In fact, this is just a new outlook on our friend the \hyperlink{sec:PROOF_OF_EQUIPARTITION_THEOREM_3}{equipartition
theorem}.

Here's another thing to consider.
While our particle in a 1d box has $2$ degrees of freedom---position and momentum---its 
energy depends on just one of these, and quadratically on that one.  So its expected energy is $\frac{1}{2}nkT$ where $n = 1$, not $n = 2$.

So here's another puzzle for you:

\begin{puzzle} 
Say we have a harmonic oscillator with
spring constant $\kappa$.  As long as $\kappa > 0$, the energy
depends quadratically on $2$ degrees of freedom so $\langle E
\rangle = kT$.  But when $\kappa = 0$ it depends on just one, and suddenly
$ \langle E \rangle = \frac{1}{2} kT$.  How is such a discontinuity
possible?  In other words: how can a particle care so much about the difference
between an arbitrarily small positive spring constant and a spring
constant that's exactly zero, making its expected energy twice as much
in the first case?
\end{puzzle}

I'll warn you: this puzzle is deliberately devilish.  In a way 
it's a trick question!   

 \vfill \eject \begin{center}\noindent\fbox{\begin{minipage}{30em}

\begin{center}
\entry{PARTICLE IN A BOX:  FREE ENERGY} 

\vskip 2em
\noindent
\textbf{\boldmath
A classical free particle of mass $m$ in a 1d box of length $L$ \\ has partition function
\[ 
 Z = \displaystyle{   \frac{L}{h} \sqrt{\frac{2 \pi m}{\beta}} }
\]
The free energy of any system is given by
$F = \displaystyle{ - \frac{1}{\beta}\ln Z  }$, so 
\[  \color{darkred} F = -\frac{1}{\beta} \ln  \frac{L}{h} \sqrt{\frac{2 \pi m}{\beta}}  
 \]
\vskip 0.5em
Using $\beta = 1/kT$ and fiddling around a bit, \\ we can rewrite this as
\vskip 0.5em
\[ \color{darkred}
F = -kT \left(\ln L + \frac{1}{2} \ln kT + \frac{1}{2} \ln \frac{2 \pi m}{h^2} \right)  
 \]
}
\end{center} 

\end{minipage}} \end{center} \vskip 1em 

From the partition function of a classical free particle in a one-dimensional box we
can also compute its free energy!

\vfill \eject \begin{center}\noindent\fbox{\begin{minipage}{30em}

\hypertarget{sec:PARTICLE_IN_BOX_ENTROPY}{}

\begin{center}
\entry{PARTICLE IN A BOX:  ENTROPY} 

\vskip 2em
\noindent
\textbf{\boldmath
We've shown that in thermal equilibrium, a classical particle of mass $m$ in a 1-dimensional box of length $L$ has
expected energy
\[   \langle E \rangle = \frac{1}{2} kT  \]
and free energy
\[ F = -kT \left( \ln L + \frac{1}{2} \ln kT + \frac{1}{2} \ln \frac{2 \pi m}{h^2} \right)      \]
But entropy $S$ is always $(\langle E \rangle - F)/T$, so
\vskip 1em
\[
\color{darkred}
S = k \left(\ln L + \frac{1}{2} \ln kT  + \frac{1}{2} \ln \frac{2 \pi m}{h^2} + \frac{1}{2} \right) 
\]
}
\end{center} 

\end{minipage}} \end{center} \vskip 1em

Having worked out the expected energy $\langle E \rangle$ and
free energy $F$ for a single classical particle in thermal equilibrium in a
1-dimensional box, it is easy to work out its
entropy.  We just subtract the free energy from the expected energy and divide by
temperature:
\[              S = \frac{\langle E \rangle - F}{T}   .\]
The formula we get is not very snappy:
\[
S = k \left(\ln L + \frac{1}{2} \ln kT  + \frac{1}{2} \ln \frac{2 \pi m}{h^2} + \frac{1}{2} \right) .
\]
We will get a better formula \hyperlink{sec:ENTROPY_AND_THERMAL_WAVELENGTH}{later}, and ponder its meaning.   For now,  let's just make these observations:
\begin{itemize}
\item When we make the length $L$ of the box larger, the entropy becomes larger.
\item  When we increase the temperature $T$, the entropy becomes larger.
\item When we increase the mass $m$ of the particle, the entropy becomes larger.
\end{itemize}

The first two facts should feel intuitively obvious.  When we increase the box's length, there is 
more unknown information about the \emph{position} of the particle in thermal equilibrium.
When we increase the particle's temperature, there is more unknown information about
its \emph{momentum}.    The third fact is less obvious.   When we
introduce the concept of `\hyperlink{sec:THERMAL_WAVELENGTH}{thermal wavelength}', we will see that increasing the particle's mass decreases its thermal wavelength, which in turn increases its entropy in thermal equilibrium.

\vfill \eject \begin{center}\noindent\fbox{\begin{minipage}{30em}

\noindent
\begin{center}
\entry{WHERE ARE WE NOW?}
\end{center}
\textbf{\boldmath
{\color{red} The mystery: why does each molecule of hydrogen have $\sim\!23$ bits of entropy at
standard temperature and pressure?}
\vskip 1em \noindent
{\color{brickred}
The goal: derive and understand the formula
for the entropy of a classical ideal monatomic gas:
\[   S = kN\left( \ln\frac{V}{N} + \frac{3}{2} \ln k T+ \gamma
\right) \]
including the mysterious constant $\gamma$.}
\vskip 1em \noindent
{\color{darkred}
The subgoal: compute the entropy of a single classical \break particle in a 1-dimensional box:
\[
\qquad \qquad \quad S = k \left(\ln L + \frac{1}{2} \ln kT  + \frac{1}{2} \ln \frac{2 \pi m}{h^2} + \frac{1}{2} \right) \qquad {\color{darkgreen} \scalebox{1.5}{\checkmark}}  \]}
\hskip -0.5 em The sub-subgoal:  explain entropy from the ground up, and compute the entropy of a classical harmonic oscillator: \break
\[ \qquad \qquad S =  k \left( \ln \frac{kT}{\hbar \omega} + 1 \right)  \qquad \qquad  \color{darkgreen} \scalebox{1.5}{\checkmark}\]  
}

\end{minipage}} \end{center} \vskip 1em 
 
Let's pause to remember where we are in our game plan.  First we computed the entropy of a classical
harmonic oscillator.  Now we've computed the entropy of a single classical particle in a
1-dimensional box.   The answer looks a bit like the entropy of an ideal gas!  That's no
coincidence---we're almost there now. 

In case you wanted to know the entropy of a particle in a \emph{3-dimensional} box, don't 
worry.  It's the same as the entropy of three particles of the same mass in three 1-dimensional
boxes of appropriate lengths: the length $L$, width $W$ and height $H$ of our 3d box.   So we can just
sum those 3 entropies and get our answer.  Since $\ln L + \ln W + \ln H = \ln V$ where $V$ is
the volume of our 3d box, we get
\[
 S = k \left(\ln V + \frac{3}{2} \ln kT  + \frac{3}{2} \ln \frac{2 \pi m}{h^2} + \frac{3}{2} \right).  \]
 Later we'll do this calculation more rigorously and more generally for a box of any shape.

But you may have another question: what's the \emph{meaning} of our formula for the 
entropy of a classical particle in a 1-dimensional box?  It's pretty complicated, after all, and we'll need
to understand it to have any chance of understanding the mysterious constant $\gamma$ in the formula for a classical ideal monatomic gas.    

We can understand our formula better if we delve
into a tiny bit of quantum mechanics, and the concept of `thermal wavelength'.  So let's do that.
 
 \vfill \eject \begin{center}\noindent\fbox{\begin{minipage}{30em}

\begin{center}
\entry{THE WAVELENGTH OF A PARTICLE}

\vskip 2em
\noindent
\textbf{\boldmath
In quantum mechanics particles are waves!   \\ A particle with momentum $p$
has wavelength 
\[     \color{darkred}        \lambda = \frac{h}{p} \]
where $h$ is the unreduced Planck's constant, exactly
\[    6.62607015 \cdot 10^{-34}\;\; \textrm{joule-seconds} \]
\vskip 1em
For example,  the wavelength of an electron moving at \\ 1 meter/second is
about 0.7 millimeters.}
\end{center} \vskip 1em

 \end{minipage}} \end{center} \vskip 1em

One of the most amazing discoveries of 20th-century physics: particles
are waves.  The wavelength of a particle is Planck's constant divided
by its momentum!  This was first realized by Louis de Broglie  in his 1924 Ph.D.\ thesis, 
so it's called the `\href{https://en.wikipedia.org/wiki/Matter_wave}{de Broglie wavelength}'.

Why am I telling you this?  Because I want to explain and simplify the
formula for the entropy of a particle in a box.  Even though I derived
it classically, it contains Planck's constant!  So, it will become
more intuitive if we think a tiny bit about quantum mechanics.

A good explanation of quantum mechanics would require
a whole other course.   But it's good to know that in quantum mechanics, a
particle with a given momentum has
a \emph{wavelength} associated to it: we shouldn't imagine it
as having a definite location; it's a bit `blurry'.

This will give a more  intuitive explanation for our complicated formula of the
entropy of a particle in a 1d box.  We'll use this intuition to
simplify our formula.   That will make it easier to generalize to $N$
particles in a 3d box---that is, a classical monatomic ideal gas!

 \vfill \eject \begin{center}\noindent\fbox{\begin{minipage}{30em}

\begin{center}
\entry{THE WAVELENGTH OF A WARM PARTICLE}

\vskip 2em
\noindent
\textbf{\boldmath
In thermal equilibrium, the average energy of \\
a classical free particle in 3d space is
\[   \langle E \rangle = \frac{3}{2} kT \]
where $T$ is the temperature 
and $k$ is Boltzmann's constant.   
\vskip 1em
 If the particle has mass $m$, 
\[    E = \frac{1}{2} mv^2 , \; p = mv \; \implies \; p = \sqrt{2mE} = \sqrt{3 m k T}  \]
\vskip 0.5em
In quantum mechanics, a particle of momentum $p$ has \\ wavelength 
$\lambda = h /p$ where $h$ is the unreduced Planck's constant.
\vskip 0.5em
\define{So, at temperature $T$, the typical wavelength of 
\\ a free particle of mass $m$ is roughly
\[      \lambda = \frac{h}{\sqrt{3 m k T}}   \]
}
}
\end{center} 

 \end{minipage}} \end{center} \vskip 1em
 
Particles are waves! Their wavelength is shorter when their momentum is bigger. And the warmer they are, the bigger their momentum tends to be. So there should be a formula for the typical wavelength of a warm particle. And here it is!   It helps us visualize the world: particles are a bit blurry,  with a characteristic wavelength that depends on temperature.
 
We get this formula from a blend of ideas.  Classical
mechanics says kinetic energy is $E = p^2/2m$.  Classical statistical mechanics
says $\langle E\rangle = \frac{3}{2} kT$.  Quantum mechanics says $\lambda = h/p$.
It's pretty optimistic to put these formulas together and see what we
get.   But the result is approximately correct, though subject to limitations.

We derived $\langle E\rangle = \frac{3}{2} kT$ using classical statistical
mechanics.   But it's close to
correct for a single quantum particle in a big enough box at high enough temperatures.  
Otherwise quantum
effects kick in.

Another problem is that $\langle E \rangle = \frac{3}{2} kT$ and $E = p^2/2m$
do not imply $\langle p \rangle = \sqrt{3mkT}$, even if $p$ here means the
magnitude of the momentum vector.  The arithmetic mean of a square is
not the square of the arithmetic mean!   Really the `\href{https://en.wikipedia.org/wiki/Root_mean_square}{root mean square}' of $p$ is
$\sqrt{3mkT}$.  Similarly, even if the root mean square of $p$ is $\sqrt{3mkT}$
and quantum mechanically $\lambda = h/p$, we \emph{cannot} conclude
that the root mean square of $\lambda$ is $h/\sqrt{3mkT}$.  Again,
you cannot pass a root mean square through a reciprocal!

So, our derivation above is dodgy---but it's okay as an order-of-magnitude
approximation for a warm enough particle in a big enough box.

\vfill \eject \begin{center}\noindent\fbox{\begin{minipage}{30em}
  
\hypertarget{sec:THERMAL_WAVELENGTH}{}

\begin{center}
\entry{THE PARTITION FUNCTION AND THE THERMAL WAVELENGTH} 

\vskip 2em
\noindent
\textbf{\boldmath
The partition function of a classical free particle \\
of mass $m$ in a 1d box of length $L$ is
\[ 
\begin{array}{ccl}
 Z &=& \displaystyle{  \int_0^L \int_{-\infty}^\infty e^{-\beta p^2/2m} \, \frac{dp\, dq}{h}} \\ \\
&=& \displaystyle{   \frac{L}{h} \,\sqrt{\frac{2 \pi m}{\beta}} } \\ \\
&=& \displaystyle{ \frac{L}{\Lambda} }
\end{array}
\]
\vskip 0.5em
where 
{\color{darkred}
\[     \Lambda =  h \,\sqrt{\frac{\beta}{2 \pi m}}   \]
}
is called the \define{`thermal wavelength'}.
}
\end{center} \vskip 1em

 \end{minipage}} \end{center} \vskip 1em
 
 Last time we saw that at temperature $T$, the typical wavelength of a free particle of mass 
 $m$ is roughly 
 \[   \lambda = \frac{h}{\sqrt{3 m k T}}  = h \sqrt{\frac{\beta}{3m}} \,.  \]
 But the partition function of a classical particle of mass $m$ in a box simplifies a lot if we
 introduce a slightly different distance scale, which people call the \define{thermal wavelength}
 \[  \Lambda = \frac{h}{\sqrt{2 \pi m k T}}  = h \,\sqrt{\frac{\beta}{2 \pi m}} \, .   \]
Then the partition function is just the length
of the box divided by $\Lambda$.    The thermal wavelength $\Lambda$ is a bit smaller
than $\lambda$: we have $\Lambda \approx 0.69 \lambda$.   But we probably shouldn't 
worry about this too much, since our calculation of $\lambda$ was so rough.  Of course
all these details are worth thinking about.  But the thermal 
wavelength will turn out to be very useful!

\vfill \eject \begin{center}\noindent\fbox{\begin{minipage}{30em}

\begin{center}
\entry{FREE ENERGY AND THE THERMAL WAVELENGTH} 
 \vskip 1em

\textbf{\boldmath
In thermal equilibrium, a classical free particle \\
of mass $m$  in a 1d box of length $L$ has free energy
\vskip 0.1em
\[ 
F = -\frac{1}{\beta} \ln  \frac{L}{h} \sqrt{\frac{2 \pi m}{\beta}}  
 \]  
 \vskip 0.1em
or 
\vskip 0.05em
\[  \color{darkred}  F = - kT \ln \frac{L}{\Lambda} \]
\vskip 0.1em
where
\vskip 0.1em
\[   \Lambda =  h \sqrt{\frac{\beta}{2 \pi m}} \]
\vskip 0.2em
\noindent is the thermal wavelength. 
}

\end{center}
 \end{minipage}} \end{center} \vskip 1em
 
Since the partition function of the classical free particle in a one-dimensional
box is
\[          Z = \frac{L}{\Lambda}  \]
and free energy is related to the partition function by 
\[         F = - \frac{1}{\beta} \ln Z  ,\]
we have
\[         F = - \frac{1}{\beta} \ln \frac{L}{\Lambda}.  \]
Expressing this in terms of temperature rather than coolness, we have
\[           F = - kT \ln \frac{L}{\Lambda}  . \]

\vfill \eject \begin{center}\noindent\fbox{\begin{minipage}{30em}

\hypertarget{sec:ENTROPY_AND_THERMAL_WAVELENGTH}{}

\begin{center}
\entry{ENTROPY AND THE THERMAL WAVELENGTH} 

\vskip 3em
\noindent
\textbf{\boldmath
In thermal equilibrium, a classical free particle \\
of mass $m$ in a 1d box of length $L$ has 
expected energy
\[  \color{darkred} \langle E \rangle = \frac{1}{2} kT  \]
and free energy
\[ \color{darkred} F = -kT \ln \frac{L}{\Lambda}       \]
where $\Lambda =  h/ \sqrt{2 \pi m k T}$ is the thermal wavelength. 
 \vskip 1 em
But entropy $S$ is $(\langle E \rangle - F)/T$, so
\[
\color{darkred}
S = k\left( \ln \frac{L}{\Lambda} \; + \frac{1}{2} \right)  
\]
}
\end{center} 

\end{minipage}} \end{center} \vskip 1em

Now that we have clean formulas for the expected energy and free energy of the classical free particle in a 1-dimensional box, we can get a nice formula for its entropy.  This is equivalent to the formula we saw 
\hyperlink{sec:PARTICLE_IN_BOX_ENTROPY}{before}, but it's easier to understand.   It's a sum of two terms:
\[  S = k\left( \ln \frac{L}{\Lambda} \; + \frac{1}{2} \right) . \] 

Let's make sure we understand this!    \hyperlink{sec:ENTROPY_COMES_IN_TWO_PARTS}{We've seen that}
for any system in thermal equilibrium, the entropy is the sum of two parts:
\begin{enumerate}
\item
The free energy part.   For the classical particle in a 1-dimensional
box, this is
\[    - \frac{F}{T} = k \ln \frac{L}{\Lambda}.\]
\item
The expected energy part.  For the classical particle in a 1-dimensional
box, this is
\[    \frac{\langle E \rangle}{T} = \frac{1}{2}k.  \]
\end{enumerate}
The free energy part is always $k$ times the logarithm of the \hyperlink{sec:MEANING_PARTITION}{number of accessible states}, and for the particle in a one-dimensional box the number of accessible states is $L/\Lambda$.   The expected energy part is $\frac{1}{2} k$, by the equipartition theorem, because the particle's expected energy depends on one degree of freedom.

Let us think a bit more about why the number of accessible states is $L/\Lambda$.   The most rigorous
approach is simply to compute the number of accessible states---that is, the partition function:
\[ 
\begin{array}{ccl}
 Z 
 &=& \displaystyle{  \int_0^L \int_{-\infty}^\infty e^{-E/kT} \, \frac{dp\, dq}{h}} \\ \\
 &=& \displaystyle{  \int_0^L \int_{-\infty}^\infty e^{-\beta p^2/2m} \, \frac{dp\, dq}{h}} \\ \\
&=& \displaystyle{   \frac{L}{h} \,\sqrt{\frac{2 \pi m}{\beta}} } \\ \\
&=& \displaystyle{ \frac{L}{\Lambda} .}
\end{array}
\]

A more hand-wavy approach is to imagine the space of states of the particle, meaning the space of
momentum-position pairs $(p,q) \in \R \times [0,L]$.  When it comes to counting accessible
states, each region of area $h$ holds one state.   The `accessible' states are those where the energy is not too big compared to $kT$, so the probability density $e^{-E/kT}$ is fairly large.  This is a bit vague, as it must be, because `accessibility' is not really a yes-or-no matter.   But let's just pretend it is, and say a state is accessible
if $E \le kT$.   Then the accessible region of state space is where $p^2/2m \le k T$,
or
\[                |p| \le \sqrt{2m/\beta} .  \]
This region is
\[       \left\{(p,q) \;\big\vert \; -\sqrt{2m/\beta} \le p \le \sqrt{2m/\beta} , \,   0 \le q \le L\right\}  \subseteq [0,L] \times \R \]
It has area $L \times 2 \sqrt{2m/\beta}$, so the number of states it holds is this divided by $h$,
or 
\[                   \frac{2L}{h} \sqrt{\frac{2m}{\beta}}  =  \sqrt{\frac{4}{\pi}} \frac{L}{\Lambda} .\]
This is just 13\% more than the exact value of $Z$.   More importantly, I hope this calculation gives you a mental picture of number of accessible states for the particle in a one-dimensional box.   A mental picture can be helpful even if it's oversimplified.  I like to imagine counting the little rectangles of area $h$ that can fit into the `accessible' region of state space.   

Sometimes these little rectangles are called `phase space cells', since `phase space' is
essentially a synonym for state space.
 
\vfill \eject \begin{center}\noindent\fbox{\begin{minipage}{30em}

\hypertarget{sec:PARTICLE_IN_3D_BOX_PARTITION_FUNCTION}{}

\begin{center}
\entry{PARTICLE IN A 3D BOX: PARTITION FUNCTION} 

\vskip 2em
\noindent
\textbf{\boldmath
The partition function of a classical free particle  of mass $m$ \\
in a 3d box $B$ of volume $V$ is
\[ 
 Z \;\; = \;\; \displaystyle{  \int_B \int_{\R^3} e^{-\beta \vec{p} \cdot \vec{p}/2m} \, 
 \frac{d^3p\, d^3q}{h^3}}
\;\; = \;\; \displaystyle{   \frac{V}{h^3} \,\left(\frac{2 \pi m}{\beta}\right)^{3/2} }
\]
where $\beta = 1/kT$ is the coolness.
\vskip 1em
This result becomes prettier using the thermal wavelength
\[   \Lambda =  h (\beta/2 \pi m)^{1/2}  \]
\vskip 1em
Then we get simply
{\color{darkred}
\[   Z =\frac{V}{\Lambda^{{}^3}}   \]}
}
\end{center} 
\end{minipage}} \end{center} \vskip 1em 

Now that we've worked out the statistical mechanics of a classical particle in a 
one-dimensional box, it's easy to copy everything for a three-dimensional box of
any shape.    We start with the partition function.   The energy of a free particle of mass $m$ is $\vec{p} \cdot \vec{p}/2m$, so the partition function is the integral of $\exp(-\beta\vec{p} \cdot \vec{p}/2m)$ over
all possible positions and momenta.    Integrate over
momentum and you get
\[   
\displaystyle{ \int_{\R^3} e^{-\beta (p_1^2 + p_2^3 + p_3^2)/2m} \,\frac{dp_1 dp_2 dp_3 }{h^3} } 
\; = \; \left(    \int_{-\infty}^\infty e^{-\beta p^2/2m} \, \frac{dp}{h}  \right)^3 \; =\; \left( h \sqrt{\frac{2 \pi m}{\beta}} \right) ^3.
\]
In terms of the thermal wavelength this is just $1/\Lambda^3$.   
Integrate over position and you multiply this by the volume of the box, say $V$.
So we get an incredibly simple final answer:
\[    Z = \frac{V}{\Lambda^{{}^3}} .\]
And this sort of calculation works 
in any dimension: there's nothing special about the number 3 here.

\vfill \eject \begin{center}\noindent\fbox{\begin{minipage}{30em}
  
 \begin{center}
\entry{PARTICLE IN A 3D BOX: ENTROPY } 

\vskip 1em
\noindent \textbf{\boldmath
In thermal equilibrium, a classical free particle of mass $m$ \\
in a 3d box of volume $V$ has
expected energy
\[   \langle E \rangle = \frac{3}{2} kT  \]
and free energy
\[ F = -kT \ln \frac{V}{\Lambda^3}       \]
where $\Lambda =  h/ \sqrt{2 \pi m k T}$ is the thermal wavelength. 
\vskip 1em
But entropy $S$ is $(\langle E \rangle - F)/T$, so
\[
\color{darkred}
S = k\left( \ln \frac{V}{\Lambda^3} \; + \frac{3}{2} \right)  
\]
}
\end{center} \vskip 1em 
\end{minipage}} \end{center} \vskip 1em 

The entropy of a particle in thermal equilibrium in
a three-dimensional box works very much like our earlier
calculation for a one-dimensional box, with a couple of adjustments due to the
dimension.   Since the particle's energy is now a quadratic function of 3
variables, the equipartition theorem now says its expected energy is
\[  \langle E \rangle = \frac{3}{2} kT .\]
We can work out its free energy from its partition function, which we computed in the last tweet:
\[    F = - k T  \ln Z = -k  \ln  \frac{V}{\Lambda^{{}^3}}   .\]
Thus its entropy is
\[   S = \frac{\langle E \rangle - F}{T} =  k\left( \ln \frac{V}{\Lambda^3} \; + \frac{3}{2} \right)  . \]

The meaning of the two terms here is very similar to that for the \hyperlink{sec:ENTROPY_AND_THERMAL_WAVELENGTH}{particle in the
one-dimensional box}.
The first term is $k$ times the logarithm of the \hyperlink{sec:MEANING_PARTITION}{number of accessible states}, as always for the Gibbs entropy of a system in thermal equilibrium.  Here the
number of accessible states is $V/\Lambda^3$.    The second term is $\frac{3}{2} k$ thanks to the equipartition theorem, since the particle's expected energy depends quadratically on 3 degrees of freedom.     When
$V \gg \Lambda^3$ this second term is a small correction to the first.   As this ceases to be true, the
second term becomes more important---and when $\Lambda^3$ is comparable to $V$, quantum corrections
to our calculation also become significant.

\vfill \eject \begin{center}\noindent\fbox{\begin{minipage}{30em}

\begin{center}
\entry{A TALE OF TWO GASES} 

\vskip 2em
\noindent
\textbf{\boldmath
The entropy of an ideal gas of $N$ \textit{distinguishable} \\ classical particles of mass $m$ 
in a box of volume $V$ is
\vskip 0.5em
\[   \color{darkred}  S_d =  kN \left( \ln V + \frac{3}{2} \ln kT +  \frac{3}{2}\ln \frac{2 \pi m}{h^2} + \frac{3}{2} \right)  \]
\vskip 0.5em
while for \textit{indistinguishable} particles it's
\vskip 0.5em
\[   \color{darkred}  S_i \approx kN \left( \ln \frac{V}{N} + \frac{3}{2} \ln kT +  \frac{3}{2}\ln \frac{2 \pi m}{h^2} + \frac{5}{2} \right)   \]
\vskip 0.5em
where the corrections are small compared to $N$ as $N \to \infty$.  
}
\end{center} \vskip 1em

 \end{minipage}} \end{center} \vskip 1em 
 
 Now we are finally ready to tackle the entropy of a gas.   We start with a `monatomic ideal gas',
 which means $N$ free point particles bouncing around in
 a box.  But there's a subtlety!   We'll get different answers depending on whether we think
 of these particles as \emph{distinguishable} or \emph{indistinguishable}.   That is: do we count the
 state of the gas as different if we switch two particles, or not?  
 
The formulas look very similar.   There are three differences:
\begin{itemize}
\item
 For distinguishable particles we'll get an exact formula, while for
indistinguishable particles we'll get an approximate one, where the corrections are small compared to 
$N$ when $N$ become large.
\item 
 The entropy for distinguishable particles has a term equal to $\frac{3}{2}kN$, while for
 indistinguishable particles it has a term equal to $\frac{5}{2} k N$.
 \item
 Most importantly, there's a huge difference in the volume dependence!  Where the distinguishable particles have a term in the entropy equal to $kN \ln V$, the indistinguishable ones have a term equal to $k N \ln \frac{V}{N}$, so their entropy is considerably smaller for large 
volumes. 
 \end{itemize}
 
 The last difference makes the entropy behave strangely for distinguishable particles, so in practice the 
 physically important case is the gas of \emph{indistinguishable} particles.    But we'll do the calculations in both cases, because the distinguishable case is easier, and interesting.
 
\vfill \eject \begin{center}\noindent\fbox{\begin{minipage}{32em}

\begin{center}
\entry{GAS OF DISTINGUISHABLE PARTICLES: PARTITION FUNCTION} 

\vskip 2em
\noindent
\textbf{\boldmath
The partition function of an ideal gas of $N$ distinguishable \\
classical particles of mass $m$ in a 3d box $B$ of volume $V$ is
\[ 
\begin{array}{ccl}
 Z_d &=& \displaystyle{ \scalebox{1.5}{$\displaystyle{\bigint_{B^N} \bigint_{\R^{3N}}}$ }
 e^{ - \beta \displaystyle{\sum_{i = 1}^N}  \frac{\vec{p}_i \cdot \vec{p}_i}{ 2m }} \;\;
 {\frac{d^3p_1 \cdots d^3 p_N\, d^3q_1 \cdots d^3q_N}{h^{3N}}  }}   \\ \\ 
 &=& \displaystyle{   \frac{V^N}{h^{3N}} \,\left(\frac{2 \pi m}{\beta}\right)^{3N/2}  }
\end{array}
\]
\vskip 1em
Thus
\[  \color{darkred} Z_d  \;\; = \;\; \frac{V^N}{ \Lambda^{3N}}   \]
where $\Lambda =  h (\beta/2 \pi m)^{1/2}$ is the thermal wavelength. 
}
\end{center} \vskip 1em

 \end{minipage}} \end{center} \vskip 1em 
 
 Suppose we have a system of $N$ distinguishable classical free particles in a three-dimensional box $B$
 of volume $V$.  The state of this system is described by $N$ positions $\vec{q}_1, \dots, \vec{q}_N \in B$ and $N$ momenta  $\vec{p}_1, \dots, \vec{p}_N \in \R^3$.    If each particle has mass $m$, the energy of the $i$th particle is equal to
 \[      E_i =   \frac{\vec{p}_i \cdot \vec{p}_i}{2 m}  \]
and the energy of the system is
 \[     E = \sum_{i = 1}^N    E_i  .\]
Let's call the partition function of this system $Z_d$.  To compute this we integrate $\exp(-\beta E)$ over the space of states, obtaining
 \[   Z_d = \int_{B^N} \int_{\R^{3N}} \;
 \exp(- \beta E)  \; \frac{d^3p_1 \cdots d^3 p_N\, d^3q_1 \cdots d^3q_N}{h^{3N}} . \]
 
 Above, I proceeded to compute $Z_d$ directly by doing the Gaussian integral over momenta and
 integrating each position over the box.    Here's a slightly different way.  Because 
 \[     \exp(-\beta E) =  \exp(-\beta E_1) \cdots \exp(-\beta E_N), \]
 the partition function $Z_d$ is a product of integrals which are all equal:
 \[     Z_d = \left(   \int_B \int_{\R^3} e^{-\beta \vec{p} \cdot \vec{p}/2m} \, 
 \frac{d^3p\, d^3q}{h^3}  \right)^N   .\] 
 The integral in the parentheses is the partition function of an \emph{single} particle
 in a box.  We \hyperlink{sec:PARTICLE_IN_3D_BOX_PARTITION_FUNCTION}{have already seen} 
 that this equals
 \[  \int_B \int_{\R^3} e^{-\beta \vec{p} \cdot \vec{p}/2m} \, 
 \frac{d^3p\, d^3q}{h^3} =   \frac{V}{\Lambda^3}   \]
 where $\Lambda$ is the thermal wavelength.   Thus we have
 \[        Z_d = \left( \frac{V}{\Lambda^3}\right)^N .\]

We can also do this calculation with a lot less work using Puzzle \ref{puzzle:product_of_energetic_sets}.
This implies that when we build a new system from $N$ identical noninteracting copies of some old
system, the partition function of the new system is the $N$th power of the partition function of the old system.  What I just did is show this in a special case.
 
 \vfill \eject \begin{center}\noindent\fbox{\begin{minipage}{30em}
  
\begin{center}
\entry{GAS OF DISTINGUISHABLE PARTICLES: ENTROPY } 

\vskip 1em
\noindent \textbf{\boldmath
In thermal equilibrium, an ideal gas of $N$ distinguishable 
classical particles of mass $m$ in 
a 3-dimensional box of volume $V$ has expected energy
\[   \langle E_d \rangle = \frac{3}{2} kNT  \]
and free energy
\[ F_d = -kT \ln \frac{V^N}{\Lambda^{3N}}       \]
where $\Lambda =  h/ \sqrt{2 \pi m k T}$ is the thermal wavelength.  
\vskip 1em
Its entropy $S_d$ is $(\langle E_d \rangle - F_d)/T$, so 
\vskip 1em
\[
\color{darkred}
S_d = kN\left( \ln \frac{V}{\Lambda^3} \; + \frac{3}{2} \right)  
\]
}
\end{center} \vskip 1em 
\end{minipage}} \end{center} \vskip 1em 

We use the subscript $d$ for a gas of $N$ distinguishable
particles.     Since the energy is a quadratic function of $3N$
variables, the equipartition theorem says the expected energy is
\[   \langle E_d \rangle = \frac{3}{2} kNT. \]
The free energy $F$ is minus Boltzmann's constant times the
 logarithm of the partition function, which we just computed:
\[    F_d = - k  \ln Z_d = -k  \ln  \frac{V^N}{\Lambda^{3N}}   .\]
Thus the entropy of the gas is
\[   S_d = \frac{\langle E_d \rangle - F_d}{T} =  kN \left( \ln \frac{V}{\Lambda^3} \; + \frac{3}{2} \right)  . \]
If we expand this out using 
\[   \Lambda =  \frac{h}{\sqrt{2 \pi m k T}} \]
 we get the 
formula I promised earlier:
\[    S_d =  kN \left( \ln V + \frac{3}{2} \ln kT +  \frac{3}{2}\ln \frac{2 \pi m}{h^2} + \frac{3}{2} \right).  \]
The only advantage of this messier formula is that it separates out the temperature dependence and
the volume dependence.

\vfill \eject \begin{center}\noindent\fbox{\begin{minipage}{30em}
\begin{center}
\entry{THE GIBBS ``PARADOX''} 

\vskip 2em
\noindent
\textbf{\boldmath
For the ideal gas of $N$ \textit{distinguishable} classical particles in a box 
of volume $V$, the entropy 
\[    S_d =  kN \left( \ln V + \frac{3}{2} \ln kT +  \frac{3}{2}\ln \frac{2 \pi m}{h^2} + \frac{3}{2} \right)  \]
more than doubles if we double both $N$ and $V$ \\
while keeping everything else the same. 
\\ This confused people for a while,  \\ so it's called the \define{`Gibbs
paradox'}.
}
\end{center} \vskip 1em

 \end{minipage}} \end{center} \vskip 1em 
 
 Start with a box $B$ containing an ideal gas of distinguishable classical particles.
 Then double the volume of the box  to get a new box $B'$,
 and double the number of particles in the box too, while keeping
 the temperature and everything else the same.  
 
 We might expect the entropy to double.  After all, we could take the doubled box and slip a thin wall down the middle to get two identical copies of the original box.  So the entropy should be twice as big now.   Right?
 
Apparently not!   Instead of just doubling the $k N \ln V$ term in the original entropy, we are replacing it with $2kN \ln (2 V)$, which is more than twice as big.   The reason is that in the doubled box $B'$ each individual particle has twice as much room to roam than if you put a wall down the middle.   Thus, it takes more information to say where all the particles are.

While there's no real paradox here, people found this result deeply counterintuitive, so they called it
the `\href{https://en.wikipedia.org/wiki/Gibbs_paradox}{Gibbs paradox}'.   And in fact they had a good reason for being suspicious of this result.
It would be correct if gas molecules were distinguishable.  But in fact molecules of the same
kind are not distinguishable---they don't have little labels on them that let you recognize which is
which.   And if we take this fact into account, our formula for the entropy changes.   Let's see how!

 \vfill \eject \begin{center}\noindent\fbox{\begin{minipage}{30em}

\begin{center}
\entry{GAS OF INDISTINGUISHABLE PARTICLES: PARTITION FUNCTION} 

\vskip 2em
\noindent
\textbf{\boldmath
The partition function of an ideal gas of $N$ indistinguishable
classical particles of mass $m$ in a 3d box $B$ of volume $V$ is
\[ 
\begin{array}{ccl}
 Z_i(\beta) &=& \displaystyle{ \frac{Z_d(\beta)}{N!} }  \\ \\ 
 &=& \displaystyle{ \frac{1}{N!} \, \frac{V^N}{h^{3N}} \,\left(\frac{2 \pi m}{\beta}\right)^{3N/2}  }
\end{array}
\]
\vskip 1em
Thus
\[  \color{darkred} Z_i(\beta) \;\; = \;\; \color{darkred} \frac{1}{N!} \, \frac{V^N}{ \Lambda^{3N}}   \]
where $\Lambda =  h (\beta/2 \pi m)^{1/2}$ is the thermal wavelength.   }
\end{center} \vskip 1em

 \end{minipage}} \end{center} \vskip 1em 
 
 The partition function $Z_i$ for a gas of $N$ indistinguishable particles is $1/N!$ times
 that for a gas of distinguishable particles.  Why?  We got $Z_d$ by integrating $\exp(-\beta E)$
 over the space of ordered $N$-tuples of position-momentum pairs.   The energy $E$ here does not
 change if we permute our $N$-tuple, so we can also think of it as a function of \emph{unordered}
 $N$-tuples.   Then we get $Z_i$ by
 integrating $\exp(-\beta E)$ over the space of such unordered tuples.  
 Notice that there are $N!$ ordered $N$-tuples for each unordered 
 $N$-tuple, except for $N$-tuple with repeated entries, which form a set of measure zero and thus contribute nothing to the integral.  Thus, we should not be surprised that 
 \[    Z_i(\beta) = \frac{Z_d(\beta)}{N!}  .\]
 But we've seen
 \[   Z_d(\beta) = \frac{V^N}{\Lambda^{3N} } \]
 where $\Lambda$ is the thermal wavelength, so
 \[    Z_i(\beta) = \frac{1}{N!} \,  \frac{V^N}{\Lambda^{3N} } .\]
Making this sketchy argument precise requires more notation.   I think carefully doing the case $N = 2$
 is the best way for you to see what's going on.
  
 \vfill \eject \begin{center}\noindent\fbox{\begin{minipage}{30em}
  
 \begin{center}
\entry{GAS OF INDISTINGUISHABLE PARTICLES: ENTROPY } 

\vskip 1em
\noindent \textbf{\boldmath
In thermal equilibrium, an ideal gas of $N$ indistinguishable 
classical particles of mass $m$ in 
a 3-dimensional box of volume $V$ has expected energy
\[   \langle E_i \rangle = \frac{3}{2} kNT  \]
and free energy
\[ F_i = -kT \ln \left( \frac{1}{N!} \frac{V^N}{\Lambda^{3N}} \right)      \]
where $\Lambda =  h/ \sqrt{2 \pi m k T}$ is the thermal wavelength.  
\vskip 1em
Its entropy $S_i$ is $(\langle E_i \rangle - F_i)/T$, so 
\[
\color{darkred}
S_i = kN\left( \ln \frac{V}{\Lambda^3} \; + \; \frac{3}{2} \right) - k \ln N!
\]
}
\end{center} \vskip 1em 
\end{minipage}} \end{center} \vskip 1em 

Suppose we have a gas of $N$ indistinguishable classical free
particles.     Since the energy is a quadratic function of the $3N$ momentum
variables, the equipartition theorem says the expected energy of this gas is
\[   \langle E_i \rangle = \frac{3}{2} kNT. \]
The free energy $F$ is minus Boltzmann's constant times the
 logarithm of the partition function, which we just computed:
\[    F_i = - k  \ln Z_i = -k  \ln  \left( \frac{1}{N!} \frac{V^N}{\Lambda^{3N}} \right)   .\]
Thus the entropy of the gas is
\[   S_i = \frac{\langle E_i \rangle - F_i}{T} =  kN \left( \ln \frac{V}{\Lambda^3} \; + \frac{3}{2}\right) - k \ln N!   \]
In short, it is $k \ln N!$ less than for the gas of distinguishable particles.   This makes beautiful intuitive sense: there are $N!$ permutations of the particles that we no longer care about in the 
indistinguishable case, so we learn $\ln N!$ less information when we learn everything we can about this
gas when our initial assumption was that it's in thermal equilibrium.

\vfill \eject \begin{center}\noindent\fbox{\begin{minipage}{30em}

\hypertarget{sec:STIRLING'S_FORMULA}{}

\begin{center}
\entry{STIRLING'S FORMULA}
\vskip 2em
\noindent
\textbf{\boldmath
Stirling's formula says
\[    N! \; \sim \; \sqrt{2\pi N} \left( \frac{N}{e} \right)^N \]
and it gives
\[   \ln N! \; \approx\; (\ln N - 1) N \, + \, \textstyle{\frac{1}{2}} \ln 2 \pi N \]
where the error goes to zero as $N \to +\infty$.
}
\end{center} \vskip 1em

 \end{minipage}} \end{center} \vskip 1em 
 
 Now we need a bit of math: Stirling's formula for the factorial
 function.   In one form this says
\[   \lim_{N \to +\infty} \frac{\sqrt{2\pi N} \left( \frac{N}{e} \right)^N}{N!} = 1 . \]
We abbreviate this fact, that the ratio of two quantities approaches $1$ as $N \to +\infty$,
by saying $N!$ is \define{asymptotic to} $\sqrt{2\pi N} \left( \frac{N}{e} \right)^N$.   We also write
\[    N! \; \sim \; \sqrt{2\pi N} \left( \frac{N}{e} \right)^N .\]
where the symbol $\sim$ means `asymptotic to'.   

If we take the logarithm of both sides we get
\[   \ln N! \; \approx\; (\ln N - 1) N \, + \, \textstyle{\frac{1}{2}} \ln 2 \pi N .\]
The symbol $\approx$ has a vaguer meaning: `approximately equal to'.    But it turns out that
in this instance the approximation is extremely good: the difference between the left and right sides
goes to zero as $N \to +\infty$.  In fact we will content ourselves with a cruder approximation:
\[   \ln N! \; \approx\; (\ln N - 1) N  \]
because in the entropy of an ideal gas $N$ is typically huge, so the term we have discarded
here is dwarfed by the others.

\begin{puzzle}
\label{puzzle:Stirling_correction}
Suppose $N$ is Avogadro's number, close to the number of atoms in 4 grams of helium:
\[    N \approx 6 \cdot 10^{23}. \]
What is the ratio of $\frac{1}{2} \ln 2 \pi N$ to $N$?
\end{puzzle}

While deriving Stirling's formula is fascinating and not at all trivial, doing so would take us rather far
afield.   So I will resist, and refer you instead to this:

\begin{itemize}
\item John C.\ Baez, \href{https://golem.ph.utexas.edu/category/2021/10/stirlings_formula.html}{Stirling's formula}, \textsl{The $n$-Category Caf\'e}, October 24, 2021.
\end{itemize}

 \vfill \eject \begin{center}\noindent\fbox{\begin{minipage}{30em}
 
 \hypertarget{SACKUR--TETRODE}{}
 
 \begin{center}
\entry{THE SACKUR--TETRODE EQUATION}
\vskip 2em
\noindent
\textbf{\boldmath
In thermal equilibrium, an ideal gas of $N$ indistinguishable classical particles 
in a 3-dimensional box of volume $V$ has entropy
\[   S_i = kN\left( \ln \frac{V}{\Lambda^3} \; + \; \frac{3}{2} \right) - k \ln N! \]
where $\Lambda =  h/ \sqrt{2 \pi m k T}$ is the thermal wavelength.   
\vskip 1em
Using Stirling's formula
\[   \ln N! \; \approx\; (\ln N - 1) N  \]
we get the \define{Sackur--Tetrode equation}:
\[   \color{darkred}  S_i \; \approx \; kN \left( \ln \frac{V}{N \Lambda^3} + \frac{5}{2} \right)    \]
}
\end{center} 

 \end{minipage}} \end{center} \vskip 1em 
 
 Taking our formula 
 \[   S_i = kN\left( \ln \frac{V}{\Lambda^3} \; + \; \frac{3}{2} \right) - k \ln N! \]
 and using a simple version of Stirling's formula, $\ln N! \sim (\ln N - 1) N$, we get the
 famous \define{Sackur--Tetrode equation}:
 \[   \begin{array}{ccl} 
 S_i 
 &\sim & \displaystyle{  kN\left( \ln \frac{V}{\Lambda^3} \; + \; \frac{3}{2} \right) - k (\ln N - 1) N } \\ \\
 &\sim&  \displaystyle{ kN \left( \ln \frac{V}{N \Lambda^3} + \frac{5}{2} \right) .}
 \end{array}
 \]
 Note that with this formula, if we multiply both $V$ and $N$ by the same constant, the
 entropy also gets multiplied by that constant.   In this situation we say the entropy is `extensive'.
 
 For a better approximation, we can use
 \[   \ln N! \; \approx\; (\ln N - 1) N  +\textstyle{\frac{1}{2}} \ln 2 \pi N  \]
 where the error goes to zero as $N \to \infty$.  This gives a correction to the Sackur--Tetrode
 equation:
 \[    S_i  \approx  kN \left( \ln \frac{V}{N \Lambda^3} + \frac{5}{2} \right)  - \frac{1}{2} \ln 2\pi N. \]
 Here if we multiply both $V$ and $N$ by a constant $c$, we don't just multiply the entropy by $c$: we also have to subtract $\frac{1}{2} \ln 2\pi c$.   So the entropy is not quite extensive---but this effect
 is tiny when you've got a mole of gas.
 
\vfill \eject \begin{center}\noindent\fbox{\begin{minipage}{30em}
 
\hypertarget{sec:ENTROPY_OF_IDEAL_MONATOMIC_GAS}{}

 \begin{center}
\entry{THE ENTROPY OF AN IDEAL MONATOMIC GAS}
\vskip 2em
\noindent
\textbf{\boldmath
In thermal equilibrium, an ideal gas of $N$ indistinguishable classical particles 
in a 3-dimensional box of volume $V$ has entropy given approximately by
the Sackur--Tetrode equation:
\[   \color{darkred}  S \; \approx \; kN \left( \ln \frac{V}{N \Lambda^3} + \frac{5}{2} \right)    \]
But the thermal wavelength $\Lambda$ is
\[    \Lambda = \frac{h}{\sqrt{2 \pi m k T}} \]
so we can rewrite this as
\vskip 0.1 em
\[   \color{darkred}  S \approx kN \left( \ln \frac{V}{N} + \frac{3}{2} \ln kT +  \frac{3}{2}\ln \frac{2 \pi m}{h^2} + \frac{5}{2} \right)    \]
}
\end{center} 

\end{minipage}} \end{center} \vskip 1em 

We've done it: we've figured out the entropy of a gas of $N$ indistinguishable
classical free particles in a 3-dimensional box of volume $V$.   Above I've written it in two different ways.   Let's mull over the meaning of each term in each formula.   

The first formula says
\[    S \approx  kN  \left( \ln \frac{V}{N \Lambda^3} + \frac{5}{2} \right). \]
Like the entropy of the classical harmonic oscillator and the classical free particle in a box, this breaks up into 
\hyperlink{sec:ENTROPY_COMES_IN_TWO_PARTS}{two parts}, thanks to the formula
\[                 S = \frac{\langle E \rangle - F}{T}  .\]
But it does so a bit subtly.  The two parts are not what you might naively think!  They are:
\begin{enumerate}
\item The free energy part:
\[  - \frac{F}{T} \approx kN \left( \ln \frac{V}{N \Lambda^3} +  1 \right) .\]
\item
The expected energy part:  
\[    \frac{\langle E \rangle}{T} = \frac{3}{2} kN.  \]
\end{enumerate}
As usual, the free energy part of the entropy is $k$ times the logarithm of the \hyperlink{sec:MEANING_PARTITION}{number of accessible states}.    The expected energy part of the 
entropy is $\frac{3}{2}N$ times $k$ by the  \hyperlink{sec:PROOF_OF_EQUIPARTITION_THEOREM_3}{equipartition theorem}, since there are $N$ particles each of whose energy depends on 3 momentum
degrees of freedom. 

The expected energy part of the entropy is small compared to the free energy part when $V/N \gg \Lambda^3$: that is, when the volume available per particle greatly exceeds the cube of its thermal wavelength.   This happens for a gas that is sufficiently warm and dilute, made of sufficiently massive particles.    We will see that this is true for helium at standard temperature and pressure.    It's even more true for the heavier monatomic gases: the noble gases like neon, argon, and krypton.

The surprise is the extra ``$+1$'' in the first part of the entropy---the free energy part.  It's telling us that the logarithm of the number of accessible states, divided by the number of particles, is 
\[   \ln \frac{V}{N \Lambda^3} +  1.\]
What's the physical origin of this mysterious extra nat?    

Mathematically it comes from \hyperlink{sec:STIRLING'S_FORMULA}{Stirling's formula}, which showed up when we switched from a gas of distinguishable particles to a gas of indistinguishable particles.  It may seem odd that indistinguishability would \emph{increase} the entropy by $1$ nat per particle, but don't be confused: as we've seen, it greatly \emph{reduces} it.   For a gas of distinguishable particles the log of the number of accessible states, divided by the number of particles, is $\ln(V/\Lambda^3)$.    When we switch to indistinguishable particles this drops to $\ln(V/N\Lambda^3) + 1$.

Here is a rough heuristic explanation of what's going on.  For a single particle in a box of volume $V$, the number of accessible states is $V/\Lambda^3$.  In a gas of distinguishable free particles, each roams independently around the whole volume $V$.    Thus, the log of the number of accessible states is $\ln(V/\Lambda^3)$ per particle.

For a gas of indistinguishable particles, the story changes.   For starters, we can crudely pretend each particle is trapped in its own tiny box of volume $V/N$.  After all, if it leaves this tiny box by trading places with another particle in another tiny box, nothing really changes.    In this approximation, the log of the number of accessible states is $\ln(V/N \Lambda^3)$ per particle.

But it's \emph{not really true} that each particle can only leave its tiny box by trading places with another.   We can have more than one particle in the same tiny box---or none.    That is, our gas can have density fluctuations.  An exact treatment of the problem gives, not $\ln(V/N\Lambda^3)$ nats per particle, but 
\[       \ln (V/\Lambda^3) - \ln N!  \]
Stirling's formula says this is approximately 
\[   \ln(V/\Lambda^3) - (\ln N - 1) \; \; = \; \; \ln (V/N \Lambda^3) + 1  .\]
This explains the mysterious extra nat.  The extra nat of entropy per particle is due to density fluctuations!

As we've seen, even this is an oversimplification.  A still better approximation, again coming
from Stirling's formula, says
\[      \ln (V/\Lambda^3) - \ln N! \; \; \approx \; \; \ln (V/N \Lambda^3) + 1 - \textstyle{\frac{1}{2}} \ln(2 \pi N)/N .\]
But as we saw in Puzzle \ref{puzzle:Stirling_correction}, this further correction is negligible for a mole of gas.   It only becomes interesting for microscopic systems.

Now let's look at our second formula for the entropy of a gas of $N$ indistinguishable classical free particles:
\[   S \approx kN \left( \ln \frac{V}{N} + \frac{3}{2} \ln kT +  \frac{3}{2}\ln \frac{2 \pi m}{h^2} + \frac{5}{2} \right)   . \]
Not only is this harder to remember, it's generally less friendly to physical intuition.   First of all, three of the terms involve the logarithm of dimensionful quantities.  Thus, when we change units they change, not by rescaling in the usual way, but by addition or subtraction.   Secondly, the important role of the thermal wavelength is concealed in this formula.   

The main advantage of this formula is that it separates out three contributors to the entropy per particle:
\begin{itemize}
\item The volume available per particle, $V/N$.  The bigger this is, the more entropy the gas
has per particle.
\item  The temperature, $T$.  The bigger this is, the more entropy per particle.
\item  The particle mass, $m$.  The bigger this is, the more entropy per particle.
\end{itemize}
The first two should be rather intuitive.   But what about the third?   We need to combine $V/N$ and $T$ with the particle  mass $m$ and some constants of nature to get a dimensionless quantity, which we can then take the logarithm of.  This leads us straight to the thermal wavelength:
\[    \begin{array}{ccl}
 \displaystyle{ \ln \frac{V}{N} + \frac{3}{2} \ln kT +  \frac{3}{2}\ln \frac{2 \pi m}{h^2}} 
 &=&  
  \displaystyle{  \ln   \frac{V (2 \pi m kT)^{3/2}}{N h^3} }  
  \\ \\
  &=&
   \displaystyle{ \ln \frac{V}{N \Lambda^3}}.
  \end{array}
\]
Thus, my best explanation of why a gas of heavier particles has more entropy per particle is that they have a shorter thermal wavelength, so we can specify their position more accurately, and it takes more information to do so.

\vfill \eject \begin{center}\noindent\fbox{\begin{minipage}{30em}

\noindent
\begin{center}
\entry{WHERE ARE WE NOW?}
\end{center}
\textbf{\boldmath
{\color{red} The mystery: why does each molecule of hydrogen have $\sim\!23$ bits of entropy at
standard temperature and pressure?}
\vskip 1em \noindent
{\color{brickred}
The goal: derive and understand the formula
for the entropy of a classical ideal monatomic gas:
\[   S = kN\left( \ln\frac{V}{N} + \frac{3}{2} \ln k T+ \gamma
\right) \]
including the mysterious constant $\gamma$:
\[  \qquad \qquad \gamma =  \frac{3}{2}\ln \frac{2 \pi m}{h^2} + \frac{5}{2}  \qquad \qquad  \color{darkgreen} \scalebox{1.5}{\checkmark}   \]
}
\hskip -0.5 em{\color{darkred}The subgoal: compute the entropy of a single classical \break particle in a 1-dimensional box:
\[
\qquad \qquad \quad S = k \left(\ln L + \frac{1}{2} \ln kT  + \frac{1}{2} \ln \frac{2 \pi m}{h^2} + \frac{1}{2} \right) \qquad {\color{darkgreen} \scalebox{1.5}{\checkmark}}  \]}
\hskip -0.5 em The sub-subgoal:  explain entropy from the ground up, and
compute the entropy of a classical harmonic oscillator: \break
\[ \qquad \qquad S =  k \left( \ln \frac{kT}{\hbar \omega} + 1 \right)  \qquad \qquad  \color{darkgreen} \scalebox{1.5}{\checkmark} \]  
}

\end{minipage}} \end{center} \vskip 1em 

Okay, now we know the entropy of a classical ideal monatomic gas!  We even know what it means.  Unfortunately we're trying
to figure out the entropy of hydrogen, which is diatomic.  But we can do helium, which is
monatomic... and then we'll do hydrogen.

\vfill \eject \begin{center}\noindent\fbox{\begin{minipage}{30em}
 
 \begin{center}
\entry{ENTROPY PER MOLE VERSUS BITS PER MOLECULE}
\vskip 1em
\noindent
\textbf{\boldmath
A nat of unknown information is $1.380649 \cdot 10^{-23}$ 
joules/kelvin of entropy: this is \define{Boltzmann's constant}.
\vskip 1em
There are $6.02214076 \cdot 10^{23}$ molecules per mole: \\ this
is \define{Avogadro's number}.
\vskip 1em
Thus, one nat of unknown information per molecule corresponds to
\[    1.380649 \cdot 10^{-23} \times 6.02214076 \cdot 10^{23} \approx 8.314463 \]
joule/kelvin of entropy per mole.
\vskip 1em
A bit is $\ln 2 \approx 0.69315$ nats, so one bit of unknown information per molecule corresponds to
about
 \[   0.69315 \times 8.314463 \approx {\color{darkred} 5.763146 }\]
 joule/kelvin of entropy per mole.
}
\end{center} \vskip 1em

\end{minipage}} \end{center} \vskip 1em 

Here is a little fact we need now: one bit of Shannon entropy per molecule equals about
$5.76$ joules/kelvin of Gibbs entropy per mole.   I apologize for the oppressively large number of decimal places above, but I want to compare our theoretical predictions of the entropy of
helium and hydrogen to experimental results, and it's not clear yet how 
closely our answers will match experiment, so it's good to be prepared.   

By the way, the values of Boltzmann's constant and Avogadro's number here are \emph{exact}, fixed by the definition of \href{https://en.wikipedia.org/wiki/International_System_of_Units#SI_base_units}{SI units}.   So there is no experimental uncertainty in any of the numbers on this page.

\vfill \eject \begin{center}\noindent\fbox{\begin{minipage}{30em}
 
 \begin{center}
\entry{THE ENTROPY OF HELIUM: THEORY}
\vskip 2em
\noindent
\textbf{\boldmath
The Sackur--Tetrode equation says that assuming helium is a classical ideal monatomic
gas, its entropy is
\[     S_i \; \approx \; kN \left( \ln \frac{V}{N \Lambda^3} + \frac{5}{2} \right)    \]
which corresponds to
\[        \ln \frac{V}{N \Lambda^3} + \frac{5}{2}   \]
 nats of unknown information per atom.  At standard temperature and pressure,
 this gives about 15.041 nats or
\[    \frac{15.041}{\ln 2} \color{darkred} \approx 21.700 \]
bits of unknown information per atom.
}
\end{center} \vskip 1em

\end{minipage}} \end{center} \vskip 1em 

Now let's calculate the entropy of helium in its gaseous state.  NIST has tabulated its entropy at standard temperature and pressure, specifically temperature $T = 298.15$ K and pressure $P = 1$ bar, so that's what we'll try to calculate.    An atom of helium has a mass of $m = 6.646477 \cdot 10^{-27}$ kg, so at standard temperature its thermal wavelength is 
\[   \begin{array}{ccl} \Lambda &=& \displaystyle{ \frac{h}{\sqrt{2 \pi m k T}}}  \\
 &\approx& \displaystyle{ \frac{6.62607 \cdot 10^{-34}\, \mathrm{J\, s}}{\sqrt{2 \pi \times 6.646477 \cdot 10^{-27} \, \mathrm{kg} \times  1.380649 \cdot 10^{-23} \,\mathrm{J/K} \times 298.15 \, \mathrm{K}}} } \\ \\
&\approx&  5.053721 \cdot 10^{-11} \, \mathrm{m}  . 
\end{array}
\]
For a mole of an ideal gas we have $N = 6.02214076 \cdot 10^{23}$
(this is Avogadro's number), and at standard temperature and pressure a mole of ideal gas has $V \approx 0.0247896 \, \mathrm{m}^3$: this is called its `molar volume'.   The molar volume of helium is actually slightly different from this, because helium is not an ideal gas: the atoms interact.  But since we're doing a calculation assuming helium is a classical ideal gas, let's ignore that for now.   We then get 
 \[   \frac{V}{N \Lambda^3} \approx \frac{0.0247896 \, \mathrm{m}^3}{6.02214076 \cdot 10^{23} \times (5.2799291 \cdot 10^{-11} \,  \mathrm{m})^3}  \approx 279663.\]
We thus have
 \[   \ln \frac{V}{N \Lambda^3} \approx \ln 279663 \approx 12.541.\]
 
 As explained \hyperlink{sec:ENTROPY_OF_IDEAL_MONATOMIC_GAS}{earlier}, this means that the logarithm of the number of accessible states of each helium atom would be $12.541$ if it were trapped in its own small box of volume $V/N$.    But density fluctuations contribute $1$ extra nat of entropy per atom.    Thus, the free energy part of the entropy per atom is $13.541$ nats.   On the other hand, the expected energy part of
 the entropy per atom is $\frac{3}{2}$, coming from the atom's 3 momentum degrees of freedom.   The total entropy per atom is thus
\[     \ln \frac{V}{N \Lambda^3}+ 1 + \frac{3}{2}\approx  15.041\] 
nats.

To impress our friends we can convert this to bits: we divide by $\ln 2$ and get about
\[   \frac{15.041}{0.69315}\; \mathbf{\color{darkred} \approx 21.700} \]
bits of unknown information per atom of helium.   

I've kept only 5 significant figures in the later stages of these calculations, since that's how precise the experimental data is.   Next let's compare the final result to experiment!   
 
\vfill \eject \begin{center}\noindent\fbox{\begin{minipage}{30em}

 \begin{center}
\entry{THE ENTROPY OF HELIUM: EXPERIMENT}
\vskip 2em
\noindent
\textbf{\boldmath
The entropy of helium at standard temperature and pressure has been measured to be \\
126.15 joules/kelvin per mole.    
\vskip 1em
One bit of unknown information per atom corresponds to
about $5.7631$ joule/kelvin of entropy per mole.
\vskip 1em
Thus, each atom
of helium at standard temperature and pressure carries about
\[   \frac{126.15}{5.7631} \approx {\color{darkred} 21.889 }\]
bits of unknown information.
}
\end{center} \vskip 1em

\end{minipage}} \end{center} \vskip 1em 


Experimentally, the entropy of helium at standard temperature and pressure is
$126.15$ joules/kelvin per mole.  Converting this to bits per atom we get $21.889$, very close to our theoretical result of $21.700$, but about $0.9\%$ higher.  

There are a couple of possible reasons for this slight discrepancy.  First, while our theoretical
calculation assumed that helium is an ideal gas of noninteracting point particles, this is not true.
The helium atoms interact! 

Second, our computation ignored quantum effects---except for using
Planck's constant to determine the thermal wavelength.   Even for an ideal gas, quantum effects become important when $V/N \Lambda^3$ ceases to be large.  This happens at high densities $N/V$, low temperatures $T$, or for particles of small mass $m$.   Helium has a low mass as molecules of gas go---and our ultimate goal, hydrogen, is even worse.

 Now let's tackle the final summit: hydrogen.   This is a diatomic gas, so it works differently.
 
 \vfill \eject \begin{center}\noindent\fbox{\begin{minipage}{30em}
  
 \begin{center}
\entry{ THE IDEAL DIATOMIC GAS } 

\vskip 1em
\noindent \textbf{\boldmath
In thermal equilibrium, a classical ideal diatomic gas of $N$ indistinguishable 
 molecules of mass $m$ in 
a 3-dimensional box of volume $V$ has expected energy
\[   \langle E \rangle = \frac{5}{2} kNT  \]
and free energy
\[ F = -kT \ln \left( \frac{1}{N!} \frac{V^N}{\Lambda^{3N}} \right)      \]
where $\Lambda =  h/ \sqrt{2 \pi m k T}$ is the thermal wavelength.  
\vskip 1em
Its entropy $S$ is $(\langle E \rangle - F)/T$, so 
\[
\color{darkred}
S = kN\left( \ln \frac{V}{\Lambda^3} \; + \; \frac{5}{2} \right) - k \ln N!
\]
and using Stirling's formula $\ln N! \; \approx\; (\ln N - 1) N $ we get 
\[   \color{darkred}  S \; \approx \; kN \left( \ln \frac{V}{N \Lambda^3} + \frac{7}{2} \right)    \]
}
\end{center} 
\end{minipage}} \end{center} \vskip 1em 

 It's easy to repeat our computation of entropy for a diatomic gas if we recall that the tumbling of the
 molecules add two degrees of freedom to the three for position, giving $\langle E \rangle = \frac{5}{2} k N T$.    Tracking the effects of this change we see the entropy is higher than for a monatomic
 gas.   To be precise, the entropy of a classical ideal diatomic gas is
 \[    S \; \approx \; kN \left( \ln \frac{V}{N \Lambda^3} + \frac{7}{2} \right)  .  \]
 So, it has one more nat of Shannon entropy per molecule than an ideal monatomic gas!
 Let's see how this plays out for hydrogen.   
 
 \vfill \eject \begin{center}\noindent\fbox{\begin{minipage}{30em}
 
 \begin{center}
\entry{THE ENTROPY OF HYDROGEN: THEORY}
\vskip 2em
\noindent
\textbf{\boldmath
Assuming hydrogen is a classical ideal diatomic
gas, its entropy is
\[     S \approx \; kN \left( \ln \frac{V}{N \Lambda^3} + \frac{7}{2} \right)    \]
which corresponds to
\[        \ln \frac{V}{N \Lambda^3} + \frac{7}{2}   \]
 nats of unknown information per molecule. \\ At standard temperature and pressure,
 this gives 15.144 nats or
\[   \frac{15.144}{\ln 2} \color{darkred} \approx 21.848 \]
bits of unknown information per molecule.
}
\end{center} \vskip 1em

\end{minipage}} \end{center} \vskip 1em 


A hydrogen molecule has $m = 3.34706 \cdot 10^{-27}$ kg, so at a temperature $T = 298.15$ K its thermal wavelength is 
\[   \begin{array}{ccl} \Lambda &=& \displaystyle{ \frac{h}{\sqrt{2 \pi m k T}}}  \\
 &\approx& \displaystyle{ \frac{6.62607 \cdot 10^{-34}\mathrm{J\, s}}{\sqrt{2 \pi \times 3.34706  \cdot 10^{-27} \, \mathrm{kg} \times  1.380649 \cdot 10^{-23}\, \mathrm{J/K} \times 298.15 \, \mathrm{K}}} } \\ \\
&\approx&  7.12156 \cdot 10^{-11}\, \mathrm{m}  . 
\end{array}
\]
For a mole of an ideal gas at standard temperature and pressure, 
$N = 6.02214076 \cdot 10^{23}$ and $V \approx 0.0247896\, \mathrm{m}^3$, so 
 \[   \frac{V}{N \Lambda^3} \approx \frac{0.0247896 \, \mathrm{m}^3}{6.02214076 \cdot 10^{23} \times (7.12156 \cdot 10^{-11} \,\mathrm{m})^3}  \approx 113971 \]
 We thus have
 \[   \ln \frac{V}{N \Lambda^3} \approx \ln 113971 \approx  11.644 \]
 
\hyperlink{sec:ENTROPY_OF_IDEAL_MONATOMIC_GAS}{Thanks to our previous work} we know this means that that the logarithm of the number of accessible states of each molecule would be $11.644$ if it were trapped in its own small box of volume $V/N$.     There is also a correction to this simplified picture due to density fluctuations, which gives $1$ extra nat of entropy.    These add up to give the free energy contribution to the entropy per molecule: $12.644$ nats.   This is less than we got for helium.   But the expected energy contribution to the entropy per molecule is larger: we again get $\frac{3}{2}$ nats from the molecule's 3 momentum degrees of freedom, but now we get $1$ extra nat due to its $2$ extra tumbling degrees of freedom.  The total number of nats of unknown information per hydrogen molecule is thus
\[     11.644 + 1 + 1.5 + 1 \approx 15.144 . \]

Finally, the number of bits of unknown information per hydrogen molecule is
\[   \frac{ 15.144}{0.69315} \; \mathbf{\color{darkred} \approx 21.848 } .\]
This is slightly more than for helium, where the number was $21.700$.

As a sanity check, let's do this calculation a different way.   A hydrogen molecule is close to half the mass of a helium atom, so its thermal wavelength should be $\sqrt{2}$ times as large.  In our calculation we're treating $V/N$ as the same for both gases, so hydrogen's $V/N\Lambda^3$ should be $2^{-3/2}$ times as large as that for helium.  Since ultimately we compute bits by taking a logarithm in base 2, this reduces its entropy per molecule by $3/2$ bits.    However, hydrogen's 2 tumbling degrees of freedom increase its entropy per molecule by $1$ nat, or $1/\ln 2$ bits.   We have
\[    - \frac{3}{2} + \frac{1}{\ln 2} \approx -1.5 + 1.443 \approx -0.057.  \]
This suggests that each hydrogen molecule should carry $0.057 $ \emph{fewer} bits of unknown
information than each helium atom.   Why did our more careful calculation say hydrogen should have  about
\[    21.848 - 21.700 \approx 0.148  \]
\emph{more} bits of unknown information per molecule?   What's the mistake?

The slight discrepancy arises solely from the fact that a hydrogen molecule is not exactly half the mass of a helium atom!  It's a bit heavier.  It's actually more like $0.50358$ times the mass of a helium.  This makes its thermal wavelength a bit smaller than our estimate in the last paragraph, which boosts its entropy.   It's nice that such subtleties, ultimately due to nuclear physics, are showing up here.

By the way, all our calculations have been for the most common isotopes of hydrogen and helium: hydrogen whose nucleus consists of a single proton, and helium whose nucleus consists of two protons and two neutrons.  Other isotopes have significantly different mass, and this changes the entropy values significantly.

\begin{puzzle}
Helium has a lighter isotope called helium-3, whose nucleus is made of two protons and just one
neutron.   The mass of helium-3 is $5.00823 \times 10^{-27}$ kg.   If we repeat our calculation of the entropy of helium at standard temperature and pressure, changing only this mass, what value do we get for the bits of entropy per atom of helium-3?
\end{puzzle}

\begin{puzzle}
Hydrogen has a heavier isotope called deuterium, whose nucleus is made of one proton and one neutron.
The mass of a hydrogen molecule made of two deuterium atoms is $3.34449 \times 10^{-27}$ kg.
 If we repeat our calculation of the entropy of hydrogen at standard temperature and pressure, changing only this mass, what do we get for the bits of entropy per molecule of this sort?
\end{puzzle}
 
\vfill \eject \begin{center}\noindent\fbox{\begin{minipage}{30em}
 
 \begin{center}
\entry{THE ENTROPY OF HYDROGEN: EXPERIMENT}
\vskip 2em
\noindent
\textbf{\boldmath
The entropy of hydrogen at standard temperature and pressure has been measured to be \\
130.68 joules/kelvin per mole.    
\vskip 1em
One bit of unknown information per molecule corresponds to
about $5.7631$ joule/kelvin of entropy per mole.
\vskip 1em
Thus, each molecule
of hydrogen at standard temperature and pressure has about
\[   \frac{130.68}{5.7631} \approx {\color{darkred} 22.675 }\]
bits of unknown information.
}
\end{center} \vskip 1em

\end{minipage}} \end{center} \vskip 1em 

Okay, let's compare our theoretical prediction to experiment.  

The experimental figure for the entropy of hydrogen at standard temperature and pressure is
130.68 joules/kelvin per mole, which translates into 22.675 bits per molecule.  This is larger than
our theoretical prediction of 21.848 bits per molecule by about $3.8\%$.   

That's not bad.   We can say we solved our original problem fairly well.   But the percentage error here is about 4 times worse than it was for calculation for helium.  Why is it worse?

I haven't studied this, but I can imagine two reasons.  First, remember that quantum effects kick in when $V/N\Lambda^3$ ceases to be large.  This quantity is a bit smaller for hydrogen
than for helium.  Remember, for helium it was $279663$ at standard temperature and
pressure, while for hydrogen it's $113971$.     But that's still very large, so I imagine quantum
effects are still quite tiny.

Second, hydrogen molecules are not chemically inert like helium atoms, and they're larger,
and diatomic rather than monatomic.  So I'd expect them to interact more, making the ideal gas approximation worse.   This feels like a more plausible explanation for the $3.8\%$
discrepancy.  

\begin{puzzle}
Do research to find more accurate calculations of the entropy of hydrogen
gas.  What are the main sources of error in the calculation we have done here?   
\end{puzzle}

\vfill \eject \begin{center}\noindent\fbox{\begin{minipage}{30em}

\noindent
\begin{center}
\entry{WHERE DID WE GO?}
\end{center}
\textbf{\boldmath
{\color{red} The mystery: why does each molecule of hydrogen have $\sim\!23$ bits of entropy at
standard temperature and pressure?} $\color{darkgreen} \scalebox{1.5}{\checkmark}$
\vskip 1em \noindent
{\color{brickred}
The goal: derive and understand the formula
for the entropy of a classical ideal monatomic gas:
\[   S = kN\left( \ln\frac{V}{N} + \frac{3}{2} \ln k T+ \gamma
\right) \]
including the mysterious constant $\gamma$:
\[  \qquad \qquad \gamma =  \frac{3}{2}\ln \frac{2 \pi m}{h^2} + \frac{5}{2}  \qquad \qquad  \color{darkgreen} \scalebox{1.5}{\checkmark}   \]
}
\noindent
{\color{darkred}
\hskip -0.5 em The subgoal: compute the entropy of a single classical \break particle in a 1-dimensional box:
\[
\qquad \qquad \quad S = k \left(\ln L + \frac{1}{2} \ln kT  + \frac{1}{2} \ln \frac{2 \pi m}{h^2} + \frac{1}{2} \right) \qquad {\color{darkgreen} \scalebox{1.5}{\checkmark}}  \]}
\hskip -0.5 em The sub-subgoal:  explain entropy from the ground up, and
compute the entropy of a classical harmonic oscillator: \break
\[ \quad \qquad S =  k \left( \ln \frac{kT}{\hbar \omega} + 1 \right)  \qquad \qquad  \color{darkgreen} \scalebox{1.5}{\checkmark} \]  
}

\end{minipage}} \end{center} \vskip 1em 

We're done!   Or at least we reached our stated goal.  But there is a lot more to say about entropy.
In a way we've scarcely scratched the surface.   For more on the mathematics of entropy, I recommend
these books:

\begin{itemize}
\item Thomas A.\ Cover and Joy A.\ Thomas, \textsl{Elements of Information Theory}, 
Wiley-Interscience, New York, 2006.
\item  Tom Leinster, \textsl{Entropy and Diversity: the Axiomatic Approach}, Cambridge U. Press, Cambridge, 2021.   Also \href{https://arxiv.org/abs/2012.02113}{free on the arXiv}.
\end{itemize}

\noindent For classical and quantum statistical mechanics, I recommend these:

\begin{itemize}
\item
Frederick Reif, \textsl{Fundamentals of Statistical and Thermal Physics}, Waveland Press, Long Grove, Illinois, 2009. 
\item 
Dirk Ter Haar, \textsl{Elements of Statistical Mechanics}, Elsevier, Amsterdam, 1995.
\end{itemize}

\noindent 
The second one has an intense focus on our friend the box of gas.  And for the principle of maximum entropy, I again recommend this insightful and
opinionated text:

\begin{itemize}
\item E.\ T. Jaynes, \textsl{Probability Theory: the Logic of Science}, Cambridge U.\ Press,
Cambridge, 2003.
\end{itemize}

\vfill \eject \begin{center}\noindent\fbox{\begin{minipage}{30em}

\begin{center}
\entry{THE FIRST LAW OF THERMODYNAMICS} 
\vskip 1em
\textbf{\boldmath
Suppose a system has some measure space $X$ of states \\
with functions called \define{energy} $E \maps X \to \R$ \\
and \define{volume} $V \maps X \to \R$.    
\vskip 1em
Consider probability distributions on $X$ \\
maximizing the Gibbs entropy $S$ subject to constraints \\
on $\langle E \rangle$ and $\langle V \rangle$.    
\vskip 1em
Then as we vary $\langle E \rangle$ and $\langle V \rangle$ we have
\[   \color{darkred} d \langle E \rangle = T d S - P d \langle V \rangle \]
where $T$ is called \define{temperature} and $P$ is called \define{pressure}.}

\end{center} \vskip 1em

 \end{minipage}} \end{center} \vskip 1em
 
I said we were done.  But what kind of course on entropy doesn't cover the
three laws of thermodynamics?  I talked a bit about the \hyperlink{sec:THIRD_LAW}{Third
Law}, but I haven't even mentioned the other two yet.   

Here's why:
this wasn't a course on thermodynamics.   In `classical thermodynamics' there's a tradition of taking concepts such as energy, work and heat as primitive, and treating the laws of thermodynamics
as axioms.   I've instead been explaining a bit of `classical statistical mechanics', where we 
start with probability theory and attempt to \emph{derive} classical thermodynamics.   In this approach the laws of thermodynamics are not fundamental.   They actually look a bit odd: they become results that hold under various conditions, so each one becomes a collection of theorems and conjectures.    

I'll  state versions of the three laws of thermodynamics in the language we've developed here.  But please be aware that my versions are idiosyncratic and will make some people raise their eyebrows.
I'm afraid you'll have to go elsewhere, like Reif's book, to learn these laws in their traditional form!

We've been maximizing entropy subject to a constraint on the expected value of \emph{one} quantity. 
What if we do two---or more?  Everything works the same way, but the fundamental relation between
temperature, energy and entropy, $d \langle E \rangle = T dS$, gets one extra term for each 
constraint.  The resulting equation is a version of the `First Law of Thermodynamics'.

I'll explain the case with one extra constraint.   Suppose we've got a measure space $X$ whose points
are states of some system.  Choose two functions on it.  They could be anything, but let's call them \define{energy} and \define{volume} and write them as  $E \maps X \to \R$ and $V \maps X \to \R$.
These terms are favored because thermodynamics arose in part from the study of steam engines, where 
you've got a cylinder of steam with some energy and some volume.   For any probability distribution 
$p \maps X \to [0,\infty)$, we can write down a formula for its Shannon entropy 
\[                H = - \int_X p(x) \ln p(x) \, dx  \]
 and also the expected values
\[          \langle E \rangle = \int_X E(x) \, dx, \qquad   \langle V \rangle = \int_X V(x) \, dx .\]
Let's not worry now about whether these integrals converge.   

Suppose we only know $\langle E \rangle$ and $\langle V \rangle$, and we are trying to 
choose the `best' probability distribution $p$ with these expected values.
What should we do?  Following the principle of maximum entropy, we seek the probability
distribution $p$ that maximizes $H$ subject to our constraints on $\langle E \rangle$ and $\langle V
\rangle$.   If we do this, we are led to a Lagrange multipliers problem, much as in the \hyperlink{sec:MAXIMIZING_ENTROPY_SUBJECT_TO_A_CONSTRAINT}{the simpler
case} of one constraint.   But now we need two Lagrange multipliers:
let's call them $\beta$ and $\gamma$.    We get this equation:
\[               d H = \beta d \langle E \rangle + \gamma d \langle V \rangle .\]
This is the First Law!  

But this isn't the way physicists usually write it.   To get the First Law in its usual form, first let's switch to using Gibbs entropy $S = kH$, and emphasize the role of energy by solving for $d \langle E \rangle$:
\[             d \langle E \rangle = \frac{1}{k\beta} dS - \frac{\gamma}{\beta} d \langle V \rangle .\]
Then, to simplify the look of this equation, let's introduce
variables called \define{temperature} and \define{pressure}:
\[                 T = \frac{1}{k \beta}, \qquad P = \frac{\gamma}{\beta}.  \]
Now the \define{First Law of Thermodynamics} looks like this:
\[                 d \langle E \rangle = T d S - P d\langle V\rangle .\]
It says that as we move around among probability distributions that maximize entropy subject to constraints on expected energy and volume, the change in expected energy is the sum of two
terms:
\begin{itemize}
\item    \define{heat}, meaning $T dS$
\item   \define{work}, meaning $- P d\langle V \rangle$.
\end{itemize}
For example, if we have a cylinder of steam with pressure $P$ and we increase its expected
volume by a little bit $\Delta \langle V\rangle$, its expected energy goes \emph{down} by about $P \Delta \langle V \rangle$: that's how we understand the minus sign.   In this situation the external world has done an amount of work $-P \Delta \langle V \rangle$ on the cylinder of steam, but most people say the cylinder of steam has done an amount of work $P \Delta \langle V \rangle $ on the external world.

Here are a few puzzles if you want to dig deeper.    In the first two, I ask you to generalize ideas from \hyperlink{sec:MAXIMIZING_ENTROPY_SUBJECT_TO_A_CONSTRAINT}{our earlier work} on maximizing entropy subject to a single constraint.

\begin{puzzle}
\label{puzzle:maximizing_under_two_constraints_1}
Let $X = \{1,\dots, n\}$ and let
$E, V \maps X \to \R$ be two functions whose values at $i \in X$ we call $E_i$ and $V_i$.
Suppose $p$ is a probability distribution maximizing the Shannon entropy $H$ on the surface where
\[                   \langle E \rangle = e, \qquad \langle V \rangle = v ,\]
and also suppose $p_1, \dots, p_n > 0$.  Show that at $p$ we have
\[                 d H = \beta d \langle E \rangle + \gamma d\langle V \rangle  \]
for some $\beta, \gamma \in \R$.   (Hint: first do the case where not all the $E_i$ are equal
and not all the $V_i$ are equal.  This guarantees that $d \langle E \rangle$ and $d \langle V \rangle$
are nonzero.   You can handle the other cases separately.)
\end{puzzle}

\begin{puzzle}
\label{puzzle:maximizing_under_two_constraints_2} 
Under the conditions of Puzzle \ref{puzzle:maximizing_under_two_constraints_1} show
that
\[     p_i = \frac{\exp(-\beta E_i - \gamma V_i)}{\displaystyle{\sum_{i=1}^n \exp(-\beta E_i - \gamma V_i)}}.
\]
\end{puzzle}

\begin{puzzle}
\label{puzzle:maximizing_under_several_constraints} 
Generalize the results of Puzzles \ref{puzzle:maximizing_under_two_constraints_1} and
\ref{puzzle:maximizing_under_two_constraints_2} to the case of any finite number of
constraints.
\end{puzzle}

\begin{puzzle}
Generalize the results of Puzzle \ref{puzzle:maximizing_under_several_constraints} to
the case of a system with a countable infinity of states, or an arbitrary measure space of
states.  You will need to add assumptions to ensure that the sums or integrals converge.
\end{puzzle}

\vfill \eject \begin{center}\noindent\fbox{\begin{minipage}{30em}

\begin{center}
\entry{THE SECOND LAW OF THERMODYNAMICS} 
\vskip 1em
\textbf{\boldmath Suppose a system has some measure space $X$ of states \\
and at any time $t$ there is a probability distribution $p(t)$ on $X$. 
\vskip 1em  
We say the \define{second law of thermodynamics} holds if
\[         t_1 \le t_2 \implies S(p(t_1)) \le S(p(t_2))  \]
\vskip 1em
This seems to be widely true, yet the conditions under which \\
it holds 
are subtle and much-argued.
}

\end{center} \vskip 1em

 \end{minipage}} \end{center} \vskip 1em
 
 The \define{Second Law of Thermodynamics}, as commonly stated, says that the entropy of a closed
system never decreases.  This appears to be a profound fact about our universe.
A huge challenge to physics is to understand where this law comes from.  Can it
be derived from some realistic assumptions?    One problem
is that the laws of classical mechanics are invariant under time-reversal.  Thus, if we 
evolve probability distributions on some space of states according to these laws, for any
probability distribution whose entropy is nondecreasing, there is a time-reversed one
whose entropy is non\emph{increasing}.    

This is called the problem of the \href{https://en.wikipedia.org/wiki/Arrow_of_time}{arrow of time}: briefly, why does the future look so different from the past?  Quantum mechanics makes the problem subtler, but does not provide an easy resolution.   The solution may be that we happen
to live in a universe---a particular solution of the laws of physics---where entropy was very low
at the Big Bang, making it easy for entropy to increase after that.      But if you get ten physicists
in a room and ask them to explain the arrow of time, you are likely to hear ten different opinions.   Thus, I will not attempt to resolve it here.   For more on that, I recommend this book:

\begin{itemize}
\item H.\ D.\ Zeh,  \textsl{The Physical Basis of The Direction of Time}, Springer, Berlin, 2010.
\end{itemize}

Instead, let's see how the Second Law sheds light on the meaning of temperature.
You'll notice that in our course I never talked about systems evolving in time, and I never
talked about two systems interacting: always just a single system.   Now let's imagine
two systems, each in thermal equilibrium, but at possibly different temperatures.  Say the
first has entropy $S_1$, expected energy $\langle E_1 \rangle$ and temperature $T_1$.
As usual, these are related by
\[                     dS_1 = \frac{ d \langle E_1 \rangle}{T_1} .\]
Say the second system works similarly, with
\[                     dS_2 = \frac{ d \langle E_2 \rangle}{T_2} .\]
We can define the total entropy of the two systems by
\[                    S = S_1 + S_2 \]
and the total expected energy by
\[                 \langle E \rangle = \langle E_1 \rangle + \langle E_2 \rangle.\]
Suppose now that the two systems can exchange energy with each other, but in a slow and gentle
way, so we can approximately treat each one as in thermal equilibrium at any moment.  
If no energy flows in or out of the combined system, the total expected energy is conserved, so
\[                \frac{d \langle E \rangle}{dt} = 0 \]
and thus
\[                \frac{d \langle E_1 \rangle}{dt} = - \frac{d\langle E_2 \rangle}{dt}  . \]
What does the Second Law give us in this situation?  It implies
\[                \frac{dS}{dt} \ge 0 \]
or
\[                 \frac{dS_1}{dt} + \frac{dS_2}{dt}  \ge 0. \]
It follows that
\[                 \frac{1}{T_1} \frac{d \langle E_1 \rangle}{dt} +  
\frac{1}{T_2} \frac{d \langle E_2 \rangle}{dt} \ge 0  \]
or 
\[             \frac{1}{T_1} \frac{d \langle E_1 \rangle}{dt} -  
\frac{1}{T_2} \frac{d \langle E_1 \rangle}{dt} \ge 0.  \]
We can rewrite this as
\[        \left(   \frac{1}{T_1} - \frac{1}{T_2} \right) \frac{d \langle E_1 \rangle}{dt} \ge 0.\]
Now suppose both $T_1$ and $T_2$ are positive.   Then we get a remarkable
consequence: \define{as two systems exchange energy, with each
staying in thermal 
equilibrium at every moment, expected energy can only flow from the system
with higher temperature to the system with lower temperature!}    

\vskip 3em
\begin{puzzle}
Suppose one or both of the temperatures $T_1, T_2$ are negative.  How does this conclusion
change?
\end{puzzle}

\vfill \eject \begin{center}\noindent\fbox{\begin{minipage}{30em}

\hypertarget{sec:THIRD_LAW_REVISITED}{}

\begin{center}
\entry{THE THIRD LAW OF THERMODYNAMICS: REVISITED} 
\vskip 1em
\textbf{\boldmath
If a system has countably many states, \\ with just one of lowest energy,  \\
and thermal equilibrium is possible for this system \\ 
for some temperature $T > 0$, \\
then its entropy in thermal equilibrium approaches zero \\
\define{exponentially fast as a function of $1/T$} \\
as $T$ approaches zero from above.
}

\end{center} 

 \end{minipage}} \end{center} \vskip 1em

In our \hyperlink{sec:THIRD_LAW}{earlier work} on the Third Law, we only studied systems with
finitely many states.   \hyperlink{sec:PARTITION_FUNCTION_KNOWS_ALL_REVISITED}{Later}
we saw how to compute entropy from the free energy and expected energy.  This makes it a bit easier 
to handle systems with a countable infinity of 
states.   In the following puzzles, which are only for the most devoted readers, let's use these ideas
to prove and \emph{improve} the Third Law for systems with countably many states.   

\hyperlink{sec:THIRD_LAW}{Earlier} we worked with temperature, but it's cooler to use coolness.   For all the following puzzles, let's
suppose we have a system with a countable infinity of states $n = 1, 2, 3, \dots$ with energies $E_n$.     Also suppose thermal equilibrium is possible for some  $\beta_0 > 0$, i.e., the sum
\[    Z(\beta_0) = \sum_{n = 1}^\infty \exp(- \beta_0 E_n)  \]
converges.   (Our arguments also apply to systems with finitely many states, where this convergence
condition is automatic.)

\begin{puzzle}
\label{puzzle:convergence}
 Show that the system's partition function, expected energy, free energy and entropy are well-defined for all $\beta > \beta_0$.  
\end{puzzle}

\begin{puzzle}
\label{puzzle:energy_shift}
Show that if we add some constant to the energy of each state
\[                      \tilde{E}_n = E_n + c  \]
we get a new `shifted' system whose partition function, expected energy, free energy and entropy 
are related to those of our original system by
\[              \tilde{Z} = \exp(-\beta c) Z, \quad \langle \tilde{E} \rangle = \langle E \rangle + c, 
\quad \tilde{F} = F + c, \quad \tilde{S} = S \]
for all $\beta > \beta_0$.  
\end{puzzle}

Now further suppose that our original system has just one state of least
energy.  \hyperlink{sec:HAGEDORN_TEMPERATURE}{Earlier} we saw that we could
 reindex the states so that $E_1 < E_2 \le E_3 \le \cdots$
and $E_n \to +\infty$.  The same is true of our new shifted system, and let's choose $c = -E_1$
so that the lowest energy of the shifted system is zero.    With this shift we have
\[           0 = \tilde{E}_1 < \tilde{E}_2 \le \tilde{E}_3 \le \cdots  \]
and $\tilde{E}_n \to +\infty$.   

\begin{puzzle}
\label{puzzle:third_law_calculation_1}
Show that for any coolness $\beta \ge \beta_0$ we have
\[    \tilde{Z}(\beta)  - 1 = \sum_{n = 2}^\infty e^{-\beta \tilde{E}_n}  .\]
Using this equation show
\[  |\tilde{Z}(\beta) - 1| \le e^{-(\beta - \beta_0) \tilde{E}_2} |\tilde{Z}(\beta_0) - 1|   \]
and thus
\[             |\tilde{Z}(\beta) - 1| < \mathrm{const} \, e^{-(\beta - \beta_0) \tilde{E}_2} \]
for some constant independent of $\beta$.  
Use the fact that $\tilde{F}(\beta) = - \frac{1}{\beta} \ln \tilde{Z}(\beta)$ to show that
for large enough $\beta$, 
\[           |\tilde{F}(\beta)|  < \mathrm{const} \,   e^{-(\beta - \beta_0) \tilde{E}_2} \]
possibly for a different constant independent of $\beta$.  
Using Puzzle \ref{puzzle:energy_shift}, conclude that
\[         |F(\beta) - E_1|    < \mathrm{const} \,   e^{-(\beta - \beta_0) (E_2 - E_1)} .\]
\end{puzzle}

Voil\`a!  This shows that for a system with countably many states and just one state of lowest energy,
if thermal equilibrium is possible at some positive temperature, then the free energy must approach
this lowest energy  exponentially fast as $\beta \to +\infty$.   Now let's show something similar for the expected energy.  Again we use the shifted system to simplify the calculations.   I'll leave more work to you
this time.

\begin{puzzle}
\label{third_law_calculation_2}
Show that at any coolness $\beta > \beta_0$ we have
\[          \frac{d}{d \beta} \tilde{Z}(\beta) = -\sum_{n=2} \tilde{E}_n e^{-\beta \tilde{E}_n} .\]
Use this to show that $\frac{d}{d \beta} \tilde{Z}(\beta)$ goes to zero exponentially
fast as $\beta \to +\infty$.   Using
\[    \langle \tilde{E} \rangle(\beta) =  - \frac{d}{d \beta} \ln Z(\beta) = - \frac{1}{Z(\beta)} \frac{d}{d \beta} Z(\beta) \]
and Puzzle \ref{puzzle:third_law_calculation_1}, show that $\langle \tilde{E} \rangle(\beta) $ goes to 
zero exponentially fast as $\beta \to + \infty$.    Using Puzzle \ref{puzzle:energy_shift}, conclude that $\langle E \rangle (\beta)$ approaches
$E_1$ exponentially fast as $\beta \to +\infty$.    Finally, since
\[    S = k \beta (F - \langle E \rangle) \]
and both $F$ and $\langle E \rangle$ approach $E_1$ exponentially fast as $\beta \to +\infty$, conclude
that $S$ approaches $0$ exponentially fast as $\beta \to +\infty$.
\end{puzzle}

Let's summarize!   Suppose we have a system with a countable infinity of states and just one
state of lowest energy.  If thermal equilibrium is possible for this system for some $T > 0$,
the \define{Third Law of Thermodynamics} says its entropy in thermal equilibrium 
goes to zero as $T$ approaches zero from above.  But in fact we can say more: for some $a,b > 0$ we have
\[          |S(\beta)| < a e^{- b \beta} \]
for all large enough $\beta$.

\vfill \eject 

\end{document}